\documentclass[twocolumn,floats,floatfix,showpacs,amsmath,amssymb,prd,superscriptaddress,nofootinbib]{revtex4-1}
\usepackage{hyperref}
\usepackage{graphicx}
\usepackage{graphics}
\usepackage{epsfig}
\usepackage{psfig}
\usepackage{bm}
\usepackage{longtable}                 
\usepackage{verbatim}
\usepackage{yfonts}
\usepackage{psfrag}
\usepackage{url}
\usepackage{subfigure}
\usepackage{color}

\newcommand{\be}{\begin{equation}}
\newcommand{\ee}{\end{equation}}
\newcommand{\msun}{M_{\odot}}

\def\beq{\begin{eqnarray}}
\def\eeq{\end{eqnarray}}
\def\f{\frac}

\begin{document}

\title{Reconstructing the massive black hole cosmic history through gravitational waves}

\author{Alberto Sesana} 
\email{alberto.sesana@aei.mpg.de}
\affiliation{Albert Einstein Institute, Am Muhlenberg 1 D-14476 Golm, Germany}
\author{Jonathan Gair} 
\email{jgair@ast.cam.ac.uk}
\affiliation{Institute of Astronomy, University of Cambridge, Cambridge, CB3 0HA, UK}
\author{Emanuele Berti} 
\email{berti@phy.olemiss.edu}
\affiliation{Department of Physics and Astronomy, The University of Mississippi,
  University, MS 38677-1848, USA}
  \affiliation{California Institute of Technology, Pasadena, CA 91109, USA}
\author{Marta Volonteri} 
\email{martav@umich.edu}
\affiliation{Department of Astronomy, University of Michigan, Ann Arbor, MI, USA}

\date{\today}

\begin{abstract} 
The massive black holes we observe in galaxies today are the natural
end-product of a complex evolutionary path, in which black holes seeded in
proto-galaxies at high redshift grow through cosmic history via a sequence of
mergers and accretion episodes.  Electromagnetic observations probe a small
subset of the population of massive black holes (namely, those that are active
or those that are very close to us), but planned space-based
gravitational-wave observatories such as the Laser Interferometer Space
Antenna (LISA) can measure the parameters of ``electromagnetically invisible''
massive black holes out to high redshift. In this paper we introduce a
Bayesian framework to analyze the information that can be gathered from a set
of such measurements. Our goal is to connect a set of massive black hole
binary merger observations to the underlying model of massive black hole
formation.  In other words, {\it given a set of observed massive black hole
  coalescences, we assess what information can be extracted about the
  underlying massive black hole population model.} For concreteness we
consider ten specific models of massive black hole formation, chosen to probe
four important (and largely unconstrained) aspects of the input physics used
in structure formation simulations: seed formation, metallicity ``feedback'',
accretion efficiency and accretion geometry. For the first time we allow for
the possibility of ``model mixing'', by drawing the observed population from
some combination of the ``pure'' models that have been simulated.  A Bayesian
analysis allows us to recover a posterior probability distribution for the
``mixing parameters'' that characterize the fractions of each model
represented in the observed distribution. Our work shows that LISA has
enormous potential to probe the underlying physics of structure formation.
\end{abstract}


\maketitle

\section{Introduction}
\label{s:intro}
In $\Lambda$CDM cosmologies, structure formation proceeds in a hierarchical
fashion \cite{wr78}, in which massive galaxies are the result of several
merging events involving smaller building blocks.  In this framework, the
massive black holes (MBHs) we see in today's galaxies are expected to be the
natural end-product of a complex evolutionary path, in which black holes
seeded in proto-galaxies at high redshift grow through cosmic history via a
sequence of MBH-MBH mergers and accretion episodes \cite{kh00,vhm03}.
Hierarchical models for MBH evolution, associating quasar activity to
gas-fueled accretion following galaxy mergers, have been successful in
reproducing several properties of the observed Universe, such as the present
day mass density of nuclear MBHs and the optical and X-ray luminosity
functions of quasars \cite{haiman2000,kauffmann2000,wyithe2003,
  vhm03,croton2005,m07,dimatteo2008}.

However, only a few percent of galaxies host a quasar or an active galactic
nucleus (AGN), while most galaxies harbor MBHs in their centers, as
exemplified by stellar- and gas-dynamical measurements that led to the
discovery of quiescent MBHs in almost all bright nearby galaxies
\cite{mago98}, including the Milky Way \cite{gil09}.  Our current knowledge of
the MBH population is therefore limited to a small fraction of MBHs: either
those that are active, or those in our neighborhood, where stellar- and
gas-dynamical measurements are possible.  Gravitational wave (GW)
observatories can reveal the population of electromagnetically ``black'' MBHs.

LISA will be capable of accurately measuring the parameters of individual
massive black hole binaries (MBHBs), such as their masses and luminosity
distance, allowing us to track the merger history of the MBH population out to
large redshifts. MBHB mergers have been one of the main targets of the LISA
mission since its conception (see e.g.~\cite{Hughes:2001ya}). Several authors
have explored how spins, higher harmonics in the GW signal and eccentricity
affect parameter estimation and in particular source localization, which is
fundamental to search for electromagnetic counterparts (see, for example, the
work by the LISA parameter estimation taskforce \cite{Arun:2008zn} and
references therein). Most work on parameter estimation has focused on inspiral
waveforms, but ringdown observations can also provide precise measurements of
the parameters of remnant MBHs resulting from a merger, and even test the Kerr
nature of astrophysical MBHs \cite{Berti:2005ys}. Initial studies using
numerical relativity waveforms suggest that mergers will improve the
signal-to-noise ratio of individual events and the localization accuracy of
LISA \cite{McWilliams:2009bg}.

While highly precise measurements for individual systems are interesting and
potentially very useful for making strong-field tests of general relativity,
it is the properties of the set of MBHB mergers that are observed which will
carry the most information for astrophysics. To date, most of the body of work
considering observations of more than one MBHB system has focused on the use
of MBHBs as ``standard sirens'' \cite{Holz:2005df} to probe the expansion
history of the Universe. For a subset of the observed binaries, LISA may have
sufficient angular resolution to make follow-up electromagnetic observations
feasible. If the host galaxy or galaxy cluster can be identified, this will
allow LISA to measure the dark energy equation of state to levels comparable
to those expected from other dark energy missions \cite{Arun:2007hu}.  The
effectiveness of LISA as a dark energy probe is limited by weak lensing
\cite{VanDenBroeck:2010fp}, but this can be mitigated to some extent
\cite{Hilbert:2010am}, and a combination of several GW detections may still
provide useful constraints on the dark energy equation of state
\cite{Babak:2010ej}.

GW observations of multiple MBHB mergers could also be combined to extract
useful astrophysical information about their formation and evolution through
cosmic history. As already mentioned, our access to the MBH population in the
Universe is limited to AGNs or to quiescent MBHs in nearby galaxies. In this
sense we are probing only the tip of the iceberg. Theoretical astrophysicists
have developed a large variety of MBH formation models
\cite{vhm03,kbd04,ln06,bvr06,bv10} that are compatible with observational
constraints. However, the natural lack of observations of faint objects at
high redshifts and the difficulties in measuring MBH spins leave a lot of
freedom in modelling MBH seed formation and mass accretion.  In the last
decade, several authors have employed different MBH formation and evolution
models to make predictions for future GW observations, focusing in particular
on LISA ~\cite{wl03,s04,s05,svh07,eno04,rw05}. This effort has been very
valuable, and established the detection of a large population of MBH binaries
as one of the cornerstones of the LISA mission.

In this paper we tackle the inverse problem:
we do not ask what astrophysics can do for LISA, but what LISA can do for
astrophysics.
In particular, we ask the following question: {\it can we discriminate among
  different MBH formation and evolution scenarios on the basis of GW
  observations only?} More ambitiously, given a set of observed MBHB
coalescences, what information can be extracted about the underlying MBH
population model?  For example, will GW observations tell us something about
the mass spectrum of the seed black holes at high redshift that are
inaccessible to conventional electromagnetic observations, or about the poorly
understood physics of accretion?  Such information cannot be gleaned from a
single GW observation, but it is encoded in the collective properties of the
whole detected sample of coalescing binaries.  In this paper we describe a
method to extract this information in order to make meaningful astrophysical
statements. The method is based on a Bayesian framework, using a parametric
model for the probability distribution of observed events.

The paper is organized as follows. Section \ref{setup} presents the general
framework of our analysis. There we review the MBH formation models considered
in this paper and explain how these models translate into a {\it theoretically
  observable} distribution via a ``transfer function'' that depends (for a
given source) on the detector characteristics and on the assumed model for the
gravitational waveform. We describe how to sample MBH distributions via Monte
Carlo methods, and how to interpret the observations in a Bayesian
framework. In Section \ref{puremodels} we apply these statistical methods to
the problem of deciding, given a set of LISA observations, whether we can
correctly tell the true model from an alternative, for each pair in our family
of MBH formation models. We focus in particular on specific comparisons that
would allow us to set constraints on the main uncertainties in the input
physics: namely, the seed formation mechanism, the redshift distribution of
the first seeds, the efficiency of accretion during each merger and the
geometry of accretion. In Section \ref{mixmodels} we describe how to go beyond
a simple catalog of pure models, either by introducing phenomenological mixing
parameters (designed to gauge the relative importance of different physical
mechanisms in the birth and growth of MBHs) between the pure models, or by
consistently implementing a mixture of different physical assumptions in a
merger tree simulation. In Section \ref{mixresults} we explore how well a
``consistently mixed'' model can be recovered as a superposition of pure
models with the phenomenological mixing parameters. In the conclusions we
point out possible extensions of our work. Appendix \ref{errprop} provides
details of our treatment of errors (due to instrumental noise, uncertainties
in cosmological parameters and weak lensing) in the MBHB observations.
Appendix \ref{app:average} compares parameter estimation calculations that do,
or do not, take into account the orbital motion of LISA. The results suggest
that angle-averaged codes that do {\it not} take into account the orbital
motion may reduce computational requirements in Monte Carlo simulations, while
still providing reasonable estimates of at least some binary parameters.

\section{\label{setup} Massive black holes: formation models,
  gravitational wave observations and their interpretation}

Our goal is to assess the effectiveness of GW observations in extracting
useful information about the evolution of the MBH population in the Universe.
Recent work by Plowman et al.  \cite{plowman09,plowman10} attempted to address
the same question. Here we use different techniques, which improve on their
analysis in several ways. Plowman et al. used the non-parametric
Kolmogorov-Smirnov (KS) test to compare distributions of model parameters
between models. This limited their comparisons to two parameters at a time, as
higher-dimensional KS tests are not known. We instead use a parametric model
by considering the number of events in any given part of parameter space to be
drawn from a Poisson probability distribution. This allows us to use a
Bayesian framework for the analysis. Such a framework will be important for
the analysis of the actual LISA observations once these have been made, and it
can be applied to a parameter space of any dimension.

In Ref.~\cite{Gair:2010bx} we used similar techniques to compare the same four
models that were considered by Plowman et al., which were the models used for
LISA parameter estimation accuracy studies in Ref.~\cite{Arun:2008zn}. In this
paper we go considerably further by considering six additional models, chosen
to probe four key aspects of the input physics used in structure formation
simulations: seed formation, metallicity ``feedback'', accretion efficiency
and accretion geometry. In addition, we consider for the first time ``model
mixing''. The idea is to assume that the observed population is drawn from
some combination of the ``pure'' models that have been simulated.  The
Bayesian framework allows us to recover a posterior probability distribution
for the ``mixing parameters'' that characterize the fraction of each model
represented in the observed distribution. Such an analysis is not possible in
the KS framework. The model mixing analysis is very important, as the real
Universe is most certainly not drawn from any of the idealized models that
currently exist. The mixing parameters will reflect the relative contributions
in the true Universe of the different input physics in the pure models.

For our analysis, we adopt the following strategy:
\begin{itemize}
\item We consider a set of MBH formation and evolution models predicting
  different coalescing MBHB {\it theoretical} distributions (Section
  \ref{mbhpop});
\item To account for detection incompleteness, we filter the distribution
  predicted by each model using a detector ``transfer function'' that produces
  the {\it observed theoretical} distributions under some specific assumptions
  about the GW detector (Section \ref{theodist}). This is basically the
  distribution one would observe assuming an infinite number of detections;
\item We generate Monte Carlo realizations of the coalescing MBHB population
  from one of the models (Section \ref{datasets}) or from a mixture of models
  (Section \ref{mixmodels}), and simulate GW observations of the inspiralling
  binaries, including errors in the source parameter estimation;
\item We then compare (in a statistical sense) the catalog of observed events
  -- including measurement errors -- with the {\it observed theoretical}
  distributions, to assess at what level of confidence we can recover the
  parent model. The statistical methods we use are detailed in Section
  \ref{stats}.
\end{itemize}
In this paper we will consider the Laser Interferometer Space Antenna (LISA)
as an illustrative case, but the strategy outlined above can easily be
generalized to other proposed space-borne GW observatories, such as ALIA,
DECIGO or BBO \cite{BBO,Crowder:2005nr,Ando:2010zz}.

An important caveat is that, for a source at redshift $z$, GW observations do
not measure the binary parameters in the source frame, but rather the
corresponding redshifted quantities in the detector frame. For this reason,
throughout the paper we shall characterize MBHBs via their redshifted
parameters.  Given a MBHB with rest-frame masses $M_{1,r}>M_{2,r}$, the masses
in the detector frame are given by $M_1=(1+z)M_{1,r}$, $M_2=(1+z)M_{2,r}$. In
terms of these masses we can also define (as is the custom in GW physics) the
total mass $M=M_1+M_2$, the mass ratio $q=M_2/M_1$, the symmetric mass ratio
$\eta=M_1M_2/M^2$ and the chirp mass ${\cal M}=\eta^{3/5}M$.  In our
calculations we assume a concordance $\Lambda$CDM cosmology characterized by
$H_0=70$km s$^{-1}$ Mpc$^{-1}$, $\Omega_M=0.27$ and $\Omega_\Lambda=0.73$.

For simplicity we will focus on the inspiral of circular, non-spinning
binaries; therefore, each coalescing MBHB in our populations will be
characterized by only three intrinsic parameters ($z$, $M$ and $q$). In terms
of gravitational waveform modelling, the results presented here can be
considered conservative. Different accretion models may result in different
MBH spin distributions. Including spin in the analysis will provide additional
information that will help to further constrain the physical mechanisms at
work in shaping the MBH population model \cite{Berti:2008af,plowman10}. The
inclusion of the merger/ringdown portion of the signal will increase the
signal-to-noise ratio (SNR) of observed binaries and allow measurements of the
parameters of the merger remnants, providing additional information on the
mechanisms responsible for MBH growth.

In the following subsections we will introduce all the elements and
methodologies relevant to our analysis.

\subsection{\label{mbhpop}Cosmological massive black hole populations}

The assembly of MBHs is reconstructed through dedicated Monte Carlo merger
tree simulations \cite{vhm03} which are framed in the hierarchical structure
formation paradigm. Each model is constructed by tracing the merger hierarchy
of $\sim$200 dark matter halos in the mass range $10^{11}-10^{15}\msun$
backwards to $z=20$, using an extended Press \& Schechter (EPS) algorithm (see
\cite{vhm03} for details).  The halos are then seeded with black holes and
their evolution is tracked forward to the present time. Following a major
merger (defined as a merger between two halos with mass ratio
$M_{h_{2}}/M_{h_{1}}>0.1$, where $M_{h_{2}}$ is the mass of the lighter halo),
MBHs accrete efficiently an amount of mass that scales with the fifth power of
the host halo circular velocity and that is normalized to reproduce the
observed local correlation between MBH mass and the bulge stellar velocity
dispersion (the $M-\sigma$ relation, see \cite{tr02} and references
therein). For each of the simulated halos, all of the binary coalescences that
occur are stored in a catalog. The results for each halo are then weighted
using the EPS halo mass function and are numerically integrated over the
observable volume shell at every redshift to obtain the coalescence rate of
MBHBs as a function of black hole masses and redshift (see, e.g., Fig. 1 in
\cite{svh07}).  We then find the theoretical distribution of (potentially)
observable coalescing binaries by multiplying the rate by the LISA mission
lifetime (here assumed to be three years) to obtain the distribution ${\cal
  N}_i\equiv d^3N_i/dzdMdq$, where the index $i$ labels the MBH formation
model.

In the general picture of MBH cosmic evolution, the MBH population is shaped
by the details of the {\it seeding process} and the {\it accretion
  history}. Both issues are poorly understood, and largely unconstrained by
present observations.  We identify four key factors that have a direct impact
on specific observable properties of the merging MBHB population:
\begin{enumerate}
\item the seed formation mechanism shapes the initial seed mass function;
\item the impact of metallicity on MBH formation determines the redshift
  distribution of the seeds;
\item the accretion efficiency determines the growth rate of MBHs over cosmic
  history;
\item the accretion geometry is crucial in the evolution of the MBH spins.
\end{enumerate}
We explore different formation scenarios by considering two different
prescriptions for each of the elements in the above list, as follows:

\begin{table*}[htb]
\begin{center}
\begin{tabular}{l|cccccc}
\hline
Name & $i$ & Seeding & Metallicity & Accretion model& Accretion geometry & $\bar{N}_i$ $[yr^{-1}]$\\
\hline
{\it VHM-noZ-Edd-co} &1& POPIII & $Z=0$ & Eddington & coherent & 86\\
{\it VHM-noZ-Edd-ch} &2& POPIII & $Z=0$ & Eddington & chaotic & 81\\
{\it VHM-Z-Edd-co} &3& POPIII & all {\it Z} & Eddington & coherent & 108\\
{\it VHM-Z-Edd-ch} &4& POPIII & all {\it Z} & Eddington & chaotic & 113\\
{\it BVR-noZ-Edd-co} &5& Quasistar & $Z=0$ & Eddington & coherent & 26\\
{\it BVR-noZ-Edd-ch} &6& Quasistar & $Z=0$ & Eddington & chaotic & 24\\
{\it BVR-Z-Edd-co} &7& Quasistar & all {\it Z} & Eddington & coherent & 22\\
{\it BVR-Z-Edd-ch} &8& Quasistar & all {\it Z} & Eddington & chaotic & 29\\
{\it BVR-noZ-MH-co} &9& Quasistar & $Z=0$ & Merloni \& Heinz & coherent & 33\\
{\it BVR-noZ-MH-ch} &10& Quasistar & $Z=0$ & Merloni \& Heinz & chaotic & 33\\
\hline				
\end{tabular}
\end{center}
\caption{The ten ``pure'' MBH population models considered in this paper. For
  convenience, in the following we will identify models by the integer, $i$, 
  listed in the second column. In the last column, $\bar{N}_i$ denotes the
  predicted coalescence rate.}
\label{tabI}
\end{table*}

\begin{enumerate}

\item{\it The seed formation mechanism}.  Two distinct families of models have
  become popular in the last decade, usually referred to as ``light'' and
  ``heavy'' seed models. Here we consider two different scenarios
  representative of the two possibilities. (i) The {\it ``VHM''} model;
  developed by Volonteri, Haardt \& Madau \cite{vhm03}, this model is
  characterized by light seeds ($M\sim 100\msun$), which are thought to be the
  remnants of Population III (POPIII) stars, the first generation of stars in
  the Universe \cite{mr01}.  (ii) The {\it ``BVR''} model; proposed by
  Begelman, Volonteri \& Rees \cite{bvr06}, this model belongs to the family
  of ``heavy seed'' models. Bar within bar instabilities \cite{sfb89}
  occurring in massive protogalactic disks trigger gas inflow toward the
  center, where a ``quasistar'' forms.  The core of the quasistar collapses
  into a seed black hole that efficiently accretes from the quasistar
  envelope, resulting in a final seed black hole with mass $M\sim$ few $\times
  10^4\msun$.

\item{\it Metallicity ``feedback''}.  Both of the black hole formation models
  described above require that a large amount of gas is efficiently
  transported to the halo center. The gas inflow has to occur on a timescale
  that is shorter than that of star formation, to avoid competition in gas
  consumption and disruption of the inflow process by supernovae explosions.
  It has been suggested that metal-free conditions are conducive to efficient
  gas inflow, as fragmentation is inhibited \cite{bl03}. If fragmentation is
  suppressed, and cooling proceeds gradually, the gaseous component can cool
  and contract before many stars form.  The gas metallicity $Z$ is therefore
  an important environmental factor to take into account, and we consider two
  cases.  (i) {\it ``noZ''} models; black hole seeding is assumed to be
  efficient at zero metallicity only, with a sharp threshold in cosmic time.
  In these models, seeds form at very high redshift ($20>z>15$).  (ii) {\it
    ``Z''} models; efficient seed formation occurs also at later times. Here
  we treat POPIII star and quasistar black hole formation differently. We
  still assume that POPIII stars can form only out of metal-free gas, but we
  track the probability that a halo at late times is still metal-free by
  adopting the metal enrichment models developed in \cite{s03}.  For the case
  of quasistars, instead, we drop the assumption of zero metallicity.  This
  choice is motivated by recent high-resolution numerical simulations of
  gas-rich galaxies at solar metallicities (e.g. \cite{mayer10}), which show
  that bar within bar instabilities can drive a significant amount of gas to
  the central nucleus before star formation quenches the inflow. These models
  are characterized by seed formation also at later times, in metal enriched
  halos. See \cite{bv10} for full details on the model and its implementation.

\item{\it The accretion efficiency}. MBHs powering AGNs exhibit a broad
  phenomenology; they accrete at different rates, with different efficiencies
  and luminosities (see \cite{mh08} and references therein). In the absence of
  a solid coherent theory for describing the accretion process, several toy
  models are viable, and we consider two of these models. (i) {\it ``Edd''}
  accretion model; the easiest possible recipe is to assume that accretion
  occurs at the Eddington rate, parametrized through the Eddington ratio
  $f_e$ (we take $f_e=0.3$ in our models).  (ii) {\it ``MH''} accretion model;
  we also use a more sophisticated scheme combining low and high accretion
  rates, as described by Merloni \& Heinz \cite{mh08}.

\item{\it The geometry of accretion}.  Standard accretion disks are unstable
  to self gravity beyond a few thousands of Schwarzschild radii
  \cite{tom64}. It is therefore not guaranteed that the supply of gas to the
  central black hole will be continuous, smooth and planar. We consider two
  different scenarios. (i) Coherent accretion ({\it ``co''} models); the flow
  of material that feeds the black hole is assumed to be continuous, smooth
  and planar. Accretion is a single steady episode lasting about a Salpeter
  time. (ii) Chaotic accretion ({\it ``ch''} models); in this scenario,
  proposed by King \& Pringle \cite{kp06}, a single accretion event is made of
  a collection of short-lived accretion episodes, and the angular momentum of
  each accreted matter clump is randomly oriented.  These accretion models
  primarily lead to different expectations for the black hole spins:
  intermediate-high, $a \sim 0.6-0.9$, in the coherent case; low, $a < 0.2$,
  in the chaotic case \cite{Berti:2008af}.  In this work we ignore black hole
  spin in the modelling of gravitational waveforms, and therefore we do not
  assess the impact of spin measurements in resolving different MBH formation
  scenarios. However, the accretion prescription also leaves an imprint on the
  component masses. The models assume that the mass-to-energy conversion
  efficiency, $\epsilon$, depends on black hole spin only, so the two models
  predict different average efficiencies of $\sim 20\%$ and $\sim 10\%$,
  respectively. The mass-to-energy conversion directly affects mass growth,
  with high efficiency implying slow growth, since for a black hole accreting
  at the Eddington rate the black hole mass increases with time as
\begin{equation}
M(t)=M(0)\,\exp\left(\frac{1-\epsilon}{\epsilon} \frac{t}{t_{\rm Edd}}\right)\,,
\end{equation}
where $t_{\rm Edd}=0.45$Gyr. 
The ``coherent'' versus ``chaotic'' models thus allow us to study how
different growth rates affect LISA observations.
\end{enumerate} 

By choosing two different prescriptions for each of the four pieces of input
physics listed above we built ten different MBH population models, which are
summarized in Table \ref{tabI}.  We shall refer to these models as ``pure'',
in the sense that we do not mix different recipes for seed formation and
accretion history (e.g., accretion is either coherent or chaotic, etc.). We
will consider ``mixed'' models in Section \ref{mixmodels}.  It is worth
emphasizing that all of these models successfully reproduce various properties
of the observed Universe, such as the present-day mass density of nuclear MBHs
and the optical and X-ray luminosity functions of quasars. GW observations may
therefore provide an invaluable tool to constrain the birth and growth of
MBHs, particularly at high redshift.

\begin{figure*}[htb]
\begin{tabular}{ccc}
\includegraphics[scale=0.33,clip=true]{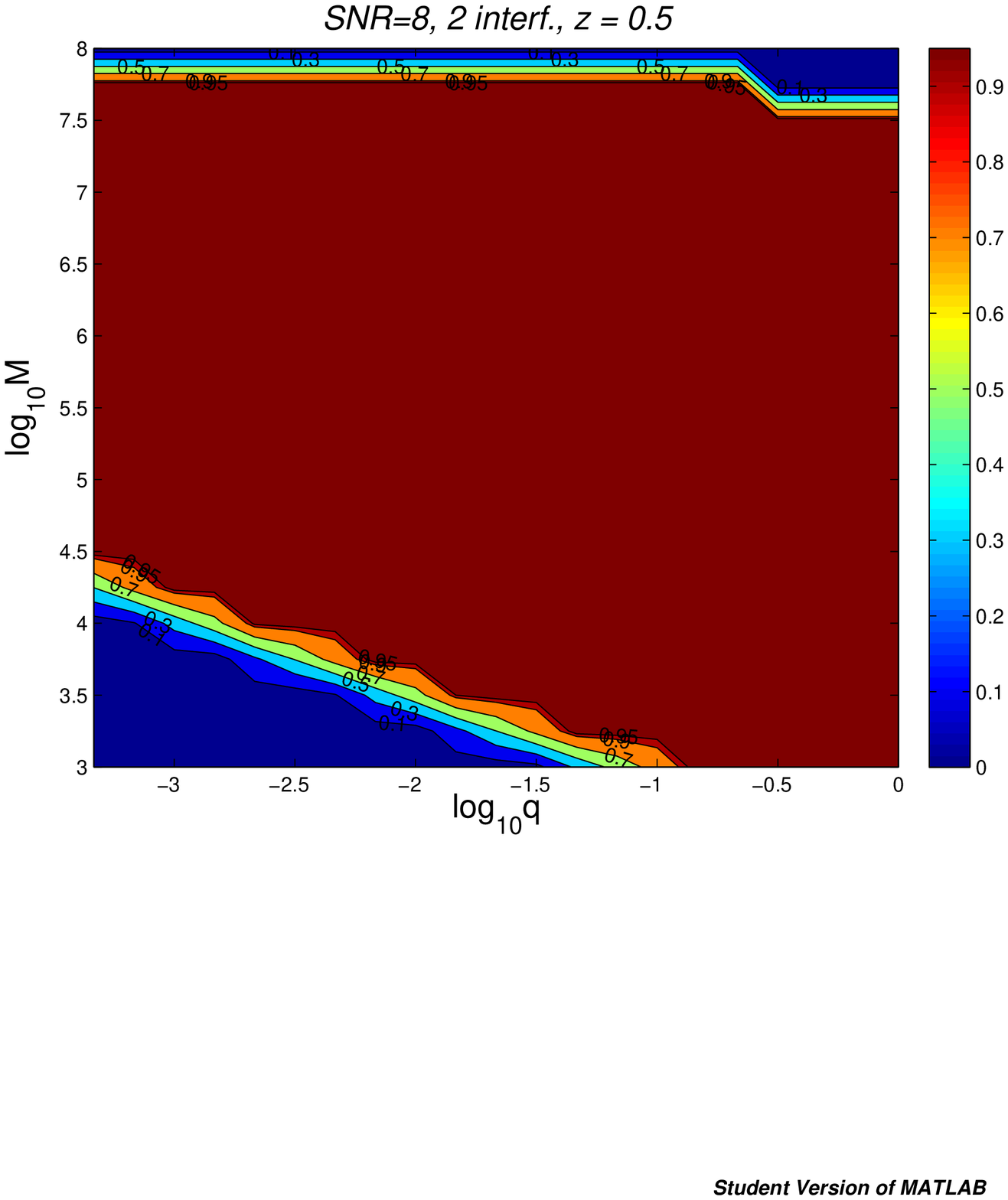}&
\includegraphics[scale=0.33,clip=true]{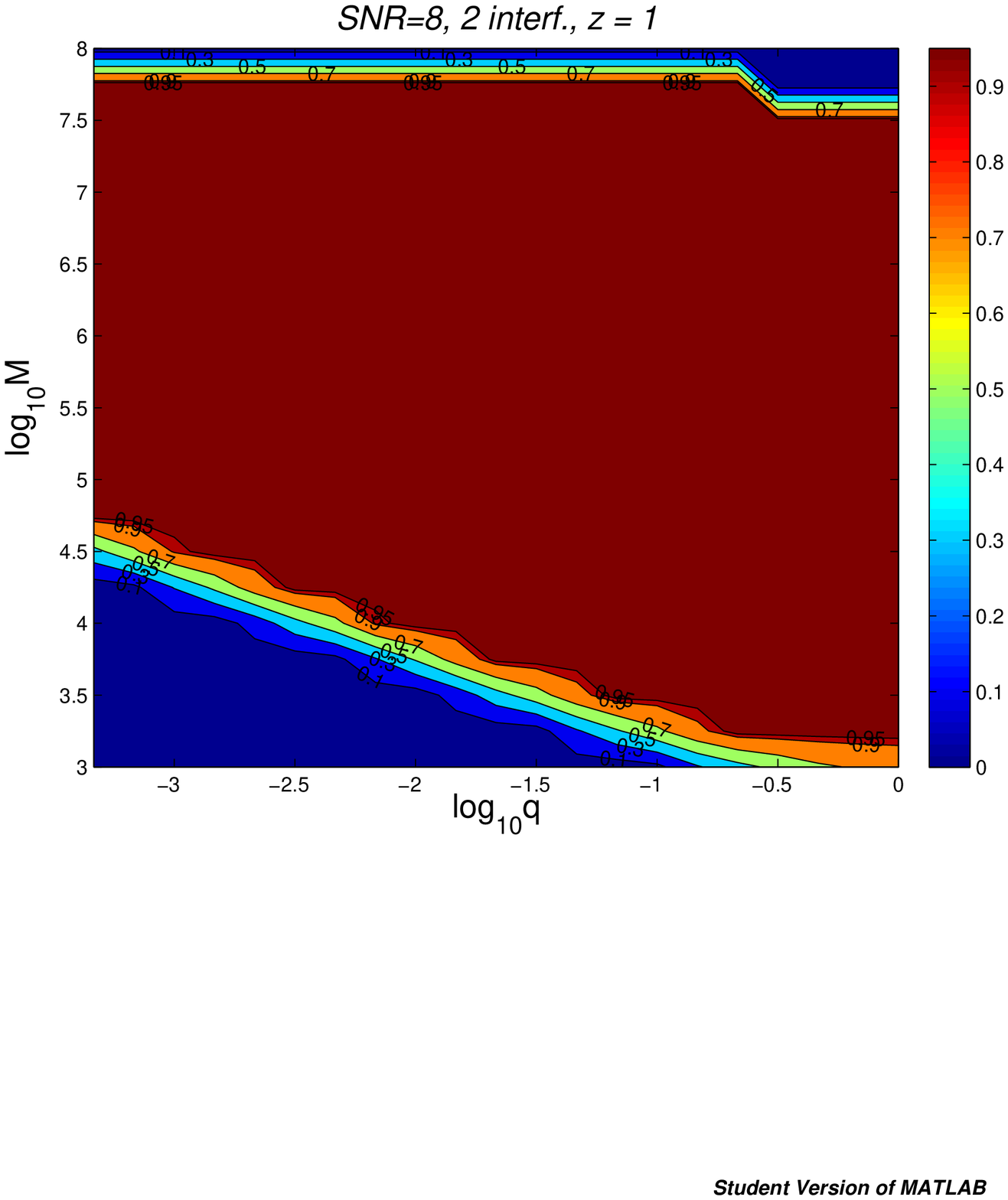}&
\includegraphics[scale=0.33,clip=true]{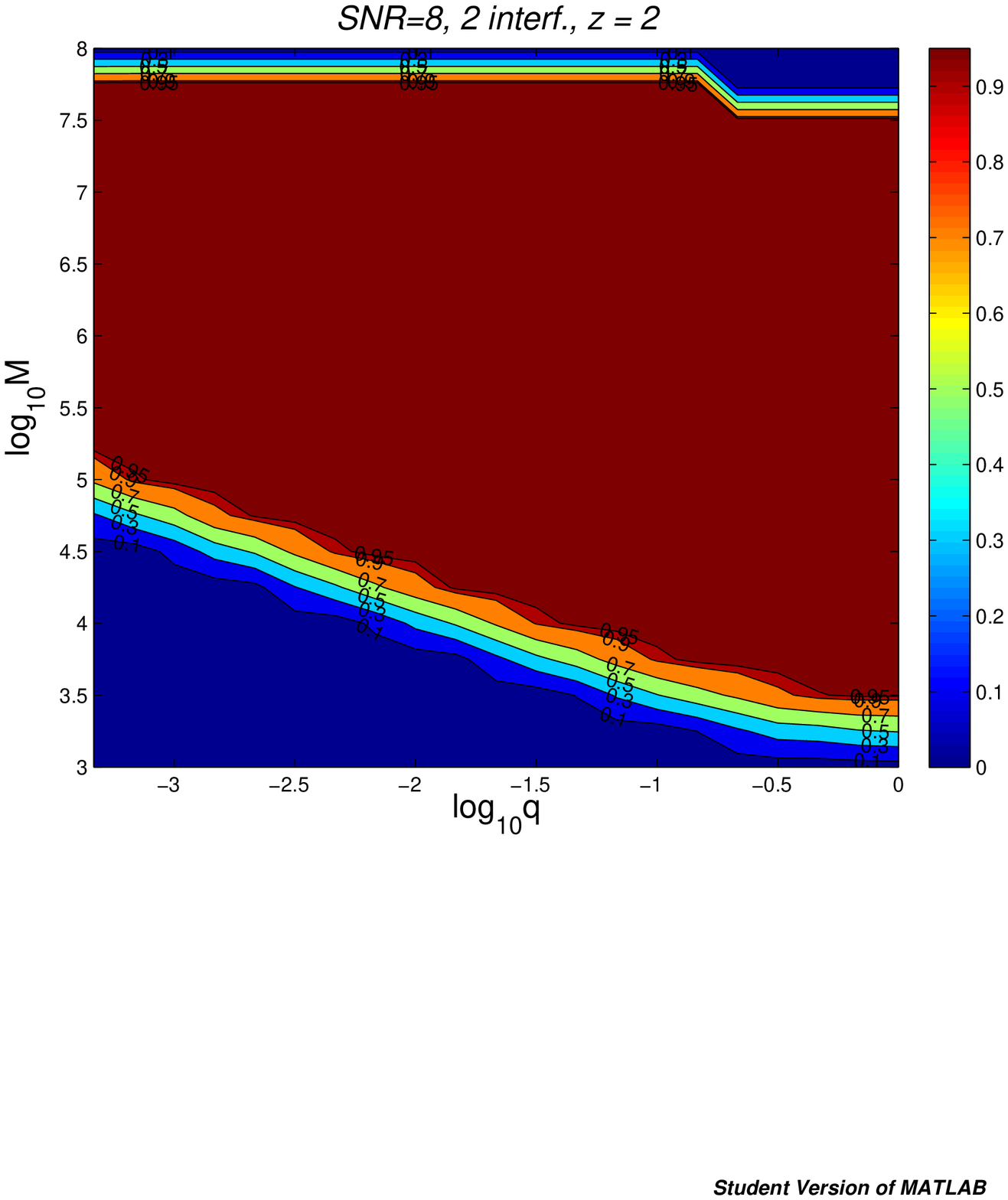}\\
\includegraphics[scale=0.33,clip=true]{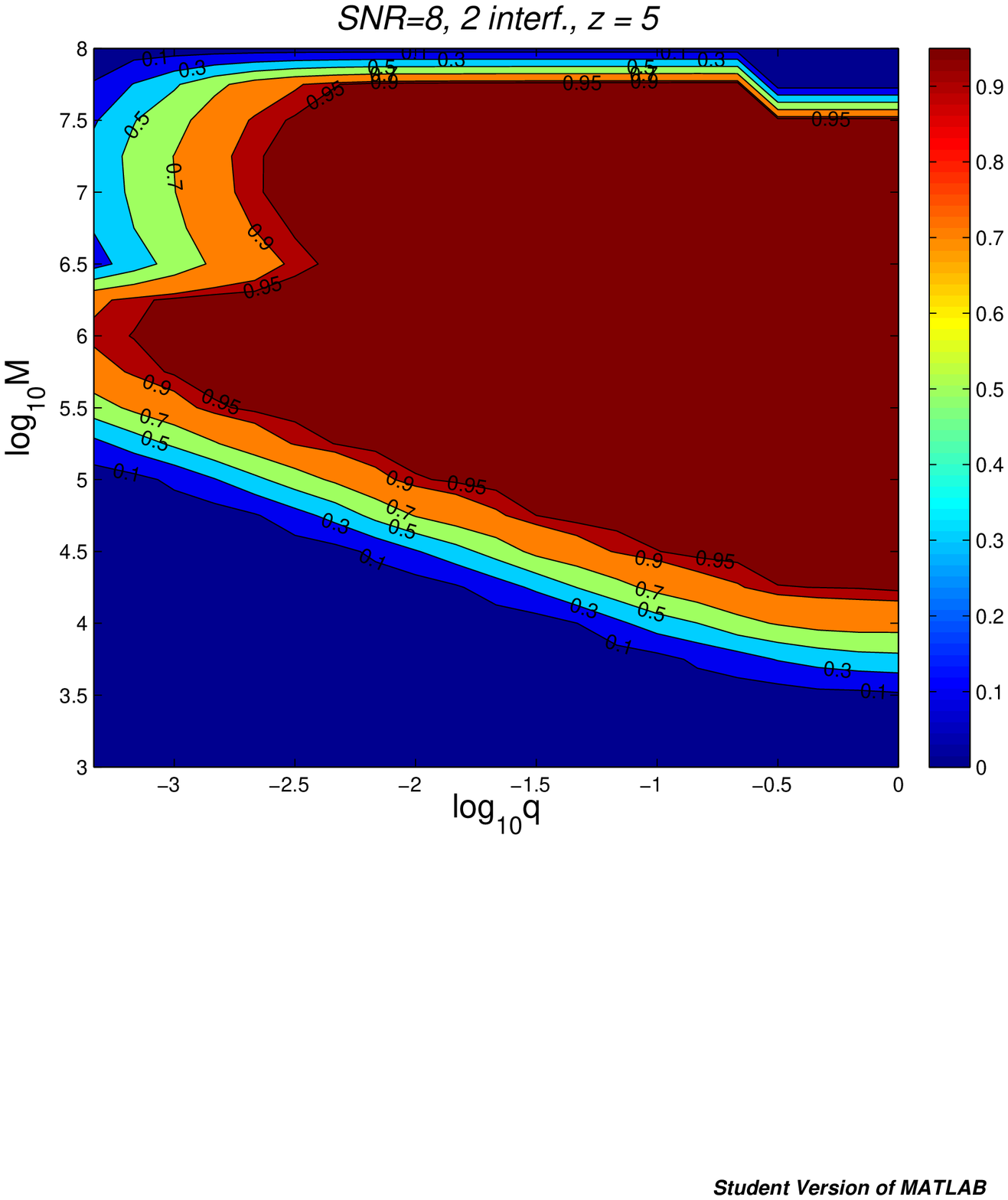}& 
\includegraphics[scale=0.33,clip=true]{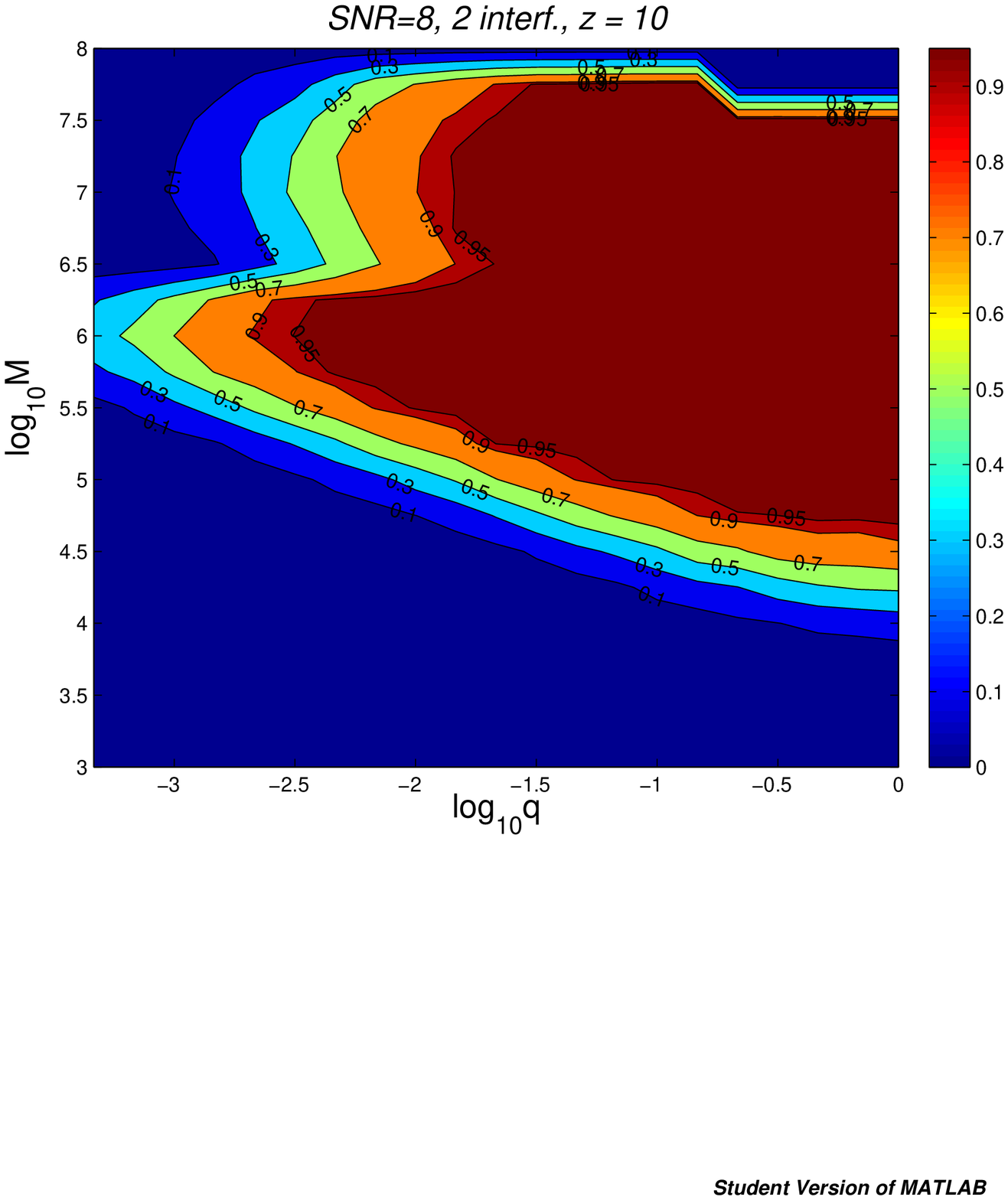}&
\includegraphics[scale=0.33,clip=true]{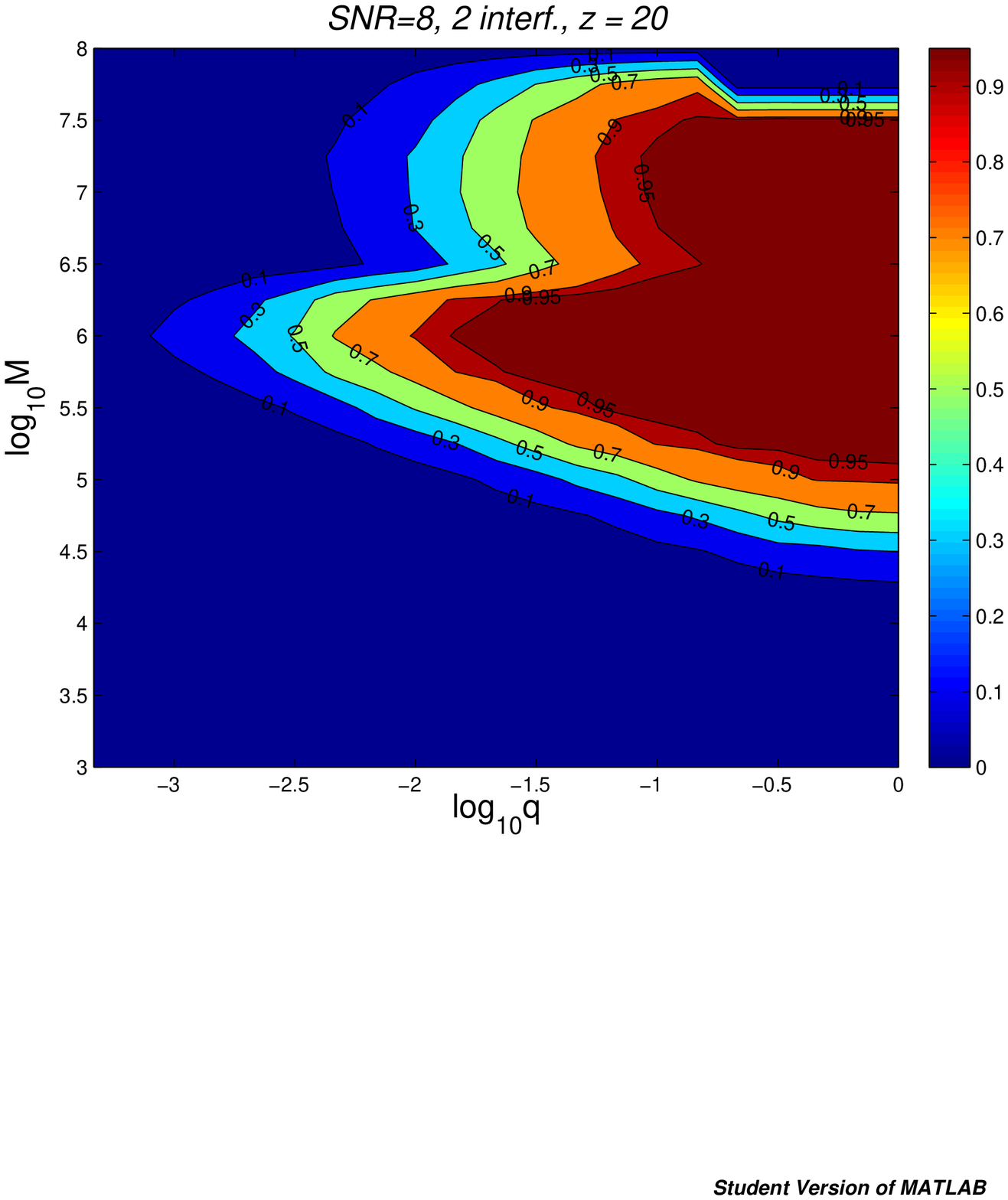}\\
\end{tabular}
\caption{Transfer function for the case $\rho_{\rm thr}=8$ and observation
  with two interferometers, for $z=0.5,\,2,\,5,\,10,\,15,\,20$.
\label{Tr83yrs2int} }
\end{figure*}

\subsection{\label{theodist}Theoretically observable distributions: the
  transfer function}

In order to compare a set of observed events to a given MBH population model,
we must map the coalescence distribution predicted by the model to a {\it
  theoretically observable} distribution which takes into account the
``incompleteness'' of the observations resulting from the limited sensitivity
of any given GW detector. This information can be encoded in a {\it transfer
  function} $T(z,M,q)$, that depends only on the detector characteristics and
on the gravitational waveform model.

We model the detector and the gravitational waveform following
Refs.~\cite{Berti:2004bd,Berti:2005qd}. The detector response is modelled
following Cutler \cite{Cutler:1997ta}: the three-arm LISA constellation is
thought of as a superposition of a pair of linearly independent two-arm
right-angle interferometers, and we can estimate the effect of ``descoping
options'' or a failure on one satellite by assuming that only one of the two
detectors is operational. The MBHB inspiral signal is modelled using the
restricted post-Newtonian approximation, truncating the GW phasing at second
post-Newtonian order -- i.e., at order $(v/c)^4$, where $v$ is the binary
orbital velocity. We also limit our analysis to circular inspirals of
nonspinning MBHs and neglect contributions to the observable signal that come
from higher harmonics in the inspiral signal and from the (gravitationally
loud) merger/ringdown phase.  The latter assumption significantly
underestimates the energy carried in the GWs \cite{Berti:2007fi}, the SNR of
the signal \cite{McWilliams:2010eq} and the accuracy in estimating the source
parameters \cite{McWilliams:2009bg}. From the point of view of studying MBH
populations, it also means that we discard all information on the mass and
spin distribution of MBHs formed as a result of each merger
\cite{Berti:2005ys}. In this sense, our assessment of the potential of LISA to
constrain MBH formation models should be considered conservative.

An important advantage of working in the frequency domain and of adopting a
simplified waveform model is that we can sample the three-dimensional space
$(z,M,q)$ by fast Monte Carlo simulations using an adaptation of the {\sc
  Fortran} code described in \cite{Berti:2004bd}. Typically, we can estimate
SNRs and parameter estimation errors of $\sim 10^6$ binaries in one day on a
single processor.  This would not be possible with a more complex time-domain
code including spin dynamics, such as that used in \cite{plowman10}.  We
consider a $21\times21\times21$ three-dimensional grid spaced logarithmically
in the intervals $q\in [10^{-3},1]$, $M \in [10^3,10^8]$, and approximately
linearly in $z$ (namely, we consider $z=0.5$ and then all values of
$z=1,\dots,20$ in steps of $\Delta z=1$), for a total of 9261 points.  At each
point, we generate 1000 binaries assuming random position in the sky and
orientation, random phase at coalescence, and coalescence time $t_c$ in the
range [0,~3yr] (i.e., we consider only events that coalesce during the LISA
mission).  
The GW signal is calculated in the Fourier domain in the stationary phase
approximation.  Our statistical analysis, which will be discussed in
section~\ref{stats}, includes parameter measurement errors, which are modelled
as described in Appendix~\ref{errprop}. The modelling of errors due to
instrumental noise relies on the computation of the so-called Fisher
information matrix \cite{Vallisneri:2007ev}. For each system we compute the
Fisher matrix and its inverse by means of the LU decomposition, as described
in \cite{Berti:2004bd}.  The accuracy of the inversion is usually worse
for certain values of the intrinsic parameters and of the angular
position/orientation of the binary. We discard ``bad'' Fisher matrix
inversions by monitoring a quantity $\epsilon_{\rm inv}$, defined as
\be 
\epsilon_{\rm inv}=
\max_{i,j}
|I^{\rm num}_{ij}-\delta_{ij}|\,, 
\ee 
where $I^{\rm num}_{ij}$ is the ``numerical'' identity matrix obtained by
multiplying the inverse matrix by the original, and $\delta_{ij}$ is the
standard Kronecker delta symbol \cite{Berti:2004bd}. We set a maximum
tolerance of $\epsilon_{\rm inv}=10^{-3}$ to accept the inversion. For the
accepted events, we compute the SNR $\rho$ and then we define the transfer
function as
\be
T(z,M,q)=\frac{N(\rho>\rho_{\rm thr})}{N}\,, 
\ee 
where $N=1000$ is the number of successful matrix inversions at any given grid
point and $N(\rho>\rho_{\rm thr})$ is the number of binaries fulfilling the
condition $\rho>\rho_{\rm thr}$, where $\rho_{\rm thr}$ is a pre-specified SNR
threshold.

We consider four transfer functions $T_j(z,M,q)$ ($j=1,2,3,4$) according to
the following prescriptions: (1) one interferometer, $\rho_{\rm thr}=8$; (2)
one interferometer, $\rho_{\rm thr}=20$; (3) two interferometers, $\rho_{\rm
  thr}=8$; (4) two interferometers, $\rho_{\rm thr}=20$.  The chosen
thresholds correspond (roughly) to the minimum SNR for which we expect to be
able to claim a confident detection ($\rho_{\rm thr}=8$) and the minimum SNR
for which we expect to obtain a decent accuracy in estimating the parameters
of the source ($\rho_{\rm thr}=20$). Note that, by definition, $0\leq T(z,M,q)
\leq 1$.

Examples of $T_3(z,M,q)$ in the $(M,q)$ plane at different redshifts are shown
in Figure~\ref{Tr83yrs2int}. As expected, the transfer function is close to
unity in the whole of the $(M,q)$ plane at low redshifts, but a smaller number
of events are observable as we consider binaries coalescing at higher
redshifts. When we remember that high redshifted masses correspond to low
observation frequencies, it is easy to understand that the characteristic
shape of the contour plots for large redshift (say, $z=20$) reflects the shape
of the LISA sensitivity curve (cf. Figure 1 of Ref.~\cite{Berti:2004bd}).
More details of the calculation of SNRs and parameter estimation errors in the
three dimensional space $(M\,,q\,,z)$ are given in Appendix \ref{app:average}.

The transfer functions are coupled to the event distributions predicted by the
models to obtain the {\it theoretically observable} distributions $N_T(z,M,q)$
for each model under the different assumptions on the transfer function;
namely,
\be
N_{T_{i,j}}(z,M,q)=\frac{d^3N_i}{dzdMdq}\times T_j(z,M,q)\,, 
\ee
where $i$ labels the MBH population model being considered and $j$ labels the
assumed detector specifics\footnote{Note that, in principle, the transfer
  function may depend on a third index $k$, which labels the waveform model
  used for matched filtering. We do not consider this problem here, but the
  impact of waveform models on constraining the MBH population is an important
  topic for future study.}.  These are the distributions which should be
compared to simulated observed catalogs of MBHBs.

\begin{figure}[thb]
\includegraphics[scale=0.43,clip=true]{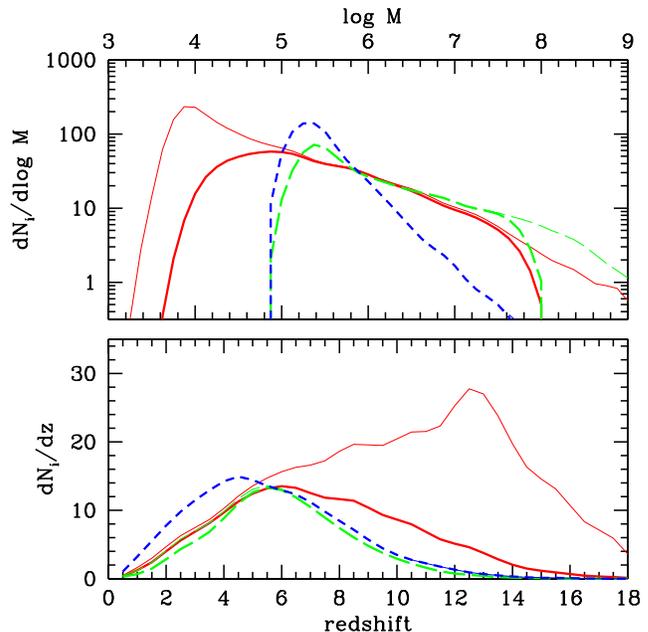}
\caption{Examples of the marginalized distributions $dN_i/dM$ (upper panel)
  and $dN_i/dz$ (lower panel) predicted by different MBH formation models. In
  each panel we plot the following models: {\it VHM-noZ-Edd-co} (solid red
  lines); {\it BVR-noZ-Edd-ch} (long-dashed green lines); {\it BVR-noZ-MH-co}
  (short-dashed blue lines). Thin lines represent the coalescence
  distributions predicted in three years, while thick lines represent the {\it
    theoretically observable} distributions after the transfer function
  $T_3(z,M,q)$ has been applied, namely $N_{T_{i,3}}(M)$ and $N_{T_{i,3}}(z)$
  (see text for details).  }
\label{fig:dist}
\end{figure}

For illustration, in Figure \ref{fig:dist} we compare the marginalized
distributions
\begin{align}
&\f{dN_i}{dM}=\int dz \int dq \, \f{d^3N_i}{dzdMdq}\quad{\rm and}
\nonumber\\
&\f{dN_i}{dz}=\int dM \int dq \, \f{d^3N_i}{dzdMdq}
\end{align}
(thin lines) with the corresponding marginalized distributions
\begin{align}
&N_{T_{i,3}}(M)= \int dz \int dq \, N_{T_{i,3}}(z,M,q)\quad{\rm and}
\nonumber\\
&N_{T_{i,3}}(z)=\int dM \int dq \, N_{T_{i,3}}(z,M,q)
\end{align}
(thick lines). Note that for some of the heavy seed models (namely, the
short-dashed line corresponding to model {\it BVR-noZ-MH-co}) the two curves
perfectly overlap: in these cases LISA observations do not miss events, i.e.
they are ``complete''.

\subsection{\label{datasets}Synthetic Monte Carlo catalogs}
To simulate LISA observations we perform 1000 Monte Carlo samplings of the
$d^3N_i/dzdMdq$ distribution predicted by each model, producing 1000 catalogs
of coalescing binaries over a period of three years.  In each catalog, the
source position in the sky and the direction of the orbital angular momentum
are assumed to be uniformly distributed. The phase at coalescence $\Phi_c$ and
the coalescence time $t_c$ are randomly chosen in the range [0, $2\pi$] and
[0, 3yrs], respectively.  Each waveform is described by the set of parameters
\be
\Theta=(\log A, \log {\cal M}, \log \eta,t_c,\Phi_c,
\theta_S,\Phi_S,\theta_L,\Phi_L),
\ee
where $(t_c,\Phi_c)$ are the phase and time of coalescence,
$(\theta_S,\Phi_S)$ represents the source location in ecliptic coordinates,
$(\theta_L,\Phi_L)$ give the orientation of the orbital angular momentum of
the binary and the GW amplitude of the signal
\be
A\propto \frac{{\cal M}_z^{5/6}}{D_L}\,,
\ee
where 
\be\label{DLz}
D_L=\frac{1+z}{H_0}\int_0^z
\frac{dz'}{(1+z')^2[\Omega_M(1+z')^3+\Omega_\Lambda]^{1/2}}
\ee
is the luminosity distance to the source. Our theoretical distributions are
not functions of $(M,\eta,D_L)$, but rather functions of ($M,q,z$), and the
mapping between the two sets of parameters is given in Appendix~\ref{errprop}.

\begin{figure}[htb]
\includegraphics[scale=0.43,clip=true]{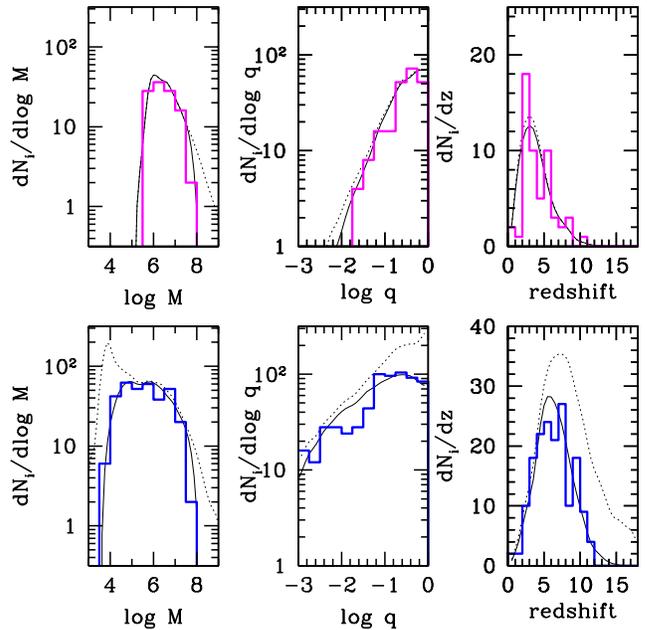}
\caption{Examples of Monte Carlo generated datasets. The left panels show the
  $dN_i/dM$ distributions, the central panels show the $dN_i/dq$ distributions
  and the right panels show the $dN_i/dz$ distributions.  The upper panels
  refer to model {\it BVR-Z-Edd-co}, the lower panels to model {\it
    VHM-Z-Edd-co}.  In each panel the dotted curves represent the theoretical
  distributions, and the solid curves represent the {\it theoretically
    observable} distribution filtered with the transfer function $T_3$. The
  thick histograms show one Monte Carlo realization of the theoretical
  distribution, as observed by LISA, under the assumption of two operational
  interferometers and $\rho_{\rm thr}=8$.  }
\label{fig:mc}
\end{figure}

LISA measurements will yield a set of data $\{D_k\}$, $k=1,\dots,N$, where $N$
is drawn from a Poisson distribution with mean $\bar{N}_i$ coincident with the
theoretical number of events predicted by the model we consider (cf. Table
\ref{tabI}). Each element in the set is described by $(\bar{z},\sigma_z;
\bar{M},\sigma_{M}; \bar{q}, \sigma_q)$, where $\bar{z}$, $\bar{M}$, $\bar{q}$
are the true parameters of the system and $\sigma_z$, $\sigma_{M}$, $\sigma_q$
are the diagonal elements of the variance-covariance matrix describing the
measurement errors. The latter are computed as described in
Appendix~\ref{errprop} and include contributions from instrumental noise, from
uncertainties in cosmological parameters and from weak lensing. We approximate
the covariance matrix as diagonal since this is conservative, and the
covariances are generally small. Strictly speaking we are not justified in
ignoring the large covariance between any two mass parameters (say, $M$ and
$\eta$); however the errors on the mass parameters are always negligible when
compared with errors on luminosity distance, cosmological parameters and weak
lensing (see Appendix \ref{errprop}). The probability density function for the
measured source parameters is then a multivariate Gaussian with these standard
deviations, centred at the true source parameters. As discussed in the next
section, the errors can be folded into the analysis in two ways. The one we
adopt is to construct the {\it theoretically observable distribution}, ${\cal
  N}_{i,j}(z,M,q)$ as described in Section \ref{theodist}, by spreading each
source over multiple bins according to the Gaussian probability distribution
for the measurement errors. We then construct data sets by assigning a unique
set of observed parameters to each event that is equal to the true parameters,
plus a random error drawn from the same probability distribution.  For each
``pure'' MBH population model (label ``$i$'') and LISA transfer function
(label ``$j$''), we produce 1000 of these ``observed'' datasets
$\{D_k\}_{i,j}$, to compare to the {\it theoretically observable}
distributions. Examples of Monte Carlo generated datasets are shown in Figure
\ref{fig:mc}.

Throughout our study, we will assume $T_{\rm obs}=3$~yrs as the fiducial LISA
mission lifetime.  However, it is interesting to study how the performance of
LISA improves as a function of the duration of the data stream used in the
analysis. This problem could be particularly relevant if, as expected, there
are gaps in the LISA data stream. For this reason we will consider increasing
observation times, $T_{\rm obs}$, of 3 months, 6 months, 1 year, 18 months, 2
years and 3 years, respectively. To construct these reduced datasets, we just
pick events from the catalog that coalesce at $t_c<T_{\rm obs}$, and then
renormalize the theoretical distributions by a factor $T_{\rm obs}/3$yr.  In
doing this, we ignore sources that coalesce outside the reduced observation
time, but which may have enough SNR to be detected in the shorter data
segment. This is conservative since we are effectively choosing only to
include the coalescing sources in our analysis. However, for MBHBs, unlike the
EMRI case (see Ref.~\cite{emrimodsel}), almost all of the source SNR (and,
consequently, the accuracy in the determination of $M$, $q$, and $z$) is
accumulated in the last month of inspiral, and so there would not be a great
deal to gain by including these sources in the analysis.
 
\subsection{\label{stats}Statistical analysis tools}
In this work we will adopt a Bayesian approach to model selection and
parameter estimation. This requires a parametric model for the distribution of
events that LISA will observe. A particular astrophysical model of MBH
formation cannot predict the actual number of events that will occur during
the LISA mission, as the mergers will occur stochastically, but instead
predicts the rate at which events with particular parameters occur. Assuming
random start times, the number of events, $n_i,$ that will be seen in a
particular bin, ${\cal B}_i$, in parameter space will be drawn from a Poisson
probability distribution with parameter $r_i$ equal to the rate integrated
over the bin:
\be
p(n_i) = \frac{(r_i)^{n_i} {\rm e}^{-r_i}}{n_i!}\,.
\label{poiss}
\ee
If we divide the parameter space up into a certain number of bins, $K$ say,
then the information that comes from LISA (the {\it data} {$D$) is the number
  of events in each bin. The overall {\it likelihood}, $p(D|\vec{\lambda},M)$
  of seeing this data under the model $M$ with parameters $\vec{\lambda}$ is
  the product of the Poisson probabilities for each bin
\be
p(D|\vec{\lambda},M) = 
\prod_{i=1}^K \frac{(r_i(\vec{\lambda}))^{n_i} {\rm e}^{-r_i(\vec{\lambda})}}{n_i!}\,.
\label{like}
\ee
The rates that enter this expression are the rates for {\it observed} events,
i.e., the product of the intrinsic rate predicted by the model with the LISA
transfer function, as discussed in the previous section. It is straightforward
to take the limit of this expression as the bin sizes tend to zero to derive a
continuum version of this equation~\cite{emrimodsel}.

LISA will not be able to measure the parameters of each system perfectly due
to instrumental noise. In addition, weak lensing will introduce errors in the
measurements of luminosity distance. Since we wish to use redshift rather than
distance as a parameter, further errors will be introduced from imperfect
knowledge of the luminosity distance-redshift relation. The modelling of these
errors was mentioned earlier, and is described in detail in
Appendix~\ref{errprop}. There are two ways in which the errors can be folded
into the statistical analysis. Once LISA observations have been made, we will
obtain posterior probability distributions for the source parameters which
account for the error-induced uncertainties. The likelihood will then be
computed by integrating the continuum version of Eq.~(\ref{like}) over the
posterior, as described in~\cite{emrimodsel}. The second approach, which we
adopt here as it is more appropriate for a priori studies of this type, is to
fold the expected errors into the computation of the observed rates, $r_i$. In
practice, we compute these rates directly from the Monte Carlo realizations
described in the preceding section. For each source in the catalog we can
assign fractional rates to every bin in parameter space, computed by
integrating the error probability distribution for the source over that
particular bin. In other words, we spread each source out into multiple bins,
as predicted by the error model described earlier. When generating
realizations of the LISA data set, we assign each source to one bin only,
according to some ``observed parameters'' (which could represent, for
instance, the maximum a posteriori parameters of the source). We take these
observed parameters to be equal to the true parameters plus an error drawn
from the same error distribution.

Given the likelihood described above, Bayes theorem allows us to assign a
posterior probability, $p(\vec{\lambda}|D,M)$, to the parameters,
$\vec{\lambda}$, of a model, $M$, given the observed data, $D$, and a prior,
$\pi(\vec{\lambda})$, for the parameters $\vec{\lambda}$:
\beq
p(\vec{\lambda}|D,M) = \frac{p(D|\vec{\lambda},M) \, \pi(\vec{\lambda})}{{\cal Z}}, \nonumber\\
{\cal Z} = \int p(D|\vec{\lambda},M) \, \pi(\vec{\lambda}) {\rm d}^N \lambda .
\label{bayes}
\eeq
When comparing two models, A and B, that could each describe the data, we can
compute the odds ratio (see, for example,~\cite{mackayinf})
\begin{equation}
O_{AB} = \frac{{\cal Z}_A \, P(A)}{{\cal Z}_B \, P(B)}\,,
\label{odds}
\end{equation}
in which $P(X)$ denotes the prior probability assigned to model $X$. If
$O_{AB} \gg 1$ ($O_{AB} \ll 1$), model $A$ (model B) provides a much better
description of the data.

In this paper, we will consider two types of model comparison. In
Section~\ref{mixmodels}, we will consider mixed models in which the observed
distribution is drawn from a superposition of two or more of the underlying
``pure'' models. In those cases, the models depend on one or more free
``mixing'' parameters for which we will obtain posterior distributions using
Eq.~(\ref{bayes}). First, however, in Section~\ref{puremodels}, we will make
direct comparisons between the pure models. In that case, the models do not
have any free parameters. The odds ratio, (\ref{odds}), then reduces to the
product of the likelihood ratio with the prior ratio
\be
O_{AB} = \frac{p(D|A)}{p(D|B)} \frac{P(A)}{P(B)} .
\ee
The models we consider have all been tuned to match existing constraints, and
so at present we have no good reason to prefer one model over the others. We
therefore assign equal prior probability to each pure model, $P(A)=P(B)=0.5$,
and the odds ratio becomes the likelihood ratio.
We assign probability $p_A=p(D|A)/(p(D|A)+p(D|B))$ to model A, and $p_B=1-p_A$
to model B.

\begin{figure*}[htb]
\begin{tabular}{cc}
\includegraphics[scale=0.3,clip=true,angle=270]{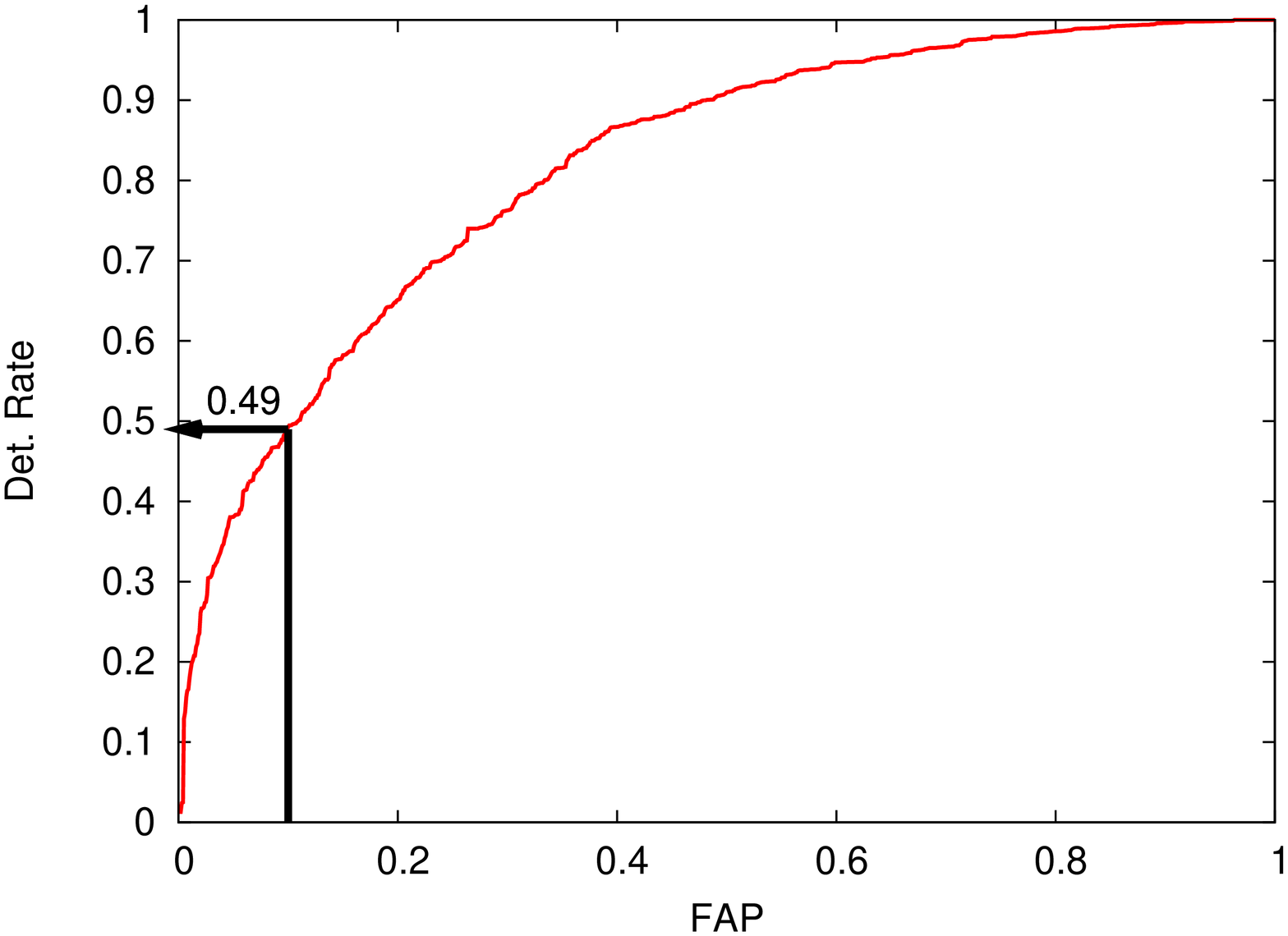}&
\includegraphics[scale=0.3,clip=true,angle=270]{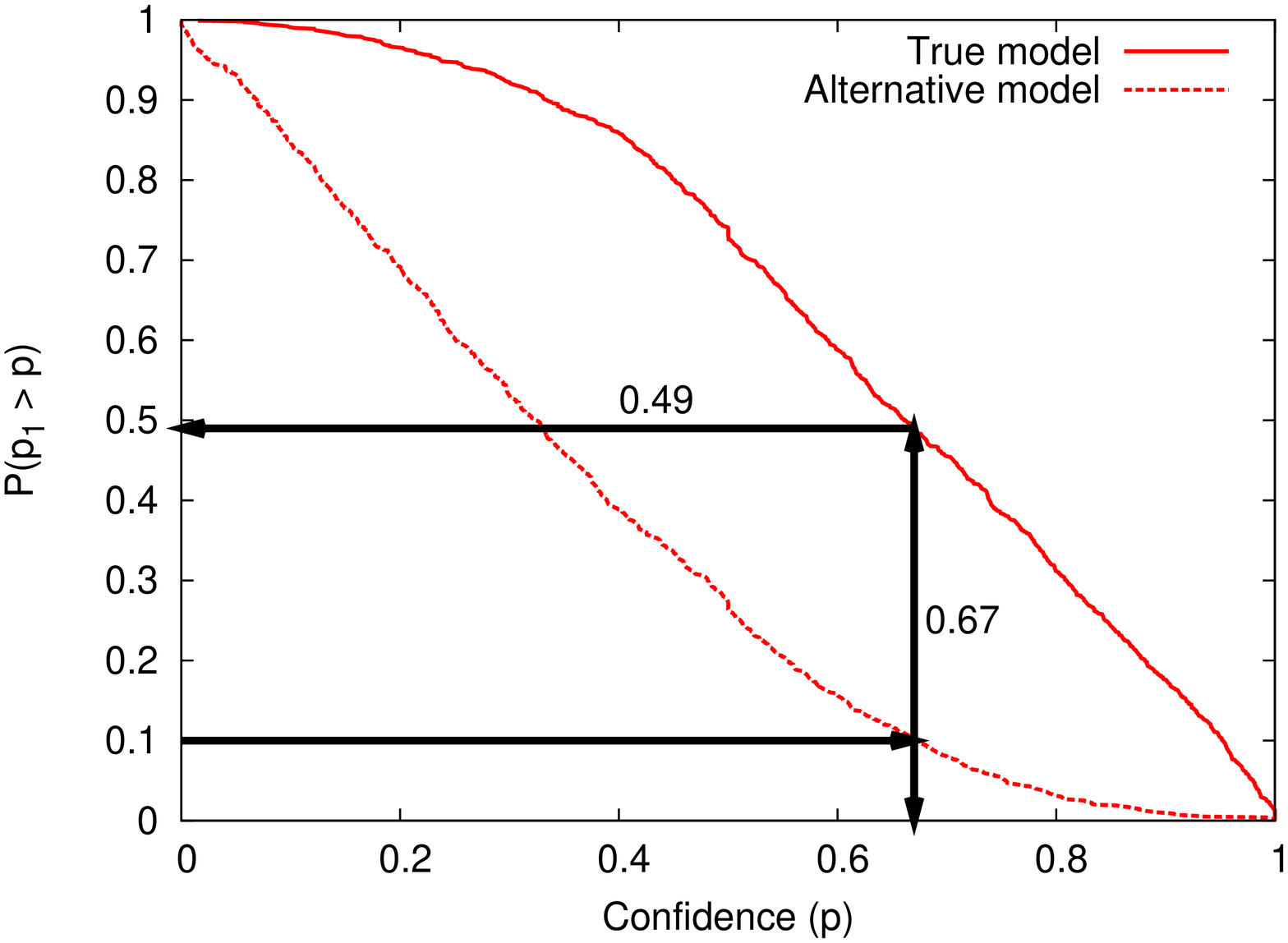}\\
\end{tabular}
\caption{Two alternative ways to represent our ability to distinguish
  models. The left panel shows an ROC curve, while the right panel shows the
  CDF of the confidence achieved over multiple realizations of the correct
  (upper curve) and wrong (lower curve) model. More details are given in the
  text. The model comparison used for this figure was {\it VHM-noZ-Edd-co} to
  {\it VHM-noZ-Edd-ch}, for a three month LISA observation and the most
  pessimistic assumption ($T_2$) on the transfer function.}
\label{fig:roc}
\end{figure*}

Once LISA data is available, each model comparison will yield this single
number, $p_A$, which is our confidence that model A is correct. Since the LISA
data is not currently available, we want to work out how likely it is that we
will achieve a certain confidence with LISA. So, we generate 1000 realizations
of the LISA data stream and look at the distribution of the likelihood ratio
and confidence over these realizations. We can represent the results of this
analysis in two alternative ways. These are illustrated in
Figure~\ref{fig:roc}, and we will refer to the two panels of this figure
extensively in the following. The left panel shows a receiver operator
characteristic (ROC) curve. This is a ``frequentist'' way to represent the
data. To generate this plot, we assume that we have specified a threshold on
the statistic, in this case the likelihood ratio, before the data is
collected. If the value of the statistic computed for the observed data
exceeds the threshold then model A is chosen, otherwise model B is chosen. For
a given threshold, the frequency with which the threshold is exceeded for
realizations of model B defines the false alarm probability (FAP), while the
frequency with which the threshold is exceeded for realizations of model A
defines the detection rate. The ROC curve shows detection rate vertically
versus FAP horizontally. In the figure, we indicate how, for an FAP of $10\%$,
we can find the detection rate, which in this case is $49\%$. This is the
format we used to present our previous results in
Ref.~\cite{Gair:2010bx}. While the ROC is a convenient way to represent the
data, it is incomplete, in that it does not tell us by how much we exceed the
threshold: the result is far more convincing if we obtain a likelihood ratio
10 times the threshold, than 1.1 times the threshold.

The right panel of Figure~\ref{fig:roc} shows an alternative representation of
the same data which contains this additional information. It shows the
cumulative distribution function (CDF) for the ``confidence'' we would have in
model A, based on our observation, i.e., the probability, $p_A$, we assign to
model A in a Bayesian interpretation of the results of an observation. The
upper curve is the CDF computed over multiple realizations of model A (i.e.,
the horizontal axis then shows our confidence in the true model), while the
lower curve shows the CDF computed from realizations of model B (i.e., the
horizontal axis then shows our confidence in the wrong model). The best way to
interpret this plot is to choose a certain confidence level, e.g., $p=0.95$
(approximately 2$\sigma$). The value on the upper curve is the frequency with
which this confidence level, or better, would be achieved in a LISA
observation when that model was correct, while the value on the lower curve at
$1-p$ is the frequency with which we would {\it not be able to rule out} model
A with that confidence, when it was not true. 

The CDF plot encodes the same information as the ROC curve. If we assign a
certain FAP, say $10\%$ as before, we draw a horizontal line at that value and
find where it intersects the lower curve. This tells us the confidence level
corresponding to that FAP, in this case $0.67$. The value on the upper curve
at this confidence level is the detection rate at that FAP, and we find that it
is $49\%$, as expected. In the current paper, we will use this second,
Bayesian, representation of the results for all the remaining plots, as it
encodes all of the information that can be gleaned from the Monte Carlo
simulations. The Bayesian approach assigns relative probabilities to the
models, rather than making a binary statement that model A is ``right'' or
model B is ``right''.

\begin{figure}[htb]
\includegraphics[scale=0.3,clip=true,angle=270]{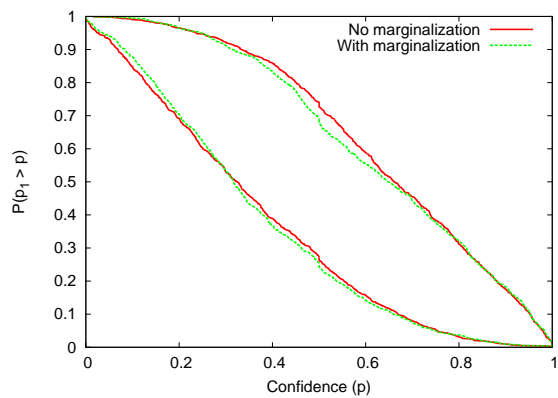}
\caption{Comparison of performance of model selection when including the total
  number of events as a parameter of the model (labeled ``no
  marginalization'') and when this parameter is marginalized over (labeled
  ``with marginalization''). The small difference indicates that the total
  number of events contains relatively little information compared to the
  shape of the parameter distributions.  We show the same model comparison as
  in Figure~\ref{fig:roc}.}
\label{fig:marg}
\end{figure}

\begin{figure*}[htb]
\begin{tabular}{cc}
\includegraphics[scale=0.3,clip=true,angle=270]{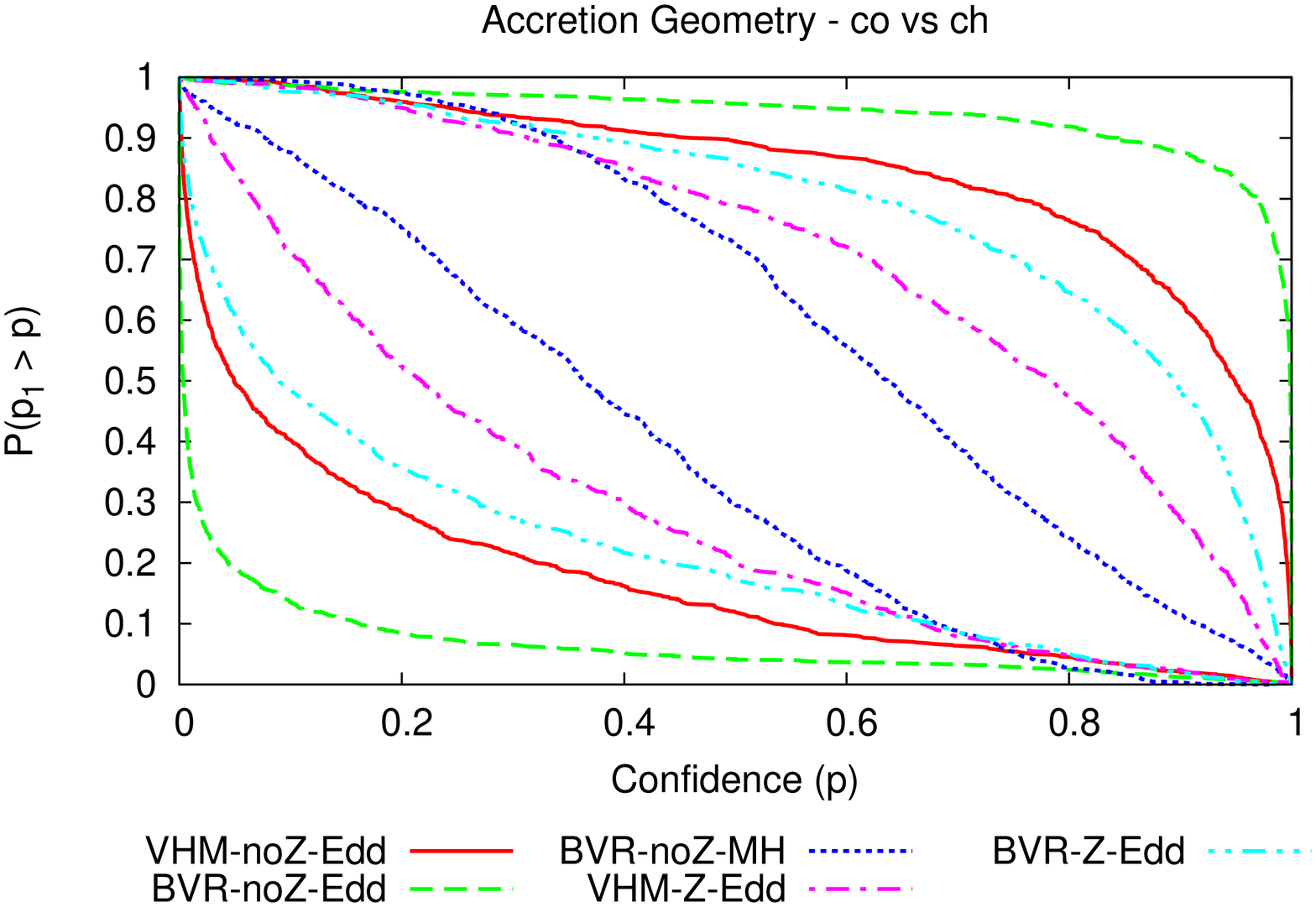}&
\includegraphics[scale=0.3,clip=true,angle=270]{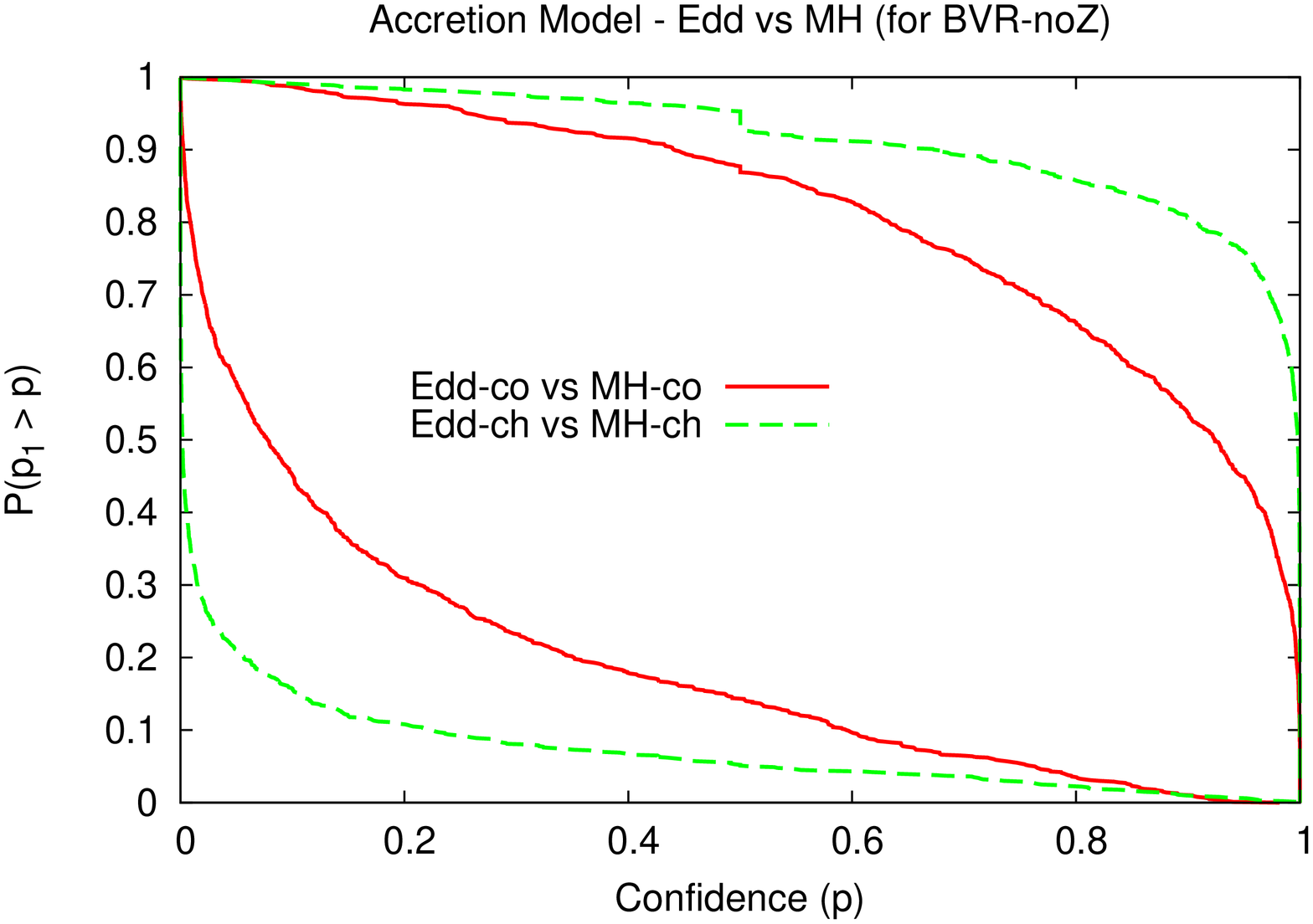}\\
\includegraphics[scale=0.3,clip=true,angle=270]{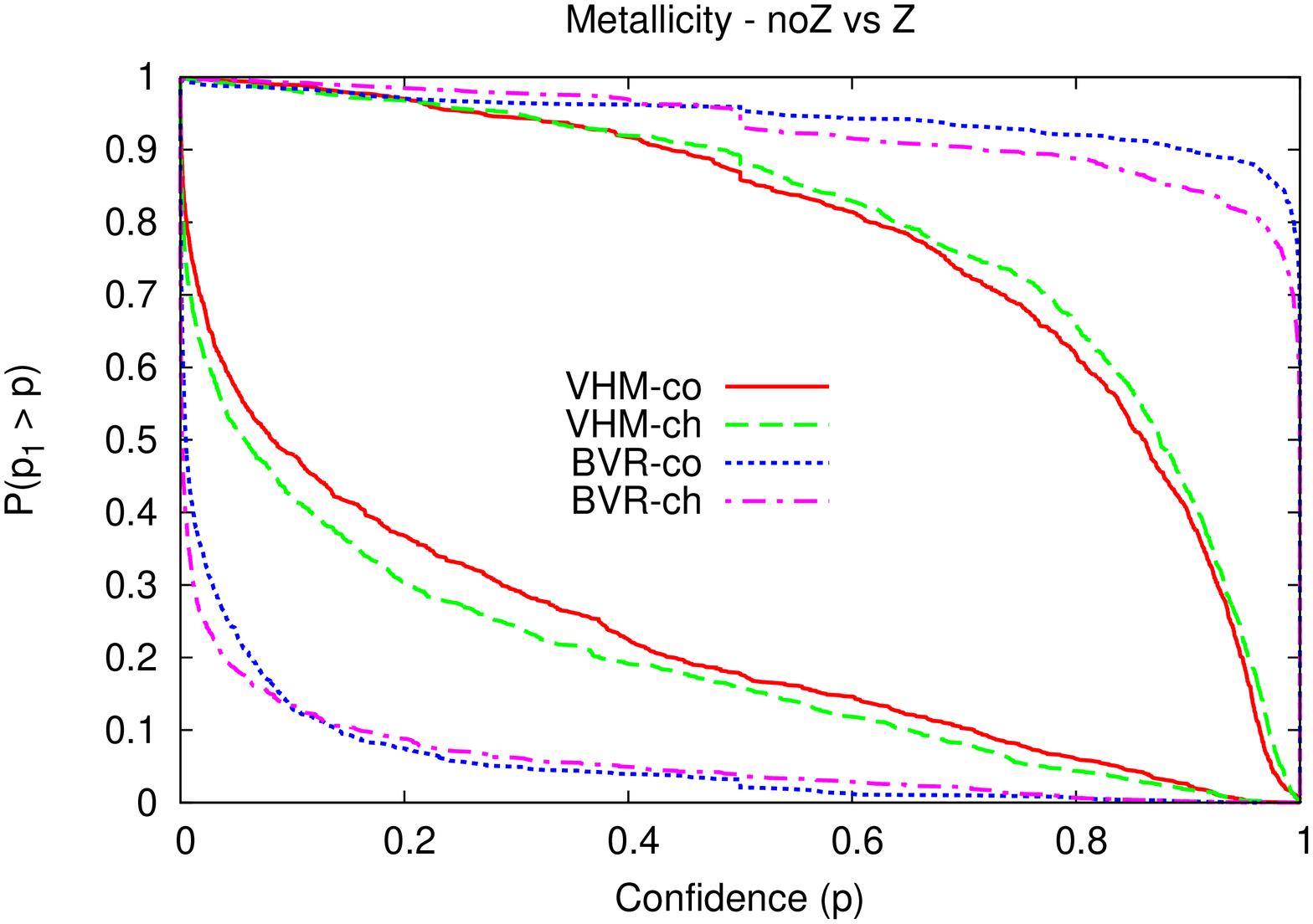}&
\includegraphics[scale=0.3,clip=true,angle=270]{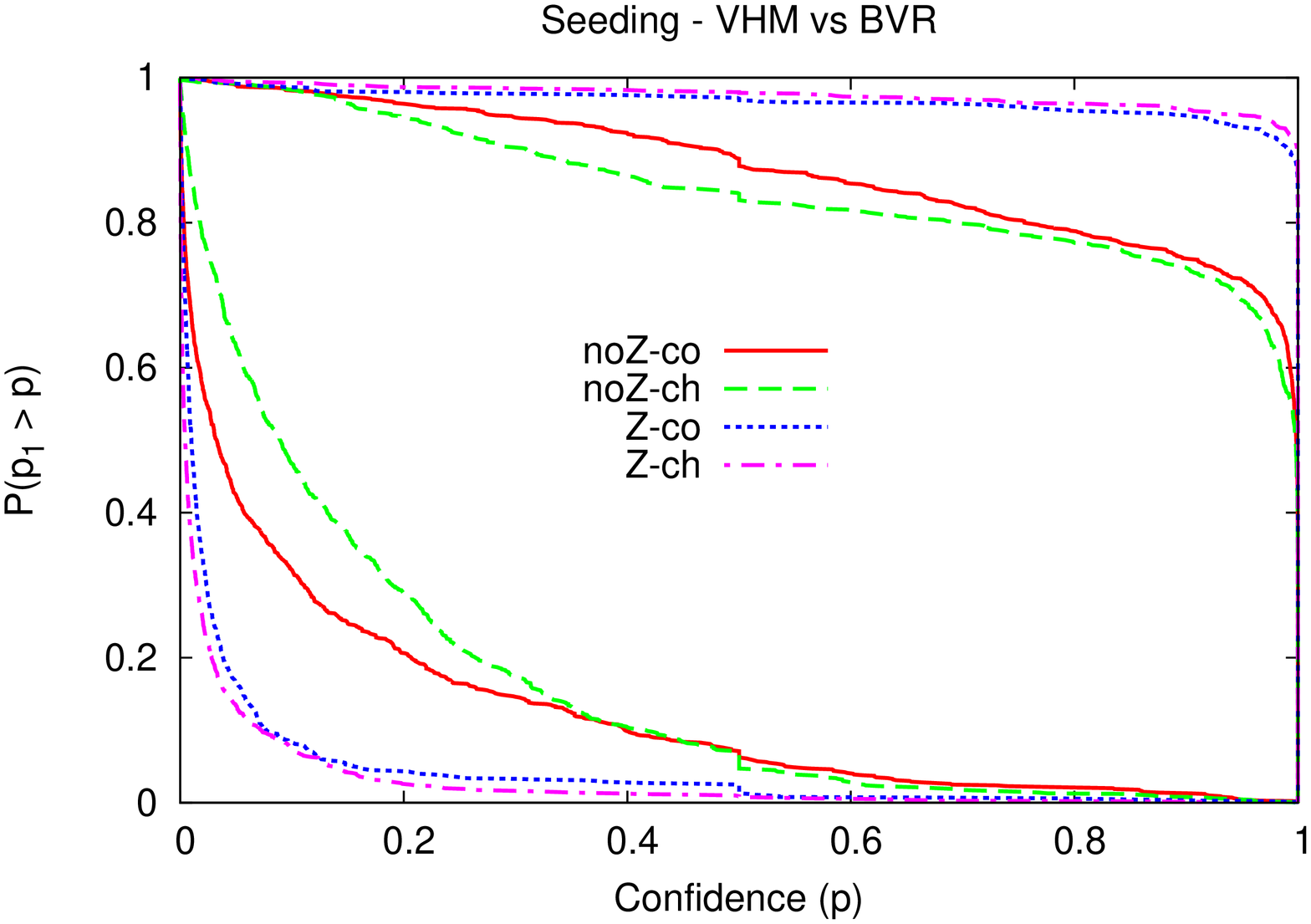}\\
\end{tabular}
\caption{Results for comparisons of the pure models. Each plot shows all
  possible comparisons varying only one of the elements listed in
  Table~\ref{tabI}. Top left panel: we consider the effect of the {\it
    accretion geometry}, comparing {\it coherent} to {\it chaotic} for each of
  the combinations of the other ingredients. Top right panel: we consider the
  effect of the {\it accretion model}, comparing {\it Eddington} accretion to
  {\it Merloni-Heinz} accretion for the {\it BVR-noZ} models. Bottom left: we
  consider the effect of {\it metallicity} by comparing the {\it noZ} to {\it
    Z} models for {\it VHM-co}, {\it VHM-ch}, {\it BVR-co} and {\it
    BVR-ch}. Bottom right: we consider the effect of the {\it seeding}
  assumption, comparing the {\it VHM} to {\it BVR} models for the four
  combinations {\it noZ-co}, {\it noZ-ch}, {\it Z-co} and {\it Z-ch}, each
  with Eddington accretion. In all panels we are making the most pessimistic
  assumptions about the detector, i.e., we use the transfer function $T_2$
  (one interferometer, $\rho_{\rm thr}=20$). These results are for a 3 month
  LISA observation, except for the top left panel which is for a one year
  observation.}
\label{fig:comp}
\end{figure*}

The models we consider differ not only in the distribution of events that they
predict, but also in the total number of events. As the latter could be
considered a less robust prediction of the models, we can ask whether it
carries much weight for model selection. This can be done by introducing a
free parameter into each model, which is an overall normalizing factor, and
then marginalizing over it, i.e., integrating the posterior probability over
this parameter. We write $r_i = N \tilde{r}_i$, where $\tilde{r}_i$ is the
rate in bin $i$ for a model that predicts $1$ event in total. The probability
marginalized over $N$ is
\be
\tilde{p}(D|M) = 
\left(\prod_{i=1}^K \frac{\tilde{r}_i^{n_i}}{n_i!} \right) 
\sum_{n=1}^\infty n^{N_{\rm obs}} {\rm e}^{-n}\,,
\ee
where $N_{\rm obs}=\sum n_i$ is the total number of events observed. The
summation in the second term is dependent only on $N_{\rm obs}$ and, as such,
is model-independent. It can thus be seen that
\be
\tilde{p}(D|M) \propto p(D|M) {\rm e}^{N_{\rm M}} N_{\rm M}^{-N_{\rm obs}}\,,
\label{Nmargeq}
\ee
where $N_{\rm M}$ is the number of events predicted by the un-normalized model
$M$. We can decouple the contribution from the total number of events and the
distribution of event parameters by replacing the likelihood $p(D|M)$ by the
marginalized likelihood $\tilde{p}(D|M)$ in the likelihood ratio. In
Figure~\ref{fig:marg} we show the effect this has on the CDFs for the Bayesian
confidence, for the comparison between model {\it VHM-noZ-Edd-co} and {\it
  VHM-noZ-Edd-ch}. We see that including marginalization makes very little
difference to the results. This implies that the number of events predicted by
a model contains little information relative to the parameter
distribution. For the remaining plots in this paper, we do not marginalize
over $N_{\rm M}$, but we have checked that in all cases the effect of the
marginalization is small.

\section{\label{puremodels}Results for the pure models}
We now describe the results of our analysis. In this section we will compare
pairs of pure models using the technique described in the previous section. We
generated 1000 realizations for each model, as described in Section
\ref{datasets}. For each pair of models $A$ and $B$, we computed the CDF of
the Bayesian confidence of model $A$ versus model $B$ over the realizations of
model $A$ and those of model $B$. We present selected results in Figure
\ref{fig:comp} using the CDF curves described in the previous section.


Each panel in Figure \ref{fig:comp} shows the results for pairs of models that
differed in only one of the four aspects of the input physics detailed in
section \ref{mbhpop} and listed in Table \ref{tabI}. To be conservative, we
consider a pessimistic scenario for the detector (transfer function $T_2$: one
interferometer, $\rho_{\rm thr}=20$).

In the upper-left panel we show all possible (five) comparisons among pairs of
models differing only in the accretion geometry (e.g., {\it BVR-noZ-Edd-co} vs
{\it BVR-noZ-Edd-ch}), assuming a one year observation. Since we ignore the
spin distributions, this is the property to which we are least sensitive, as
clearly shown by the relatively small separation of some pairs of curves in
the panel.  In most cases the models are barely distinguishable at any
reasonable confidence level.

In the upper-right panel we compare models differing in their accretion model
({\it Edd} vs {\it MH}, two comparisons), assuming a three month observation.
Our models are clearly more sensitive to this parameter, and they can be
clearly discriminated with only three months of data.

In the lower-left panel we investigate the impact of metallicity ({\it Z} vs
{\it noZ}, four comparisons). Pairs of models are generally well separated
even under pessimistic assumptions (a three month observation), and we can put
forward an interesting astrophysical interpretation of the results.  This
panel shows that metallicity ``feedback'' is better discriminated in high-mass
seed models ({\it BVR}). This is because the effect of metallicity is to
change the redshift distribution of the seeds. If seeds are massive, we can
clearly detect this redshift difference by directly observing the first
coalescing seeds in the Universe (recall that LISA observations are basically
complete for massive seed models, as shown in Figure
\ref{fig:dist}). Unfortunately, LISA is deaf to coalescences of a few hundred
solar mass binaries at high $z$. Therefore, in low-mass seed models, we can
only measure the redshift distribution of the seeds indirectly (by observing
the distribution of mergers at a later cosmological epoch), and models are
consequently harder to discriminate.

Finally, in the lower-right panel, we look at the seeding process (left: {\it
  VHM} vs {\it BVR}, four comparisons). Here the result is very similar to the
effect of metallicity.  Pairs of models are typically well separated,
especially if seeds form even at later times (the {\it Z} models). If seeds
form at high redshift only, then the mass distribution of coalescences at
lower redshift tends to be more similar, as mass growth by accretion erases
the differences in the initial seed masses.

\begin{figure}[htb]
\includegraphics[scale=0.3,clip=true,angle=270]{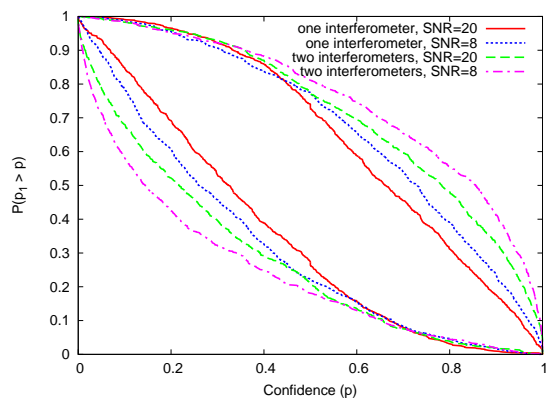}
\caption{Effect of the transfer function on the pure model selection
  results. We compare {\it VHM-noZ-Edd-co} and {\it VHM-noZ-Edd-ch}, assuming
  a fixed LISA mission duration of 3 months.}
\label{fig:filter}
\end{figure}

\begin{figure}[htb]
\includegraphics[scale=0.3,clip=true,angle=270]{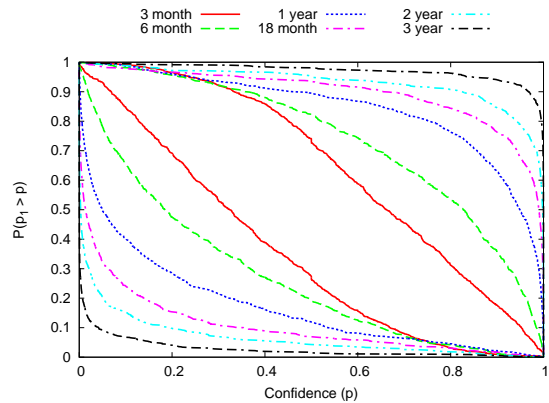}
\caption{Effect of the LISA observation duration on the pure model selection
  results. We compare {\it VHM-noZ-Edd-co} and {\it VHM-noZ-Edd-ch}, under the
  most pessimistic choice ($T_2$) for the transfer function.}
\label{fig:time}
\end{figure}

We emphasize that the results discussed so far have made the most pessimistic
assumptions about the detector performance, i.e., three months of observation
with a single interferometer and $\rho_{\rm thr}=20$. Under such assumptions
only a handful of sources will be detected, but this is already sufficient to
discriminate among most of the models. In Figures \ref{fig:filter} and
\ref{fig:time} we consider a specific model comparison (namely {\it
  VHM-noZ-Edd-co} versus {\it VHM-noZ-Edd-ch}) to display the effect of
relaxing these assumptions.

Figure \ref{fig:filter} shows that the detector performance does not affect
the results substantially. Lowering $\rho_{\rm thr}$ from 20 to 8 for two
operational interferometers only adds a few, low SNR, sources to the detected
sample, and the gain in discrimination power is limited. On the other hand,
Figure \ref{fig:time} shows that the observation duration is crucial. With an
observation time of three months, we would achieve a 2$\sigma$ confidence
level ($p_A=0.95$) with only $\sim 10\%$ probability (i.e., if we repeated an
independent 3 month LISA observation $10$ times, we would expect one of these
to reach 2$\sigma$ confidence). However, with an observation time of three
years, the probability that we will achieve 2$\sigma$ confidence in the
underlying model is more than $90\%$ (upper dashed-black curve). There
is a similar trend in all model comparisons, although the three month result
is particularly bad for this particular comparison, since these models differ
only in the accretion geometry which we have seen is the most difficult aspect
to distinguish. The trend with observation duration arises simply because the
number of detected sources increases linearly with the observation time, and
so we have a much better sampling of the underlying model for longer mission
durations.

\begin{table*}[htb]
\begin{center}
\begin{tabular}{cc}
\begin{tabular}{|l|cccccccccc|}
\hline
\hline
\multicolumn{11}{c}{Three-month observation}\\ \hline
 & 1 & 2 & 3 & 4 & 5 & 6 & 7 & 8 & 9 & 10\\
\hline
1 & $\times$ &0.10 &0.72 &0.68 &0.86 &0.88 &0.19 &0.17 &0.91 &0.92\\
2 & 0.93& $\times$ &0.75 &0.69 &0.91 &0.91 &0.17 &0.22 &0.93 &0.93\\
3 & 0.42& 0.32& $\times$ &0.24 &0.45 &0.42 &0.72 &0.69 &0.88 &0.89\\
4 & 0.65& 0.63& 0.83& $\times$ &0.77 &0.76 &0.48 &0.49 &0.80 &0.81\\
5 & 0.13& 0.08& 0.58& 0.19& $\times$ &0.03 &0.93 &0.92 &0.98 &0.99\\
6 & 0.12& 0.07& 0.58& 0.21& 0.97& $\times$ &0.94 &0.92 &0.98 &0.98\\
7 & 0.57& 0.57& 0.16& 0.20& 0.05& 0.04& $\times$ &0.01 &0.93 &0.94\\
8 & 0.58& 0.51& 0.16& 0.19& 0.07& 0.07& 0.98& $\times$ &0.94 &0.95\\
9 & 0.16& 0.12& 0.23& 0.31& 0.03& 0.04& 0.17& 0.15& $\times$ &0.01\\
10 & 0.09& 0.07& 0.15& 0.18& 0.02& 0.03& 0.14& 0.13& 0.95& $\times$\\
\hline
\hline
\end{tabular}&
\begin{tabular}{|l|cccccccccc|}
\hline
\hline
\multicolumn{11}{c}{One-year observation}\\ \hline
 & 1 & 2 & 3 & 4 & 5 & 6 & 7 & 8 & 9 & 10\\
\hline
1 & $\times$ &{\bf 0.49} &0.99 &0.99 &1.00 &1.00 &0.91 &0.88 &1.00 &1.00\\
2 & {\bf 0.50}& $\times$ &0.99 &0.99 &1.00 &1.00 &0.92 &0.93 &1.00 &1.00\\
3 & 0.00& 0.00& $\times$ &{\bf 0.83} &0.97 &0.97 &1.00 &1.00 &1.00 &1.00\\
4 & 0.02& 0.02& {\bf 0.19}& $\times$ &0.99 &0.99 &0.99 &0.99 &0.99 &0.99\\
5 & 0.00& 0.00& 0.03& 0.00& $\times$ &{\bf 0.07} &1.00 &1.00 &1.00 &1.00\\
6 & 0.00& 0.00& 0.03& 0.00& {\bf 0.93}& $\times$ &1.00 &1.00 &1.00 &1.00\\
7 & 0.07& 0.06& 0.00& 0.00& 0.00& 0.00& $\times$ &{\bf 0.16} &1.00 &1.00\\
8 & 0.09& 0.05& 0.00& 0.00& 0.00& 0.00& {\bf 0.85}& $\times$ &1.00 &1.00\\
9 & 0.00& 0.00& 0.00& 0.01& 0.00& 0.00& 0.00& 0.00& $\times$ &{\bf 0.31}\\
10 & 0.00& 0.00& 0.00& 0.00& 0.00& 0.00& 0.00& 0.00& {\bf 0.61}& $\times$\\
\hline
\hline
\end{tabular}
\end{tabular}
\end{center}
\caption{Summary of all possible comparisons of the pure models. The table on
  the left (right) assumes a LISA observation time of three months (one year),
  respectively. Models are labeled by an integer $i$, as listed in
  Table~\ref{tabI}. We take a fixed confidence level of $p=0.95$. The numbers
  in the upper-right half of each table show the fraction of realizations in
  which the {\it row} model will be chosen at more than this confidence level
  when the {\it row} model is true (in the Bayesian figures, this would be the
  point where a vertical line at $x=p$ intersects the upper curve). The
  numbers in the lower-left half of each table show the fraction of
  realizations in which the {\it row} model {\it cannot be ruled out} at that
  confidence level when the {\it column} model is true (in the Bayesian
  figures, this would be the point where a vertical line at $x=1-p$ intersects
  the lower curve). These results are for the pessimistic transfer function
  ($T_2$).
 }
\label{tabII}
\end{table*}

Comparisons between all possible pairs of models are given in Table
\ref{tabII}, where we assume a pessimistic detector performance and three
months (left) or one year of observation (right), respectively. Even though it
is difficult to discriminate among some specific pair of models in the three
month observation case, model discrimination is almost perfect in most cases
for a one year observation. The exception are the models differing in their
accretion geometry only (bold numbers in the table), for which discrimination
is difficult. However, even for such similar models we will obtain a high
confidence level with probability close to unity if we assume a standard LISA
configuration with two operational interferometers observing for three years.
  
\section{\label{mixmodels}Mixed models}
In the preceding section we (successfully) demonstrated the potential of LISA
to discriminate among a discrete set of ``pure'' models given a
priori. However, the true MBH population in the Universe will probably result
from a mixing of the physical processes described in Section \ref{mbhpop}, or
even from a completely unexplored physical mechanism. It is therefore
important to test whether we will be able to extract useful information when
the distribution of observed events comes from a mixture of the different
models, as an approximation to possible unknowns.  For this case study, we
will concentrate on the details of the seeding mechanism (mass function and
redshift distribution), deferring the more complicated details related to
accretion to a future study. Recall in this context that accretion will leave
a trace in the spin distribution of MBHs, but our simplified analysis neglects
the MBH spins by construction.

We tested two mixing procedures: (i) we generated {\it artificially} mixed
models that were a linear combination of the pure model distributions
presented in Section \ref{mbhpop}; (ii) we constructed two {\it consistently}
mixed models, in which seeds were generated according to a mixing of two
prescriptions, and their evolution was followed self-consistently in the halo
merger tree realizations. The goal here is to assess whether artificial models
can reproduce the salient features of the consistently mixed models, and to
estimate the amount of mixing necessary to ``best fit'' the consistently mixed
models. This procedure mimics the analysis of a ``real'' LISA datastream, for
which the data is unlikely to match exactly any one of the pure model
predictions.

In this section we describe details of the artificially and consistently mixed
models. In Section \ref{mixresults} that follows we will present the results
of the ``reconstruction experiment''.

\begin{figure}[htb]
\includegraphics[scale=0.43,clip=true]{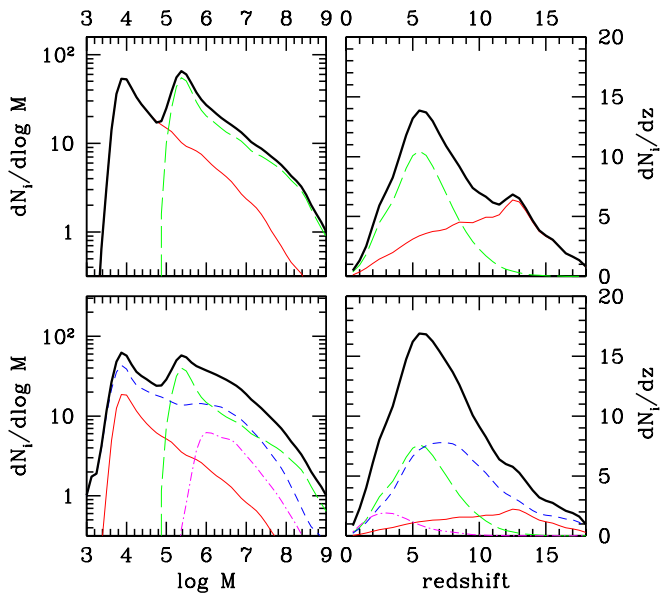}
\caption{Examples of mixed models. In the upper panels we show marginalized
  $dN_i/dM$ (left) and $dN_i/dz$ (right) distributions for the model ${\cal
    N}$-I (thick solid black lines), in which we mix models {\it
    VHM-noZ-Edd-co} (thin solid red line) and {\it BVR-noZ-Edd-co} (thin
  long-dashed green line).  The relative contribution of the models is given
  by Eq.~(\ref{mix2}) with $f_1=0.23$, $f_3=0.77$ (cf.~Table \ref{tabII}). In
  the lower panels we show the same distributions for the model ${\cal N}$-IV
  (thick solid black lines) in which we mix four ``pure'' models.  The thin
  lines represent the relative contribution of the individual models {\it
    VHM-noZ-Edd-co} (solid red), {\it VHM-Z-Edd-co} (short-dashed blue), {\it
    BVR-noZ-Edd-co} (long-dashed green), and {\it BVR-Z-Edd-co} (dot-dashed
  magenta).  The relative contribution of the models is given again by
  Eq.~(\ref{mix2}) with $f_1=0.08$, $f_3=0.22$, $f_5=0.56$ and $f_7=0.14$
  (cf.~Table \ref{tabII}).  }
\label{fig:mix}
\end{figure}

Artificial mixing simply consists in drawing coalescences from a linear
combination of the theoretical coalescence distributions predicted by the pure
models. Here, for concreteness, we fix the accretion to be Eddington-limited
and coherent, and we mix different seeding recipes. We therefore consider
models {\it VHM-noZ-Edd-co}, {\it VHM-Z-Edd-co}, {\it BVR-noZ-Edd-co}, and
{\it BVR-Z-Edd-co} ($i=1$, $3$, $5$ and $7$ respectively, in the notation of
Table~\ref{tabI}).

Each model $i$ is characterized by a mean number of predicted coalescences
$\bar{N}_i$ and a probability distribution for the parameters of the
coalescing binaries $p_i(M,q,z)$. The predicted event distribution
${\cal N}_i=d^3N_i/dzdMdq$ (see Section \ref{mbhpop}) can therefore be
factorized as
\be
{\cal N}_i={\bar{N}}_i\times p_i(M,q,z)\,.
\ee
The mixing is described by four parameters $f_i$ ($i=1\,,3\,,5\,,7$) which
determine the fraction of model $i$ included in the mixed distribution. These
fractions are constrained to add up to $1$.

\subsection{Artificial mixing}
\begin{table*}
\begin{center}
\begin{tabular}{l|ccccc|cccc|c}
\hline
NAME & Mixing &  $f_1$ & $f_3$ & $f_5$ & $f_7$&$f_1$ fit & $f_3$ fit & $f_5$ fit& $f_7$ fit& $f_1+f_5$ fit\\
\hline
p-I & $p$ &0.15&--&0.85&--&$0.15\pm0.05$&--&$0.85\pm0.05$&--&--\\
p-II & $p$ &0.54&--&0.46&--&$0.55\pm0.1$&--&$0.45\pm0.1$&--&--\\
p-III & $p$ &0.41&0.13&0.12&0.34&$0.3\pm0.2$&$0.25\pm0.25$&$0.1\pm0.05$&$0.35\pm0.1$&$0.6\pm0.05$\\
p-IV & $p$ &0.11&0.49&0.22&0.18&$0.29\pm0.29$&$0.3\pm0.3$&$0.21\pm0.05$&$0.2\pm0.05$&$0.4\pm0.05$\\
${\cal N}$-I & ${\cal N}$ &0.23&--&0.77&--&$0.2\pm0.1$&--&$0.8\pm0.1$&--&--\\
${\cal N}$-II & ${\cal N}$ &0.61&--&0.39&--&$0.6\pm0.15$&--&$0.4\pm0.15$&--&--\\
${\cal N}$-III & ${\cal N}$ &0.31&0.16&0.23&0.3&$0.4\pm0.2$&$0.1\pm0.1$&$0.2\pm0.05$&$0.3\pm0.05$&$0.5\pm0.05$\\
${\cal N}$-IV & ${\cal N}$ &0.08&0.22&0.56&0.14&$0.15\pm0.15$&$0.15\pm0.15$&$0.5\pm0.1$&$0.2\pm0.1$&$0.3\pm0.05$\\
\hline				
\end{tabular}
\end{center}
\caption{``Artificially mixed'' models. Columns 3--6 list the mixing
  parameters used to generate the models. Columns 7--10 list the best-fit
  values recovered by our analysis (see Section \ref{mixresults}).}
\label{tabIV}
\end{table*}

We tried two different mixing prescriptions.  In the first case we ignored the
number of coalescences predicted by each specific model, by mixing the
respective $p_i(M,q,z)$ distributions ($p$ mixing) and normalizing the mixed
distribution to some arbitrary number:
\be
{\cal N}_{p}=\bar{N}_{\rm m}\left\{f_1 p_1+f_3 p_3 + f_5 p_5 + f_7p_7\right\},
\label{mix1}
\ee
where $\bar{N}_{\rm m}$ was fixed to 200 coalescences in three years.  In the
second case we considered the number of predicted events to be an intrinsic
property of each individual model, and we simply mixed the $N_i(M,q,z)$
distributions (${\cal N}$ mixing) in the same way:
\be
{\cal N}_{\rm {\cal N}}=
f_1 {\cal N}_1+f_3 {\cal N}_3 + f_5{\cal N}_5 + f_7{\cal N}_7\,.
\label{mix2}
\ee
The total number of coalescences is now automatically determined by the values
of the mixing parameters. In practice, in order to enforce the constraint that
the fractions add up to $1$, we actually use a ``nested'' prescription based
on three parameters $\alpha$, $\beta$ and $\gamma$, which are allowed to take
any value in the range $[0,1]$. We then set
\be
{\cal N}_{\rm {\cal N}}=\alpha {\cal N}_1+(1-\alpha)\{\beta {\cal N}_3 + (1-\beta)[\gamma{\cal N}_5 + (1-\gamma){\cal N}_7]\}\,.
\label{nestmix2}
\nonumber
\ee
We quote our results in terms of the model fractions $f_i$, as these are the
physically relevant quantities.

Table \ref{tabIV} lists eight mixed models that we investigated. Examples of
${\cal N}$-mixed model (model ${\cal N}$-I and ${\cal N}$-IV) are also shown
in Figure \ref{fig:mix}.  The {\it theoretically observable} distributions are
generated in the same way as for the pure models, by multiplying the ${\cal
  N}$ distributions by the appropriate transfer function. Observed datasets
${D_k}$ are then generated from the mixed distribution as outlined in Section
\ref{datasets}. Given a dataset ${D_k}$, the idea is to parametrize the
distribution as a mixture of the available ``pure'' distributions according to
Eqs.~(\ref{mix1}) or (\ref{mix2}), and obtain a posterior distribution for the
mixing parameters given the observed data. These posterior distributions allow
us to assess which models were mixed, and at what mixing level. To make the
test ``realistic'', the theoretical mixing and the simulated LISA observations
were performed by A. Sesana. The observed datasets ${D_k}$ were then analyzed
blindly by J. Gair, who did not know which models were mixed nor the amount of
mixing.

\subsection{Consistent mixing}
We used the consistently mixed models (hybrid models, henceforth labelled
``{\it HY}'') described in Ref.~\cite{bv10}. The seeding process was a mixture
of the {\it VHM-Z} and {\it BVR-Z} mechanisms, and the MBH mass growth assumed
Eddington limited, coherent accretion. We considered two models with fixed
POPIII seeding efficiency, but different quasistar seeding efficiencies. The
quasistar seeding efficiency is related to the maximum halo spin parameter,
$\lambda$, that allows efficient transfer of gas to the center to form a
quasistar (see Ref.~\cite{bv10} for details). We test an inefficient quasistar
seeding model ($\lambda=0.01$, {\it HY-I}) and an efficient quasistar seeding
model ($\lambda=0.02$, {\it HY-II}) that predict MBH population observables
(local mass function, quasar luminosity function, and so on) bracketing the
current range of allowed values.

To check the effectiveness of our analysis tools in extracting information
about the parent MBH population, we try to recover the hybrid model
distributions as a mixing of the {\it VHM-Z} and {\it BVR-Z} ``pure'' models,
of the form given by either Eq.~(\ref{mix1}) or (\ref{mix2}).
The procedure is the same as detailed in the previous section. 

Let us stress again that the MBH evolution through cosmic history is followed
self-consistently in the hybrid models. This means that the predicted {\it
  theoretical distribution} is not, in general, described as a simple mixing
of the form given by Eqs.~(\ref{mix1}) or (\ref{mix2}). This is a crucial
point: {\it the success of this experiment will tell us that we can extract
  valuable information on complex MBH formation scenarios by mixing a set of
  ``pure'' models based on simple recipes}.

\begin{figure*}[htb]
\begin{tabular}{cc}
\includegraphics[scale=0.3,clip=true,angle=270]{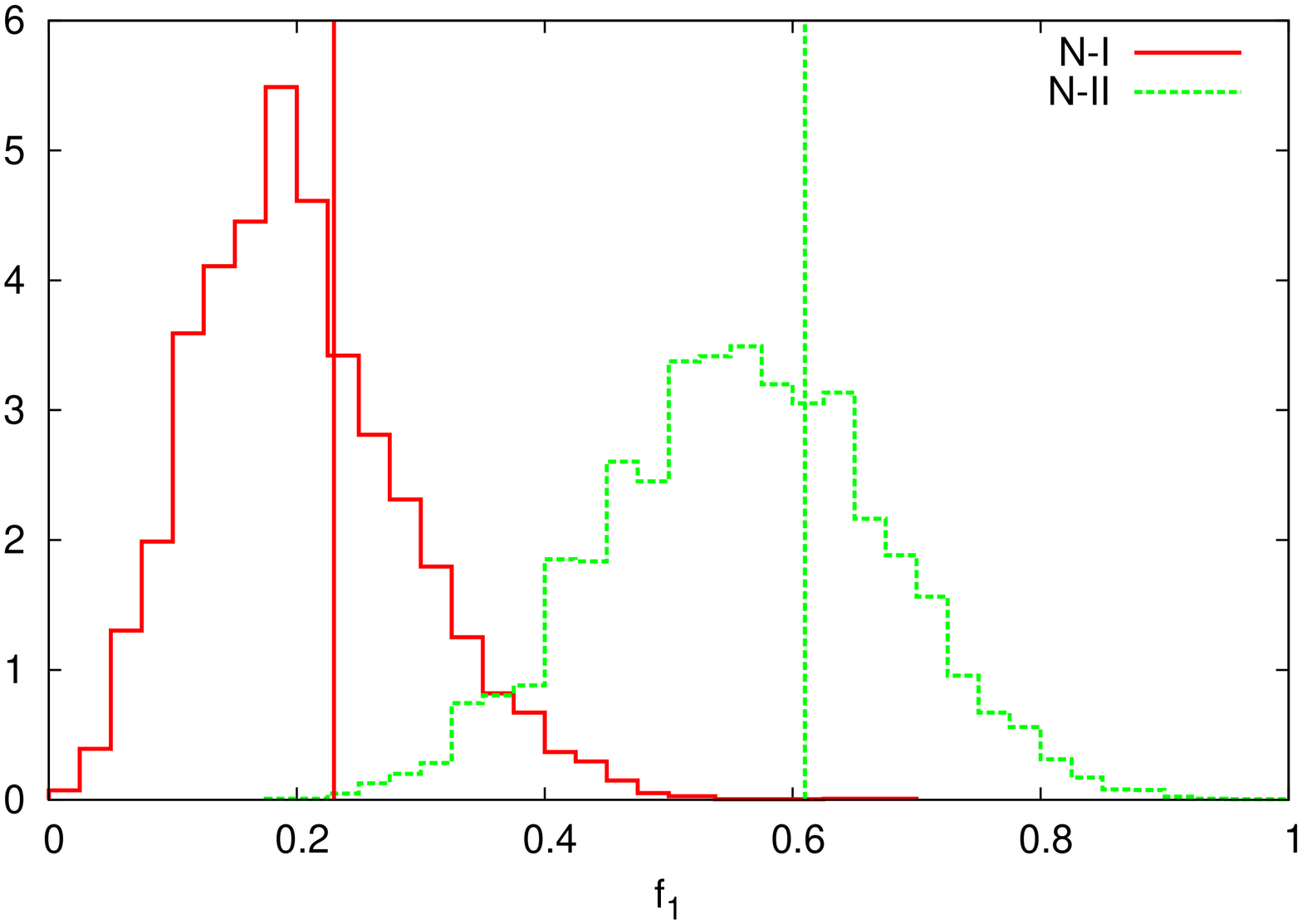}&
\includegraphics[scale=0.3,clip=true,angle=270]{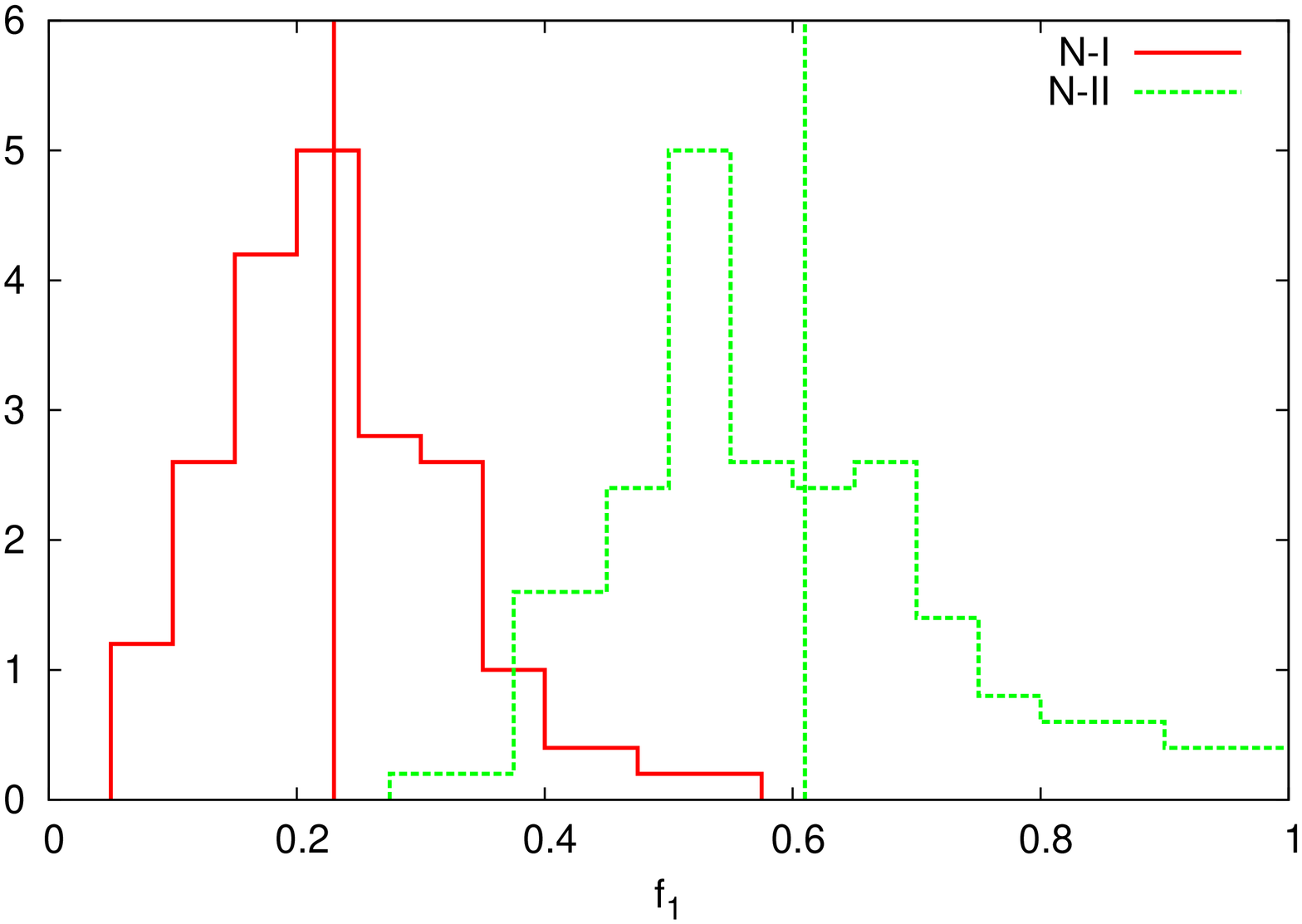}
\end{tabular}
\caption{Summary of the model mixing results for models ${\cal N}$-I and
  ${\cal N}$-II. In both panels, the horizontal axis shows the mixing fraction
  for model {\it VHM-noZ-Edd-co}. The left panel shows the posterior
  probability distribution function for this mixing fraction as found in one
  particular realization of each model. The right panel shows how the peak of
  the posterior was distributed over 100 different realizations of each of the
  two models. The vertical lines show the true values of the mixing fraction.
}
\label{fig:post6566}
\end{figure*}

\begin{figure*}[ht]
\begin{tabular}{ccc}
\includegraphics[scale=0.24,clip=true,angle=270]{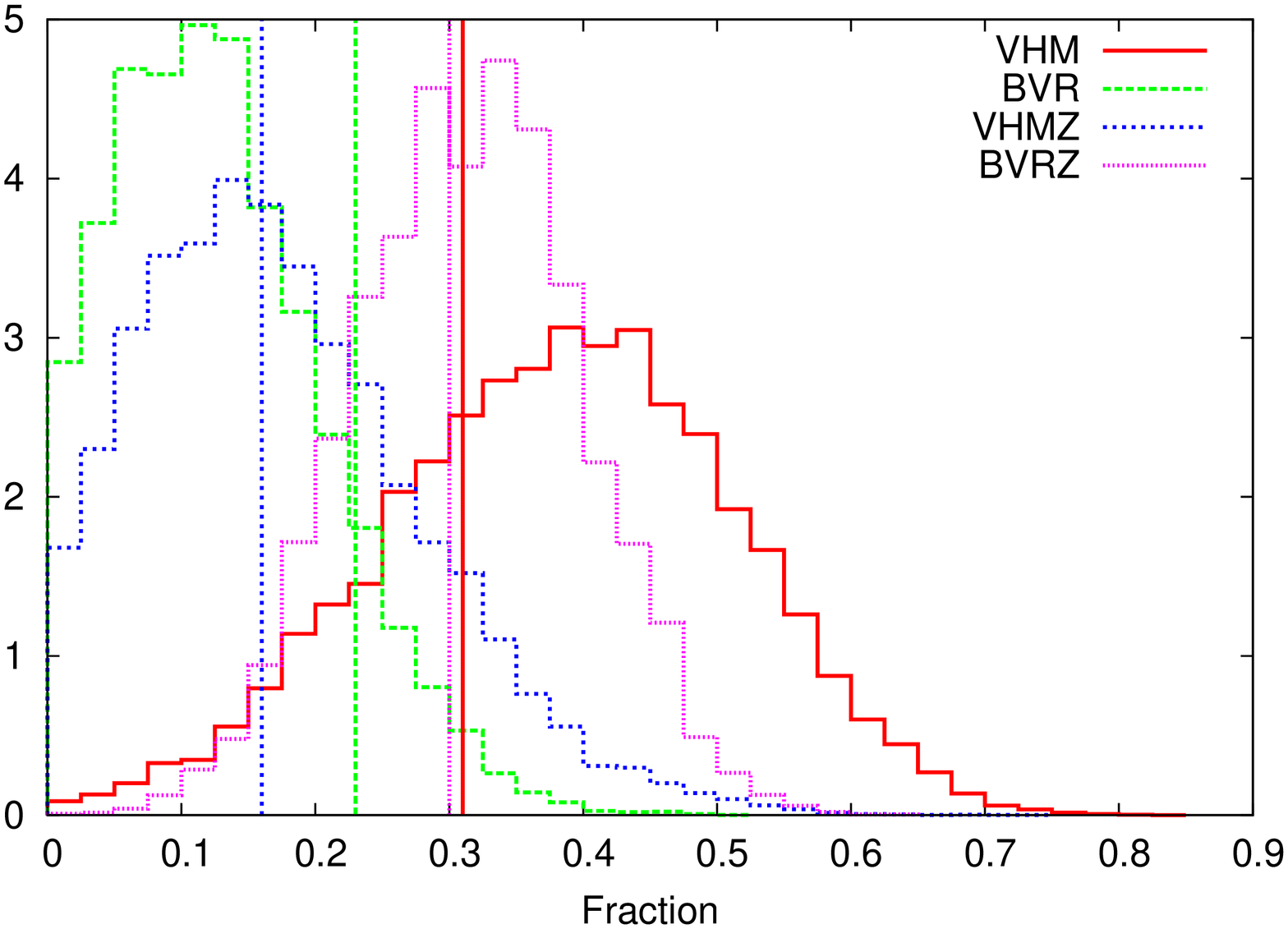}&
\includegraphics[scale=0.24,clip=true,angle=270]{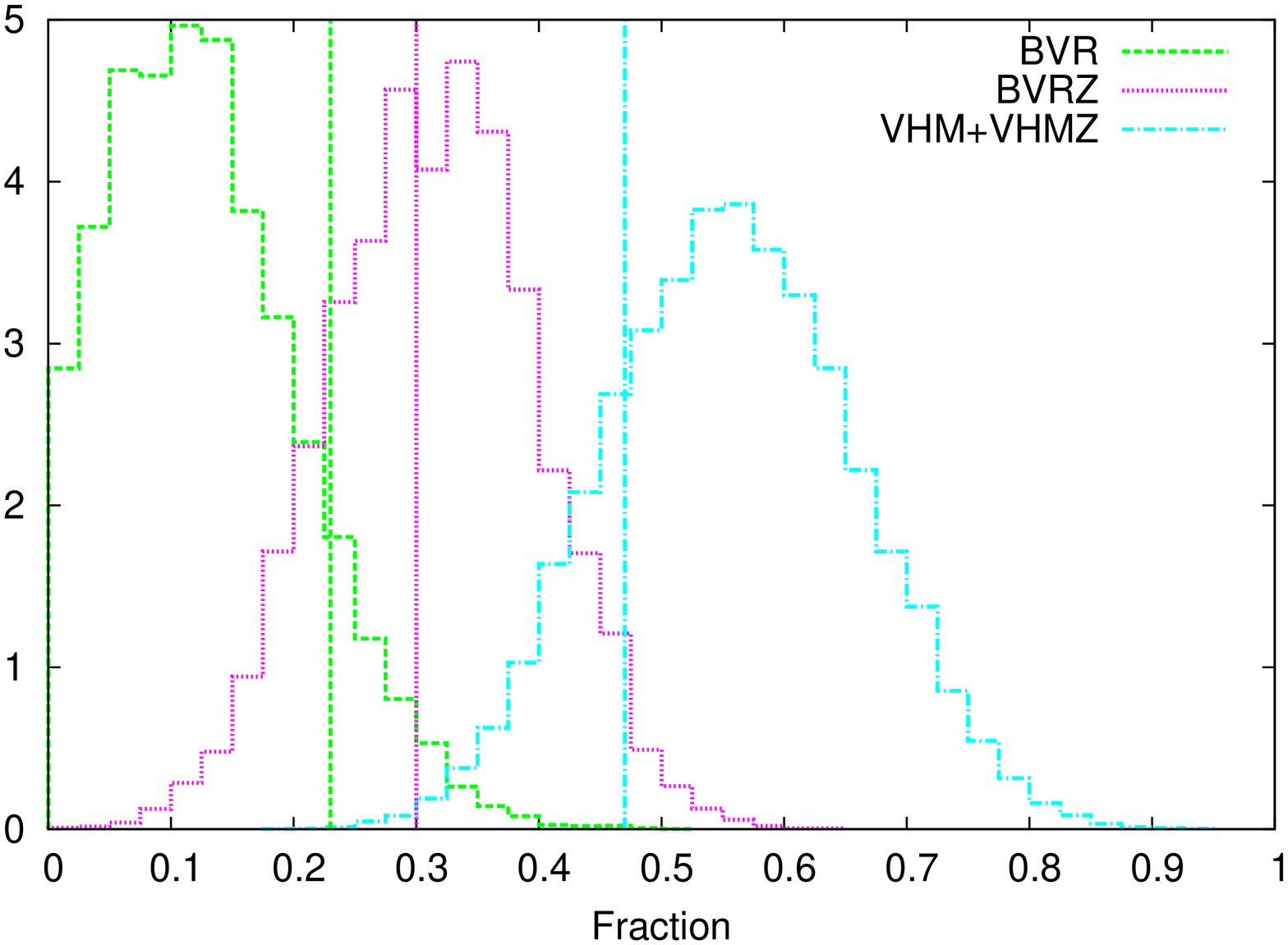}&
\includegraphics[scale=0.24,clip=true,angle=270]{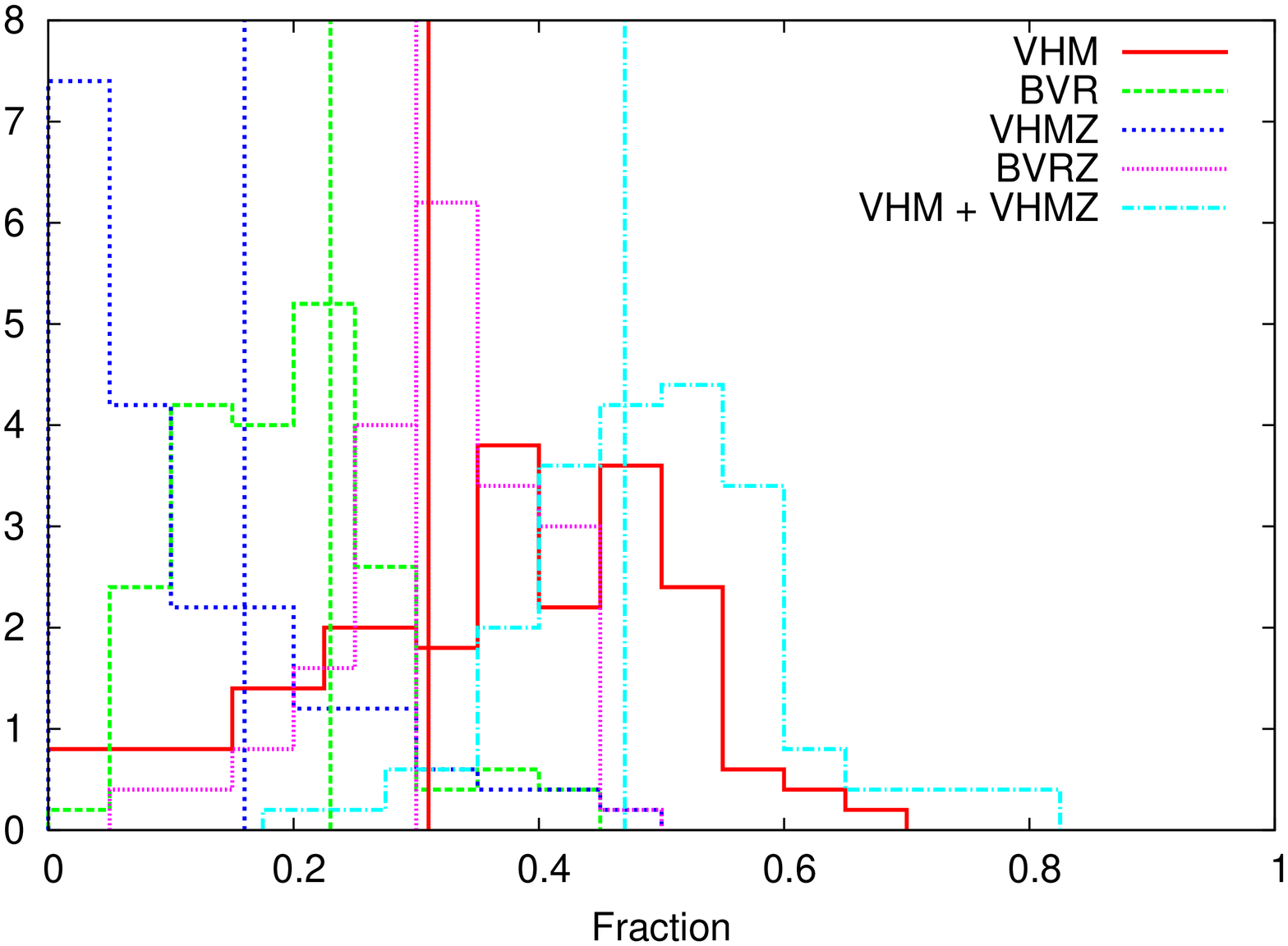}
\end{tabular}
\caption{Summary of model mixing results for model ${\cal N}$-III. We show the
  mixing fraction of the given model on the horizontal axis. The left panel
  shows the posterior probability distribution of the mixing fraction for each
  of the four models, {\it VHM}, {\it BVR}, {\it VHM-Z} and {\it BVR-Z}, found
  when analyzing a single realization of model ${\cal N}$-III. The central
  panel shows the same thing, but now considering the fractions of {\it BVR},
  {\it BVR-Z} and the sum {\it VHM}+{\it VHM-Z} in the mixed model. The right
  panel shows the distribution of the peak of the posterior pdf found over 100
  different realizations of the ${\cal N}$-III model.}
\label{fig:post67}
\end{figure*}

\section{\label{mixresults}Results for the mixed models}
In the context of mixed models, we are no longer comparing two pre-assigned
models A and B as descriptions of the observational data. We deal instead with
a single, continuous parameter space of models, where the parameters are the
mixing fractions of some subset of ``pure'' models. For example, if we mix
models 1 and 3, we have a one-dimensional parameter space given by the
contribution of model 1 ($f_1$) to the total population (the contribution of
model 3 is fixed by the constraint $f_1+f_3=1$). Given a particular
observation, we can then compute the posterior probability distribution
function (PDF) given by Bayes theorem, Eq.~(\ref{bayes}), for the mixing
fractions. The computation of the posterior can be done either over a grid of
points in the parameter space, or by exploring the parameter space by means of
Markov Chain Monte Carlo simulations (which become much more practical as the
dimension of the parameter space increases).

For each mixed model, 100 different realizations of the LISA data were
generated and a posterior probability distribution for the mixing fractions
was obtained for each one. The width of the posterior in a single realization
reflects how well that particular data set can constrain the mixing
fractions. The location of the peak of the posterior will change from
realization to realization, but we would expect the width to remain
approximately the same. We also expect that the distribution of the location
of the peak of the posterior over many realizations should resemble the
posterior for the mixing fractions computed in any single realization.

We considered a total of eight different mixed models, as listed in
Table~\ref{tabIV}, mixing either just {\it VHM-noZ-Edd-co} and {\it
  BVR-noZ-Edd-co} or these two models plus {\it VHM-Z-Edd-co} and {\it
  BVR-Z-Edd-co}. For each case, we assumed that we were using three years of
LISA data, but made pessimistic assumptions ($T_2$) for the transfer
function. While this latter assumption is slightly conservative, we checked
that there was not much difference in performance when using the most
optimistic assumptions (i.e., the transfer function $T_3$).

\begin{figure*}[htb]
\begin{tabular}{ccc}
\includegraphics[scale=0.33,clip=true]{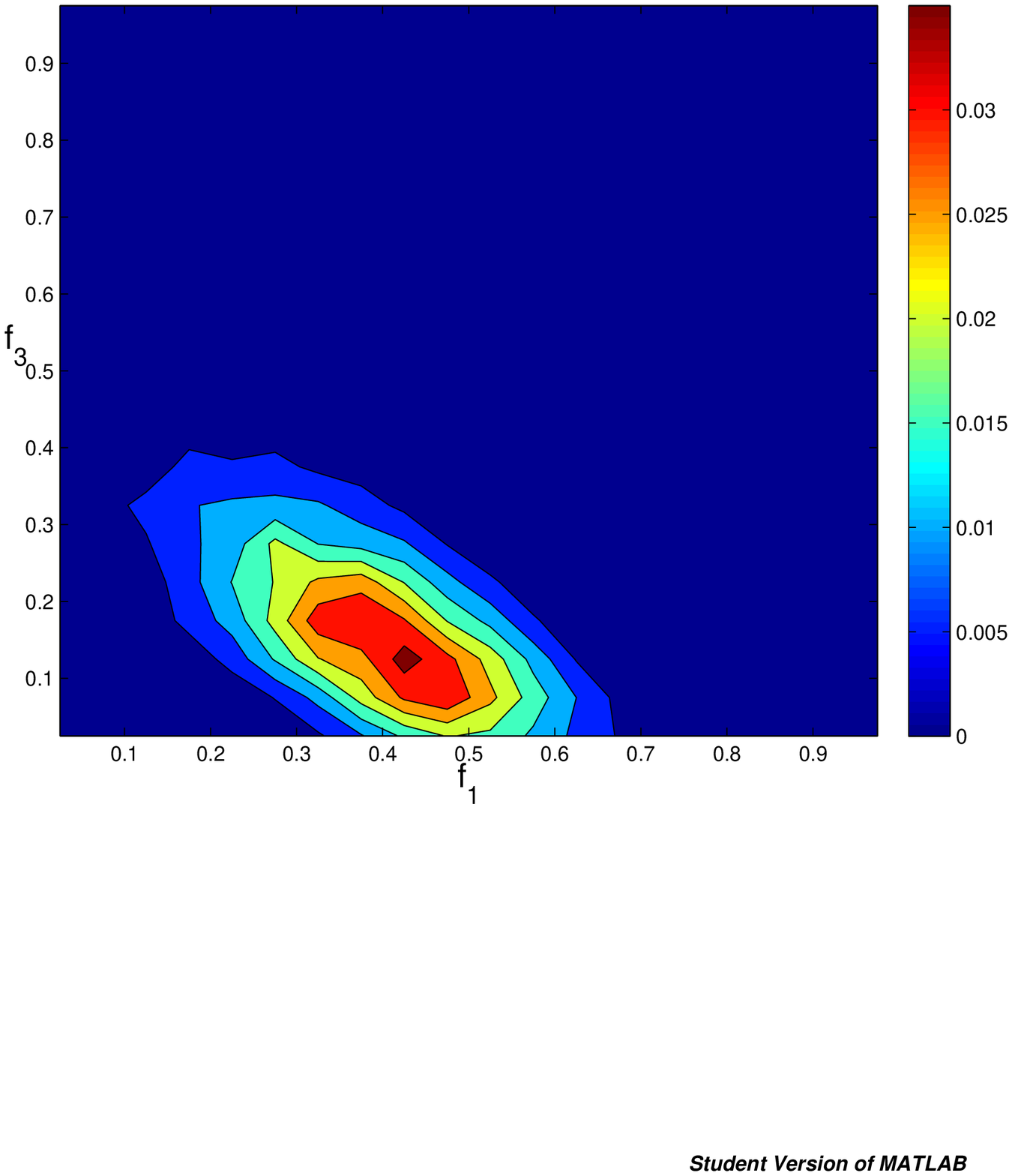}&
\includegraphics[scale=0.33,clip=true]{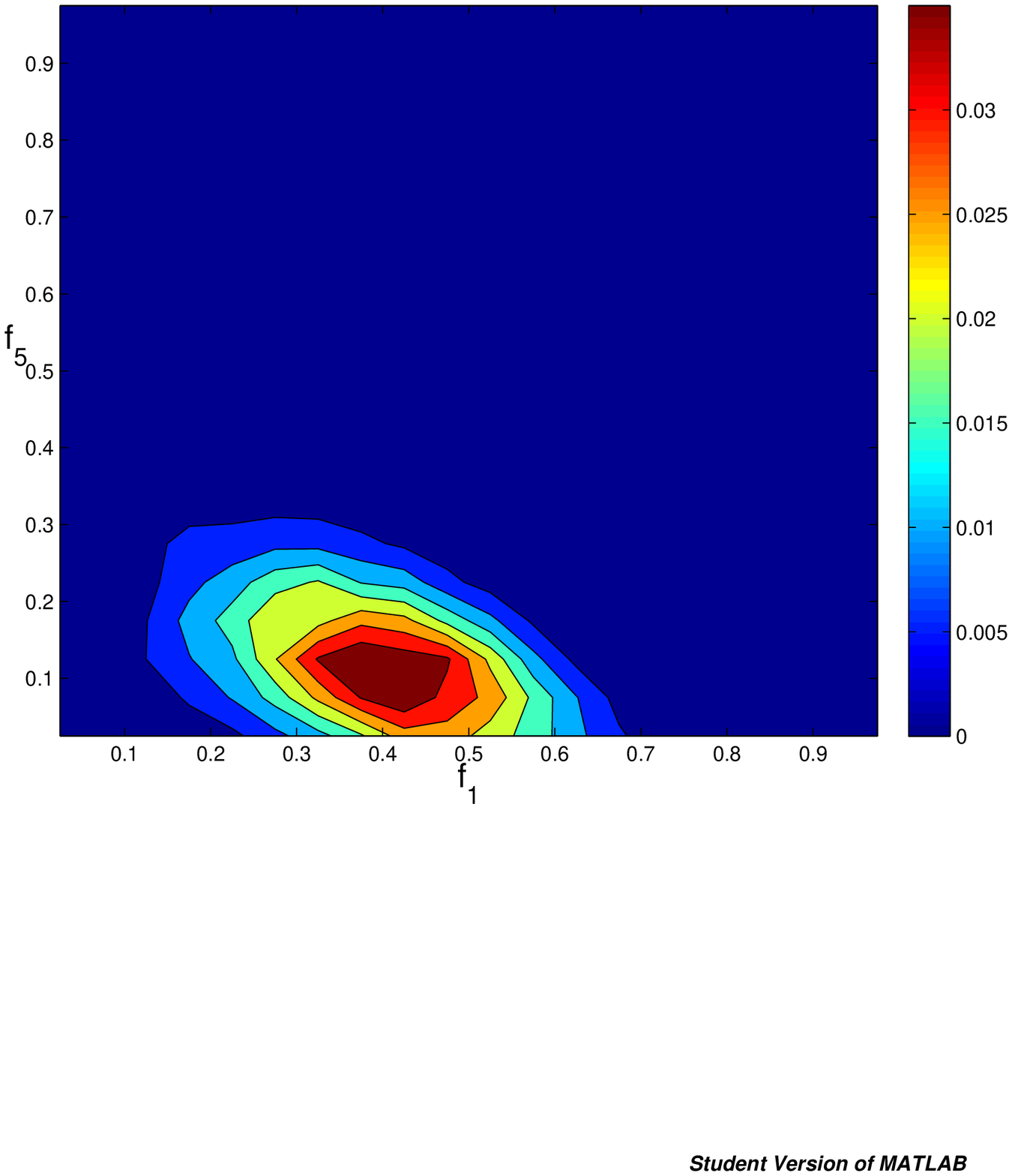}&
\includegraphics[scale=0.33,clip=true]{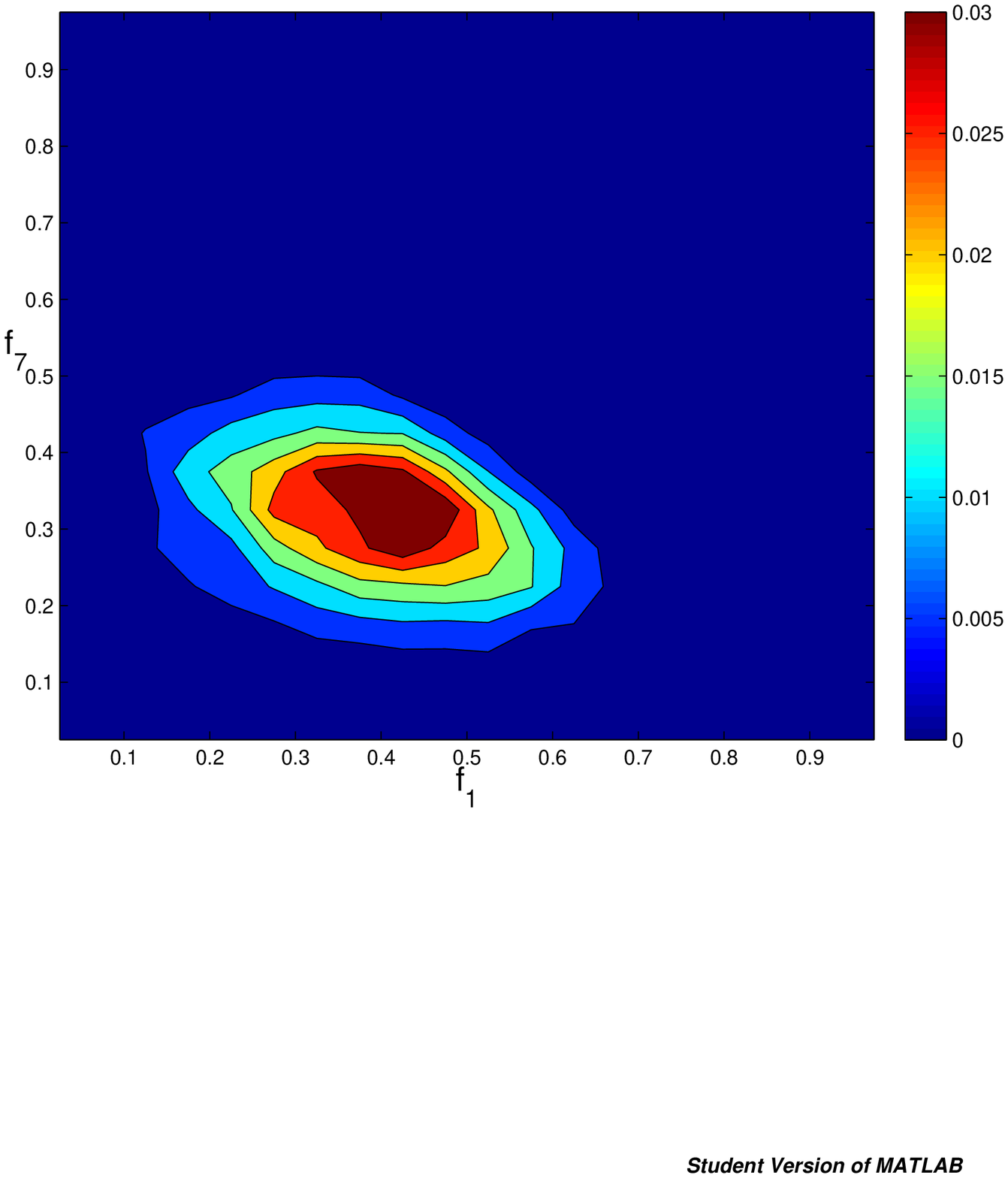}\\ 
\includegraphics[scale=0.33,clip=true]{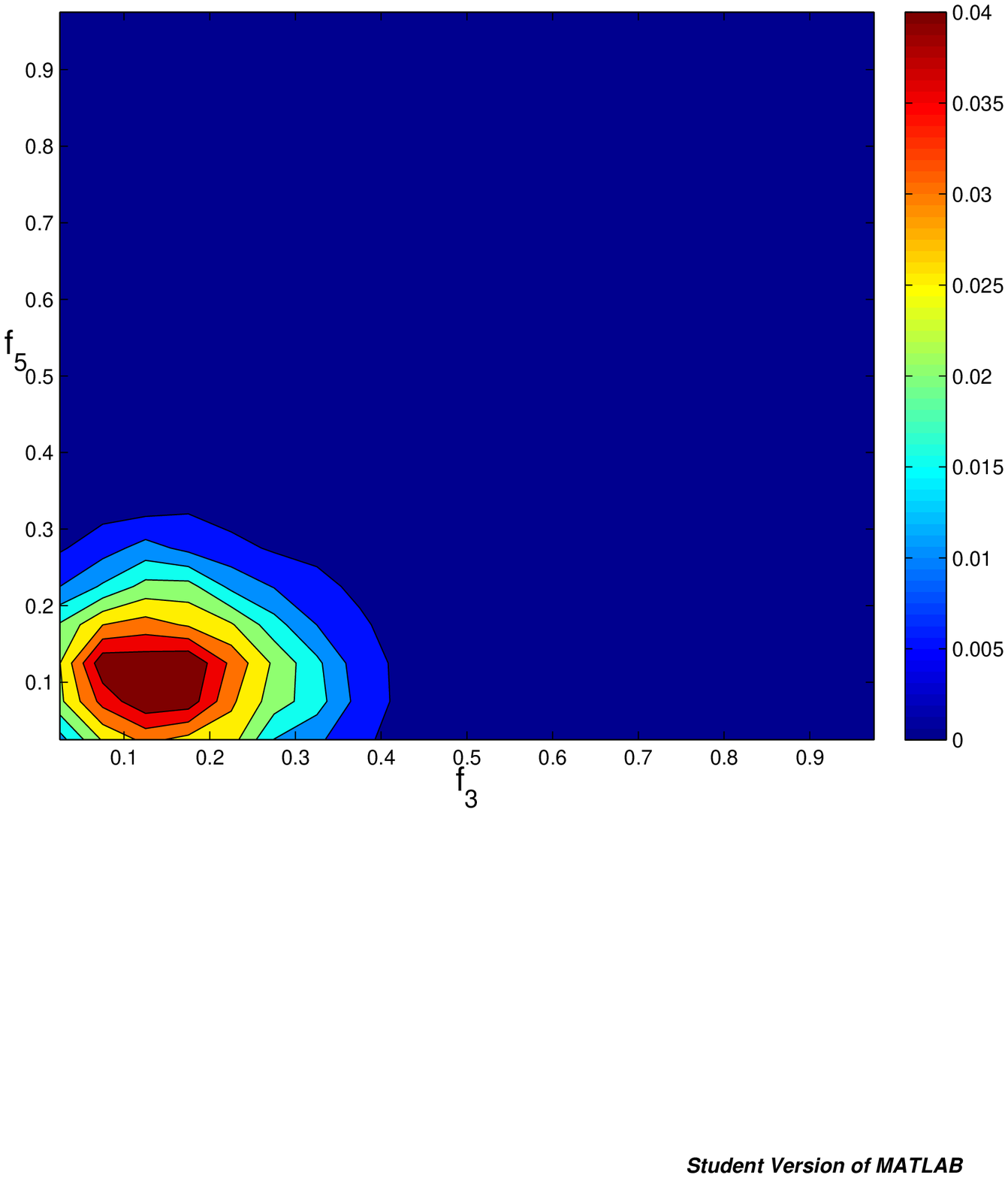}&
\includegraphics[scale=0.33,clip=true]{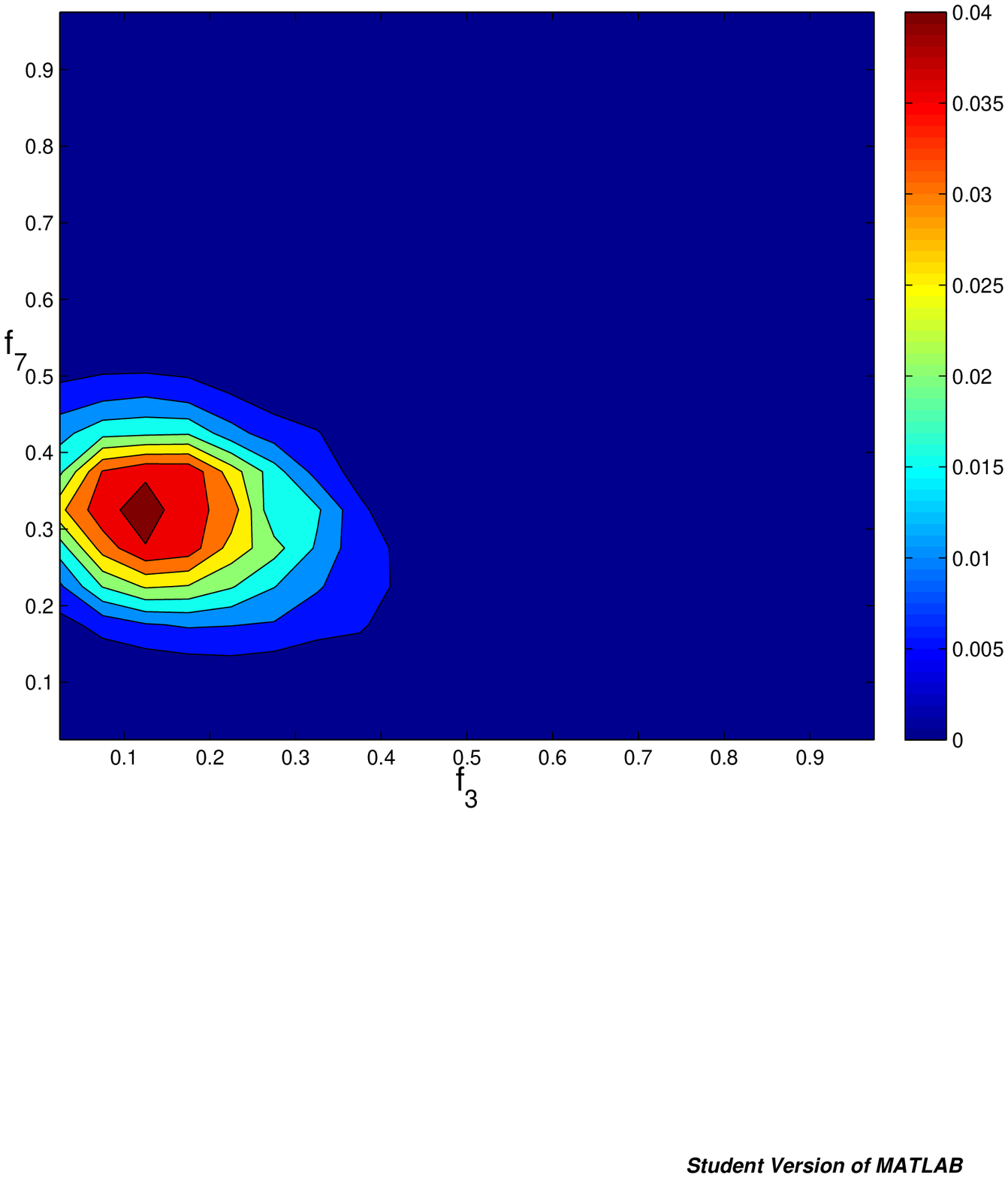}&
\includegraphics[scale=0.33,clip=true]{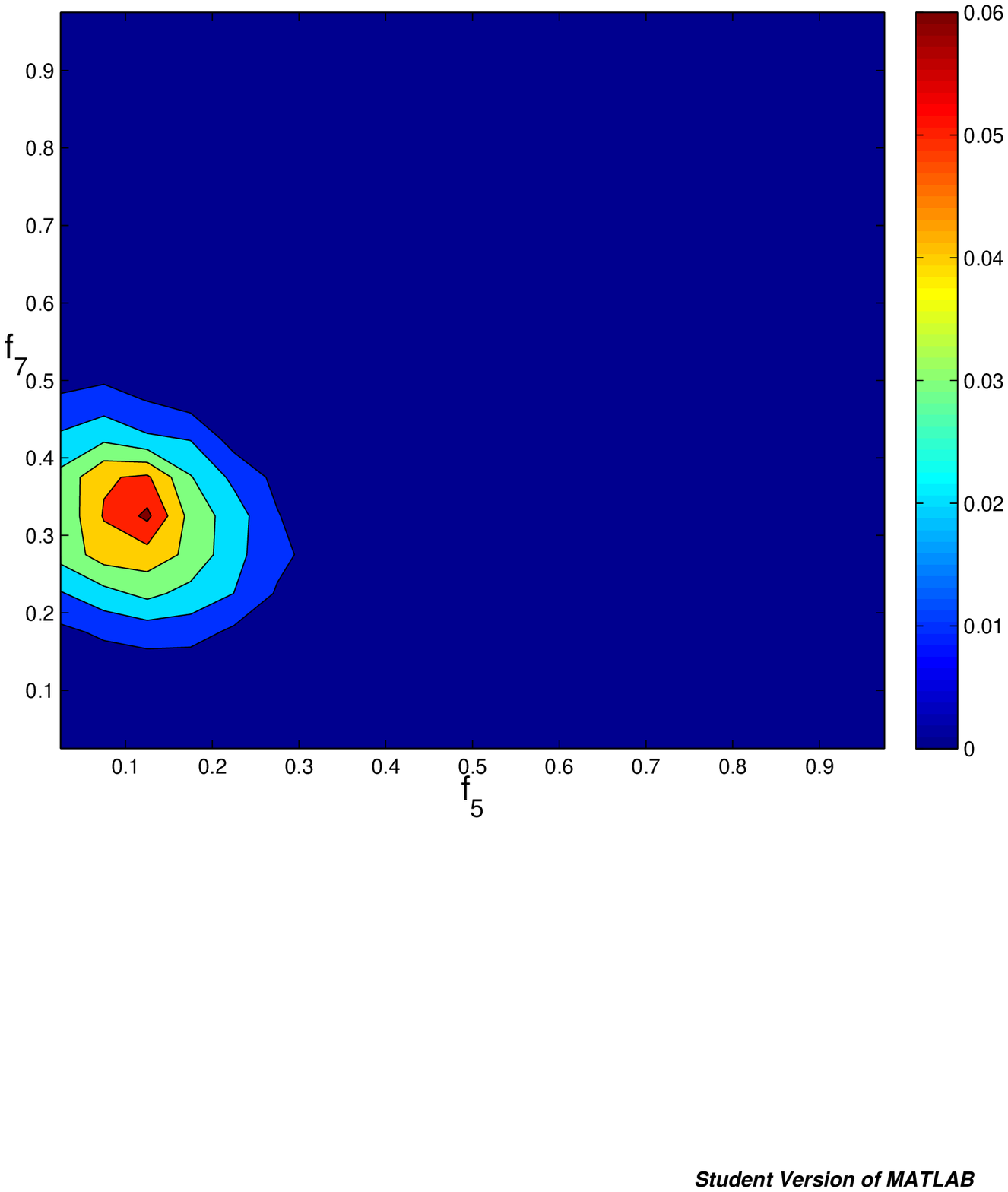}
\end{tabular}
\caption{Two-dimensional marginalized posterior PDFs obtained from a single
  realization of model ${\cal N}$-III. Each plot shows the mixing fraction of
  one model component against the mixing fraction of a second component. The
  models are numbered from $1$, $3$, $5$ and $7$, corresponding to {\it VHM},
  {\it VHM-Z}, {\it BVR} and {\it BVR-Z} respectively, as in
  Table~\ref{tabI}. The top row shows comparisons between model 1
  (horizontally) and models 3, 5 and 7, respectively. The bottom row shows
  comparisons 3 to 5, 3 to 7 and 5 to 7, respectively. Note that the
  individual components of {\it VHM} and {\it VHM-Z} are poorly constrained,
  which is why the plots involving {\it VHM} models have larger correlation
  contours than the {\it BVR} to {\it BVR-Z} comparison (bottom
  right).\label{fig:contmixed} }
\end{figure*}

\begin{figure}[htb]
\includegraphics[scale=0.43,clip=true]{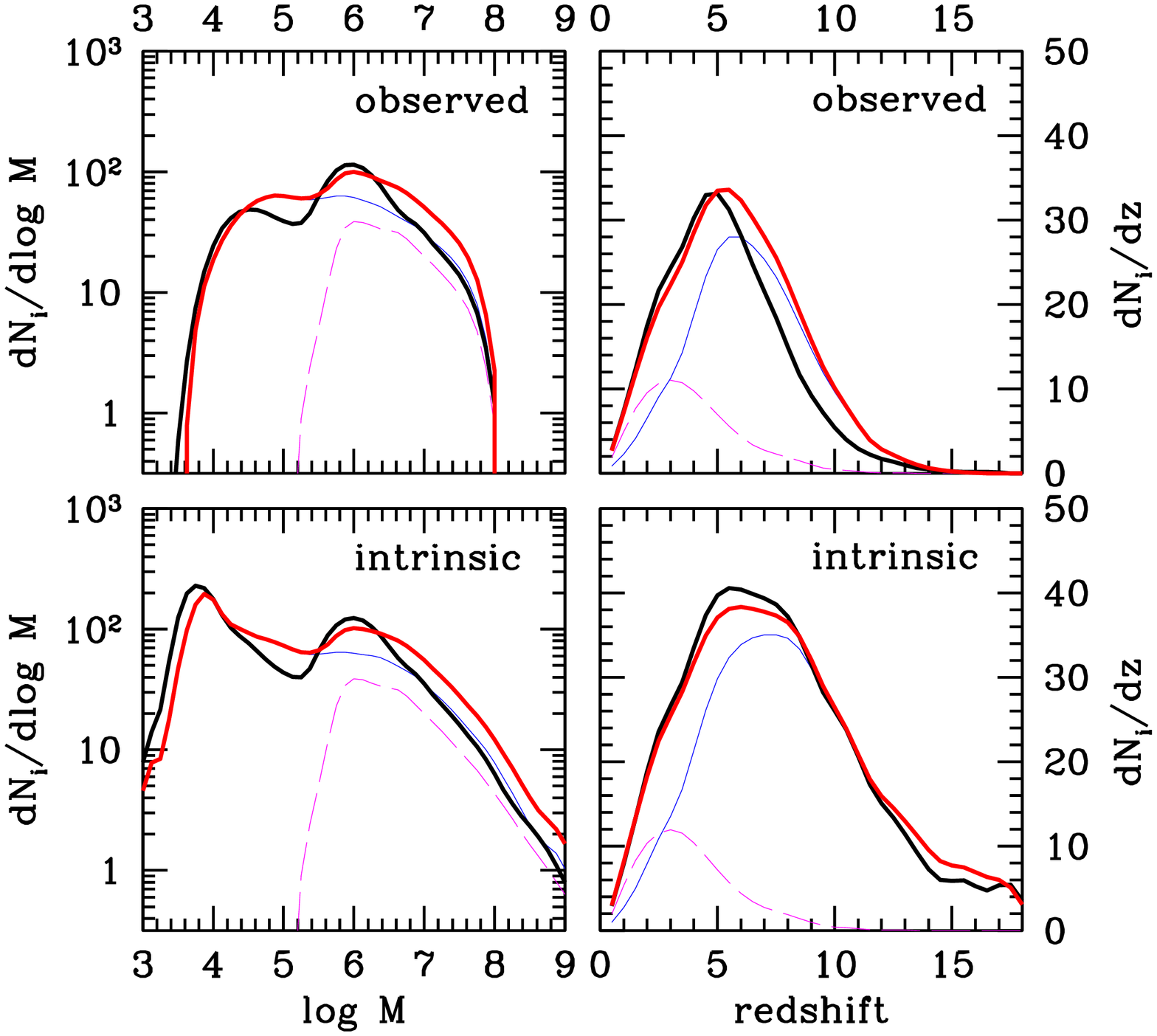}
\caption{Comparison of intrinsic and ``recovered'' distributions for the
  consistently mixed models. In each panel, the thick solid black curve shows
  the intrinsic distribution of mergers in the model, the red solid line shows
  the ``recovered'' distribution, which is a sum of the {\it VHM-Z} and {\it
    BVR-Z} models, weighted by the best-fit mixing parameter. The thin lines
  show the contributions to this recovered distribution from the {\it VHM-Z}
  (solid blue) and {\it BVR-Z} (dashed magenta) models.  The left panels show
  the distributions of the masses of mergers, while the right panels show the
  distributions of the redshifts of mergers. The upper panels show the
  distribution for observed merger events, while the lower panels show the
  intrinsic distribution of mergers.}
\label{fig:mix17}
\end{figure}

The posterior distributions of the mixing fraction found in one particular
realization each of models ${\cal N}$-I and ${\cal N}$-II are shown in the
left panel of Figure \ref{fig:post6566}. The PDFs peak around $f_1=0.2\pm0.05$
for model ${\cal N}$-I and $f_1=0.55\pm0.1$ for model ${\cal N}$-II, which is
consistent with the injected fractions listed in Table \ref{tabIV}. 

As mentioned above, we expect the peak and width of the posterior PDF to
fluctuate from realization to realization. To assess the statistical
robustness of this result we therefore repeat the experiment. In the right
panel of Figure \ref{fig:post6566}, we plot the distribution of the location
of the peak of the posterior PDF found in each of 100 realizations of the
models. As we expect, the widths of these distributions are very similar to
the PDFs found in each of the individual realizations. The distribution peaks
around $f_1=0.2\pm0.05$ for model ${\cal N}$-I and around $f_1=0.55\pm0.1$ for
model ${\cal N}$-II, in agreement with the true injected fractions listed in
Table \ref{tabIV}. This experiment shows that {\it most of the time} we can
correctly infer the relative contribution of the two models, but there is
still the possibility to draw erroneous conclusions from a single
observation. For example, in two realizations of model ${\cal N}$-II we would
prefer an almost pure {\it VHM} model, while the underlying distribution is in
fact 45\% {\it BVR}. However, in these cases the posterior PDF is also very
wide, which would be an indicator that the data set was not placing
particularly good constraints on the model in that specific case.

Figure \ref{fig:post67} shows the results for the more complex case of model
${\cal N}$-III, where all four of the pure models were mixed and the mixing
parameter space is three dimensional. Again, both the posterior PDFs given by
a specific realization (left panel) and the distribution of the peak values of
the posterior PDFs in a sample of 100 realizations return mixing fractions
which are consistent with the injected values (see Table
\ref{tabIV}). However, in this case the width of the {\it VHM-noZ} and {\it
  VHM-Z} posterior PDFs is large ($\sim 0.2$), indicating a certain degree of
degeneracy between those two models.  If we consider the sum, then the
posterior PDF is much narrower and peaks at the right value ($f_1+f_5=0.47$,
see Table \ref{tabIV}), showing that there is much less degeneracy between the
{\it VHM} and the {\it BVR} models. This is also nicely shown by the two
dimensional posterior PDFs plotted in Figure \ref{fig:contmixed}. All the
ellipse contours have principal axes more or less directed along the $x$ and
$y$ axes, with the exception of the {\it VHM-noZ} versus {\it VHM-Z} PDF,
which shows a clear anticorrelation between those two fractions.

Although we focused on the ${\cal N}$ models, the same level of accuracy in
the determination of the mixing fractions is achieved for $p$ models. The
results are collected in Table \ref{tabIV}.  The results shown in the table
refer to the most pessimistic transfer function; slightly better constraints
on the mixing fractions can be obtained if we assume two operational
interferometers and $\rho_{\rm thr}=8$.

As a final step we present our results for the consistent mixing model.  In
the two hybrid models {\it HY-I} and {\it HY-II}, {\it VHM-Z} and {\it BVR-Z}
seeding prescriptions are simultaneously employed in a consistent way in the
merger trees, and we do not expect the resulting binary population to be
perfectly reproduced by any combination of our pure models. The test here is
to combine the two ``pure'' {\it VHM-Z} ($i=3$) and {\it BVR-Z} ($i=7$) models
to see if we can mimic the true distribution with some combination of the
two. We proceed exactly as for the artificial mixing, by recovering the
posterior PDF of the mixing parameter. In this case we used only the $p$-type
mixing model, and included maginalization over the total number of events as
given by Eq.~(\ref{Nmargeq}). The rationale for this was that we thought a
consistent mixed model of this type would not necessarily have the same number
of events as the underlying models, and so we wanted to eliminate bias that
would be introduced by using the number-of-event information. We also computed
results using the ${\cal N}$-type mixing and/or not marginalizing over the
number. These results were also reasonable, but the match between the
intrinsic and recovered distributions was not as good.

For $p$-type mixing with marginalization over number, we find that model {\it
  HY-I} and {\it HY-II} are best reproduced by setting $f_3=0.85$ and
$f_3=0.45$, respectively. The marginalized mass and redshift distributions of
the best-fitting model are shown as red lines for model {\it HY-I} in Figure
\ref{fig:mix17}. As expected, we can not perfectly match the true model
distribution, but the overall agreement is good. Even though there is no
``true answer'' in this case, we can still extract useful information about
the models. For example, we can confidently say that in model {\it HY-II} the
contribution of the heavy seeding process is much higher than in model {\it
  HY-I}. This is consistent with the fact that model {\it HY-II} assumes a
much more efficient quasistar formation prescription.

\section{\label{conclusions}Conclusions}

In this paper we explored the ``inverse problem'' for GW observations, which
is fundamental in assessing the possible astrophysical impact of GW
astronomy. The question we addressed in this paper was: {\it given a sample of
  observed MBHB coalescences (with relative parameter estimation errors), what
  astrophysical information about the physical processes governing their
  formation and cosmological evolution can we extract from the observations?}
More informally: are GW observations a valuable tool for astrophysics?  We
answered this question by applying the statistical framework of Bayesian model
selection to simulated LISA observed datasets. We chose LISA as a case study,
but the analysis could straightforwardly be generalized to any other GW
detector.

We considered ten different ``pure'' MBH formation and evolution models (see
Table \ref{tabI}) differing in certain key aspects of the input physics,
specifically: (i) the seed formation mechanism, (ii) the redshift distribution
of the first seeds, (iii) the accretion efficiency during each merger and (iv)
the geometry of accretion (see Section \ref{mbhpop}). For each model we
computed the {\it intrinsic} coalescence distributions $d^3N_i/dMdqdz$. We
then constructed the {\it theoretically observable} distributions by filtering
the intrinsic distributions with four transfer functions $T_j$. These transfer
functions account for different levels of completeness of the LISA
observations according to four different sets of assumption about the
performance of the LISA detector (Section \ref{theodist}). For each model we
generated 1000 observed datasets (including observational errors), and we
analyzed them using a Bayesian model selection framework to assess their
distinguishability as a function of the detector performance and of the
duration of the dataset used for the model comparison. We find that:
\begin{itemize}

\item LISA will be able to discriminate among {\it almost} any pair of such
  ``pure'' models, even under pessimistic assumptions about the detector
  performance, after only one year of operation (see Table \ref{tabII}). In
  particular, it will be easy to identify the mass and redshift distribution
  of the seeds, and the efficiency of the accretion mechanism.

\item Models differing only by their accretion geometry are more difficult to
  discriminate. However, this was partly a consequence of our choice to
  consider measurements of only three parameters for each inspiralling binary
  (mass, mass ratio and redshift), i.e., we ignored the information encoded by
  MBH spins and in the merger/ringdown. Including spins in the analysis will
  probably make such models easily distinguishable, as demonstrated in a
  similar study by Plowman et al. ~\cite{plowman10}. In any case, even without
  the extra information carried by the spins, we can discriminate between
  these models if we consider the optimal LISA performance and three years of
  observation.

\item The impact of the detector performance on the analysis is relatively
  mild. This is because lowering the threshold to $\rho_{\rm thr}=8$ and
  considering two interferometers only adds a small number of sources to the
  detected sample, and only slightly improves parameter estimation.

\item Not surprisingly, the length of the observation is important, as the
  expected number of MBHBs in the sample increases linearly with the
  observation time. To give a specific figure of merit, {\it with a three-year
    observation window we have more than a $90\%$ probability that the parent
    model of an observed sample will be safely identified at a two-sigma
    confidence level (95\%).}

\end{itemize} 

To go beyond the pure model analysis, we considered the possibility of model
mixing. First we created new, ``artificial'' models by mixing the coalescence
distribution functions of different ``pure'' models (namely, models 1, 3, 5
and 7, see Tables \ref{tabI} and \ref{tabIV}). We used pure models differing
in their seeding mechanism and in the redshift distribution of the seeds
(different metallicity ``feedback'').  The new models are characterized by the
fractions $f_i$ of the ``pure'' models used in the mixing, with the constraint
$\sum_i f_i=1$. Then we considered two hybrid models, where halos were
simultaneously seeded according to the {\it VHM-Z} ($i=3$) and the {\it BVR-Z}
($i=7$) prescription, and the evolution of the seeds was self-consistently
followed in the halo hierarchy.  We produced several LISA observed datasets
for both the artificial and the hybrid models. We then tried to recover the
combination of ``pure'' models that best reproduces each mixed model by
maximizing over the posterior probability distribution function of the
likelihood (Section \ref{mixresults}). We find that:
\begin{itemize}

\item When the ``pure'' models used in the mixing differ only in their seeding
  prescription ({\it VHM} vs {\it BVR}), so that we have only a single mixing
  parameter (since $\sum_i f_i=1$), we can correctly infer this mixing
  parameter with an accuracy of about $10\%$.

\item When we mix four different models we can still infer the mixing
  fractions with the same accuracy, but there is a certain degree of
  degeneracy between the two {\it VHM} models ($i=1$ and $i=5$); i.e., the
  effect of metallicity ``feedback'', as the detectable MBH population does
  not differ much between these two models.  However, the fraction $f_1+f_5$
  is very well constrained, and we can clearly distinguish the relative
  contribution from the different seeding mechanisms.

\item Finally, we can also get a fairly good match to the hybrid models by
  combining ``pure'' models.  This is probably the most important result of
  our analysis. The formation and merger history of MBHs is a complex process,
  involving several physical ingredients which are poorly understood, and it
  is difficult to imagine that we will have a comprehensive theoretical
  understanding of the underlying physics before LISA flies. However, we will
  certainly be able to construct a set of models based on simple physical
  prescriptions that can be tested against the observations. Our experiment
  with the hybrid models demonstrates that {\it we can extract valuable
    information about more complex MBH formation scenarios by mixing a set of
    ``pure'' models based on simple recipes}.
\end{itemize} 

The use of a Bayesian framework is crucial for the model mixing results, since
it allows us to recover a posterior probability distribution for the ``mixing
parameters'' that characterizes the fraction of each model contributing to the
observed data set. In this respect, our analysis goes considerably beyond the
work recently presented in Ref.~\cite{plowman10}, where only pure models were
considered and the statistical analysis was based on two dimensional
Kolmogorov-Smirnov tests performed on the distributions of pairs of measured
parameters.

Despite this improvement, the building blocks of the present work can be
improved in many ways. The set of distinct ``pure'' models can not be
representative of all the physical complexity of the problem. A more powerful
approach to MBH population modelling would be to describe the relevant physics
using a set of continuous parameters representing the critical features of the
models (seed mass function, accretion efficiency and so on), and then attempt
to measure those parameters by performing a similar Bayesian analysis. We have
also adopted several simplifying assumptions about the GW observations, which
can be refined by developing a more realistic model for the GW signal,
including spins, higher harmonics, merger and ringdown. We can then attempt a
more sophisticated analysis and explore the posterior probability distribution
function in a larger and more complex parameter space, to maximize the
recovered information. All these issues should be explored in the future.

Besides the scientific impact of a GW detection in and by itself, the
ambitious goal of doing GW astronomy requires that we maximize the
astrophysical information that will be extracted from such detections. In this
respect, addressing the ``inverse problem'' in GW astronomy is extremely
important. In this paper we have made a first, small step in this
direction. We hope that our work will encourage relativists and GW astronomers
to consider in greater depth the astrophysical impact of GW detections. At the
same time, we hope to convince ``ordinary'' astronomers that GWs can be an
important tool, not only for tests of general relativity and as a laboratory
for fundamental physics, but also in astrophysics.

\begin{acknowledgments}
JG's work is supported by the Royal Society. EB's research was supported by
NSF grant PHY-0900735. MV was supported by NASA ATP Grant NNX07AH22G and a
Rackham faculty grant.
\end{acknowledgments}

\appendix
\section{\label{errprop}Error modelling}

We describe here how measurement errors are included in the analysis. Errors
arise due to instrumental noise in the LISA detector, and from the
transformation between different coordinates. The error propagation
expressions described below are probably not new, and the end result is
expected, but we include the derivation here for completeness and to clarify
the underlying assumptions.

LISA observations will determine the luminosity distance to a given source,
but we want to characterize the source by the redshift instead. The conversion
can be done using the concordance cosmology at the time LISA flies, but this
introduces additional errors, since the cosmological parameters will be known
imperfectly. Suppose we want parameters $\vec{x}$ which are given by the
measured parameters, $\vec{y}$, and a transformation
$\vec{x}(\vec{y};\vec{\lambda})$ that depends on some imperfectly known
parameters, $\vec{\lambda}$.  Suppose further that the probability
distribution for $\vec{\lambda}$ is $L(\vec{\lambda})$ and that for $\vec{y}$
is $Y(\vec{y})$. The probability distribution for $\vec{x}$ is then
\be
X(\vec{x}) = \int L(\vec{\lambda}) Y(\vec{y}(\vec{x};\vec{\lambda})) J(\vec{x}; \lambda) {\rm d}^{n_\lambda} \lambda\,,
\label{errtrans}
\ee
in which $J(\vec{x}; \lambda)$ is the Jacobian for the transformation between
$\vec{y}$ and $\vec{x}$ and $n_{\lambda}$ is the dimensionality of the
$\vec{\lambda}$ parameter space. We now make two simplifying assumptions: (i)
the distributions of the errors in $\vec{y}$ and $\vec{\lambda}$ are
multi-variate Gaussians with inverse variance-covariance matrices $\Gamma^y$
and $\Gamma^\lambda$ respectively; (ii) the errors are small, so that the
distributions are peaked near the true values of $\vec{y}_0$ and
$\lambda_0$. This latter assumption means that we can use a linear
approximation in the interesting region of the distributions
\be
y_i(\vec{x};\vec{\lambda}) \approx y_i(\vec{x}_0;\lambda_0) + \frac{\partial y_i}{\partial x_j} (x_j-x_{0j}) +  \frac{\partial y_i}{\partial \lambda_j} (\lambda_j-\lambda_{0j})\,,
\label{errderiv}
\ee
where the derivatives are evaluated at $\vec{y}_0$, $\vec{\lambda}_0$. We can
also ignore the Jacobian in the integrand of Eq.~(\ref{errtrans}), since it
will be approximately constant across the domain of integration and therefore
it plays the role of a normalization factor. Using the notation
\be
\tilde{x}_i = x_i - x_{0i}, \;\; \delta{\lambda}_i=\lambda_i - \lambda_{0i},\;\;
D^x_{ij}=\frac{\partial y_i}{\partial x_j}, \;\;
D^\lambda_{ij}=\frac{\partial y_i}{\partial \lambda_j}\,,
\ee
we see that the integrand is proportional to the exponential of
\beq
&-\frac{1}{2}&\left\{ \left((D^x {\bf \tilde{x}})^T + \delta{\bf \lambda}^T D^\lambda \right)\Gamma^y \left(D^x {\bf \tilde{x}}+(D^\lambda)^T\delta{\bf \lambda} \right) \right.\nonumber\\&&\left.+\delta{\bf \lambda}^T \Gamma^\lambda \delta{\bf \lambda}\right\}\,,
\eeq
where matrix notation is being used. This can be rearranged to give
\beq
&-\frac{1}{2}&\left\{(D^x {\bf \tilde{x}})^T \Gamma^y D^x {\bf \tilde{x}}-{\bf a}^T (\Gamma^\lambda+D^\lambda\Gamma^y(D^\lambda)^T {\bf a}\right.\nonumber\\&&\left.+\left(\delta{\bf \lambda}+{\bf a}\right)^T \left(\Gamma^\lambda+D^\lambda\Gamma^y(D^\lambda)^T\right)\left(\delta{\bf \lambda}+{\bf a}\right)\right\} \label{errint}\,,
\eeq
where
\be
{\bf a} = \left(\Gamma^\lambda+D^\lambda\Gamma^y(D^\lambda)^T\right)^{-1} D^\lambda \Gamma^y D^x {\bf\tilde{x}}\,.
\ee
The term on the second line of Eq.~(\ref{errint}) is just a Gaussian, whose
centre has been shifted to ${\bf a}$, and with variance-covariance matrix that
is independent of ${\bf \tilde{x}}$. When we integrate over the distribution
of $\vec{\lambda}$, i.e., over $\delta {\bf \lambda}$, we find the probability
distribution is proportional to the exponential of
\be
-\frac{1}{2}\left\{(D^x {\bf x})^T \Gamma^y D^x {\bf x}-{\bf a}^T (\Gamma^\lambda+D^\lambda\Gamma^y(D^\lambda)^T {\bf a}\right\}\,,
\ee
which is a multi-variate Gaussian with variance-covariance matrix $\Gamma^x$ equal to
\be
(D^x)^T \left( \Gamma^y - \Gamma^y(D^\lambda)^T\left(\Gamma^\lambda+D^\lambda\Gamma^y (D^\lambda)^T \right)^{-1}D^\lambda\Gamma^y\right)D^x.
\ee
Although this expression looks complicated, the inverse of this matrix takes
the simple form
\be
(\Gamma^x)^{-1} = (D^x)^{-1} \left((\Gamma^y)^{-1} + (D^\lambda)^T(\Gamma^\lambda)^{-1}D^\lambda\right) ((D^x)^T)^{-1} .
\ee
As it is this inverse matrix which determines the width of the distributions,
we see that it takes the form we might expect, i.e., the error is the sum of
the error contributions from the instrumental noise, $\Gamma^y$, and that from
the uncertainty in the cosmological parameters, $\Gamma^\lambda$. The remaining
terms just propagate the errors through the transformation in the standard
way.

In this paper, we estimate the errors in the observed parameters, $\Gamma^y$,
using the Fisher matrix formalism of Ref.~\cite{Berti:2004bd}. These errors
are given in terms of the chirp mass, ${\cal M}$, the amplitude, $A$, and the
symmetric mass ratio, $\eta$, so we must transform these coordinates to mass, $M$,
luminosity distance, $D_L$, and mass ratio, $q$. We convert luminosity distance
to redshift by inverting the standard cosmological relation of
Eq.~(\ref{DLz}). We assume that there are errors in $H_0$ and $\Omega_\Lambda$
but enforce flatness (i.e., $\Omega_M+\Omega_\Lambda=1$).

The diagonal components of the total error matrix in the new variables,
$(\Gamma^x)^{-1}$, are then
\beq
\left(\Gamma^x\right)^{-1}_{\ln q\ln q}&=&\frac{1}{1-4\eta}\left(\Gamma^y\right)^{-1}_{\ln\eta\ln\eta}\,, \nonumber\\
\left(\Gamma^x\right)^{-1}_{\ln M\ln M}&=&\left(\Gamma^y\right)^{-1}_{\ln{\cal M}\ln{\cal M}}+
\frac{9}{25}\left(\Gamma^y\right)^{-1}_{\ln\eta\ln\eta}\nonumber\\ &&-
\frac{6}{5}\left(\Gamma^y\right)^{-1}_{\ln{\cal M}\ln\eta}, \nonumber\\
\left(\Gamma^x\right)^{-1}_{zz}&=&\left(\frac{\partial D_L}{\partial z}\right)^{-2}\left[D_L^2 \left(\frac{5}{6}\left(\Gamma^y\right)^{-1}_{\ln{\cal M}\ln{\cal M}}\right.\right.\nonumber\\&&
+\left(\Gamma^y\right)^{-1}_{\ln A\ln A}+
\frac{5}{3}\left(\Gamma^y\right)^{-1}_{\ln{\cal M}\ln A}\nonumber\\&&\left.\left.+
\frac{\Delta H_0^2}{H_0^2}\right)+\Delta\Omega_\Lambda^2\left(\frac{\partial D_L}{\partial \Omega_\Lambda}\right)\right].
\eeq
where $(\Gamma^y)^{-1}_{ij}$ denotes the components of the inverse Fisher
matrix as given in Ref.~\cite{Berti:2004bd}, and $\Delta H_0$, $\Delta
\Omega_\Lambda$ are the errors in the cosmological parameters at the time of
the LISA mission, which we take to be $\Delta H_0/H_0 = \Delta
\Omega_{\Lambda}/\Omega_{\Lambda}= 0.01$. The off-diagonal components in the
total error matrix come only from the rotation of the noise error matrix,
$(D^x)^{-1} (\Gamma^y)^{-1} ((D^x)^T)^{-1}$, but in practice we ignore these
and draw errors based on the diagonal variance-covariance matrix with
components as above. This is conservative, in that it approximates the error
ellipse by a bounding circle, but we have also checked that our results did not
significantly change when they were recomputed using the full error model
including correlations.

\begin{figure*}[htb]
\begin{tabular}{ccc}
\includegraphics[scale=0.33,clip=true]{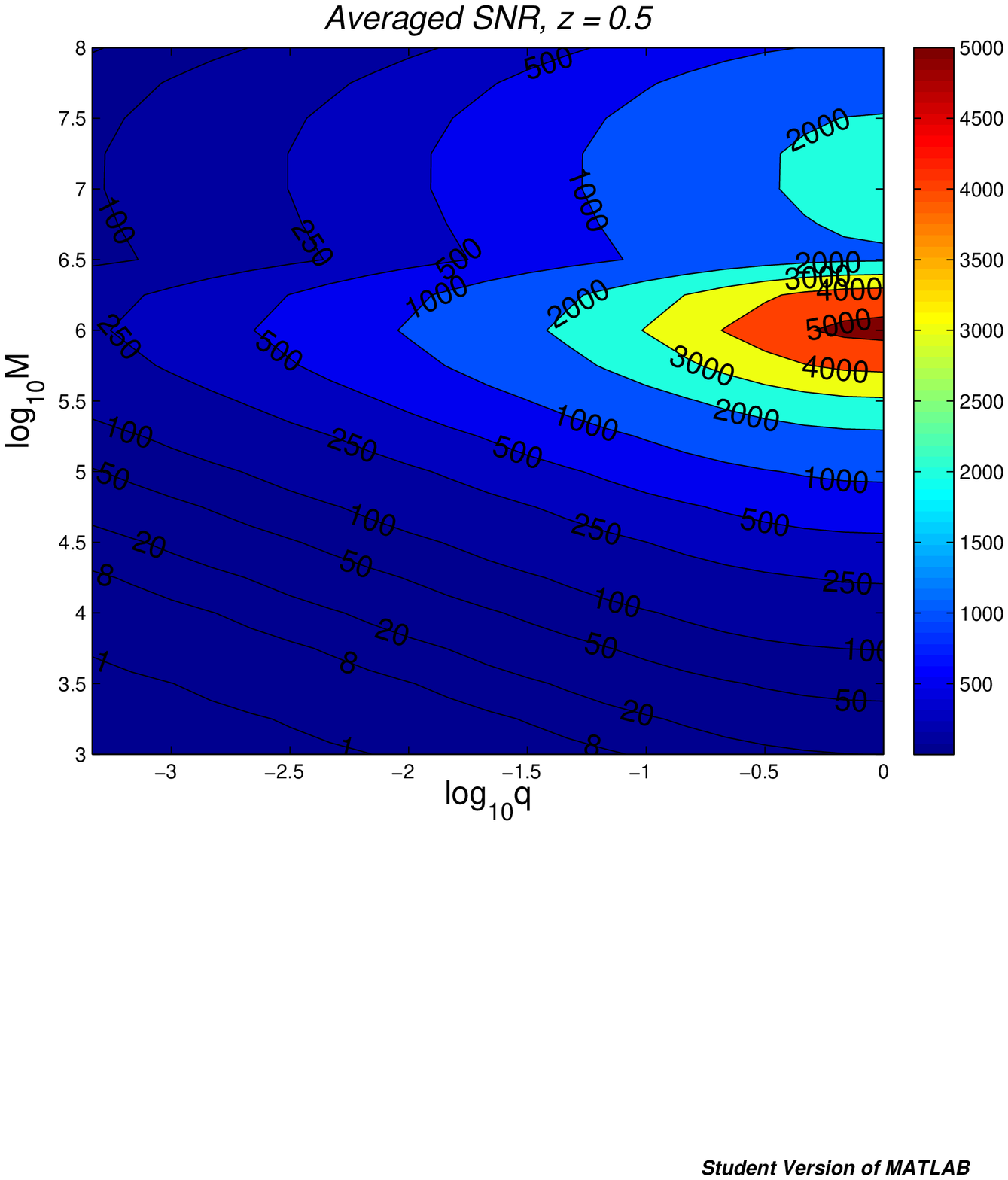}&
\includegraphics[scale=0.33,clip=true]{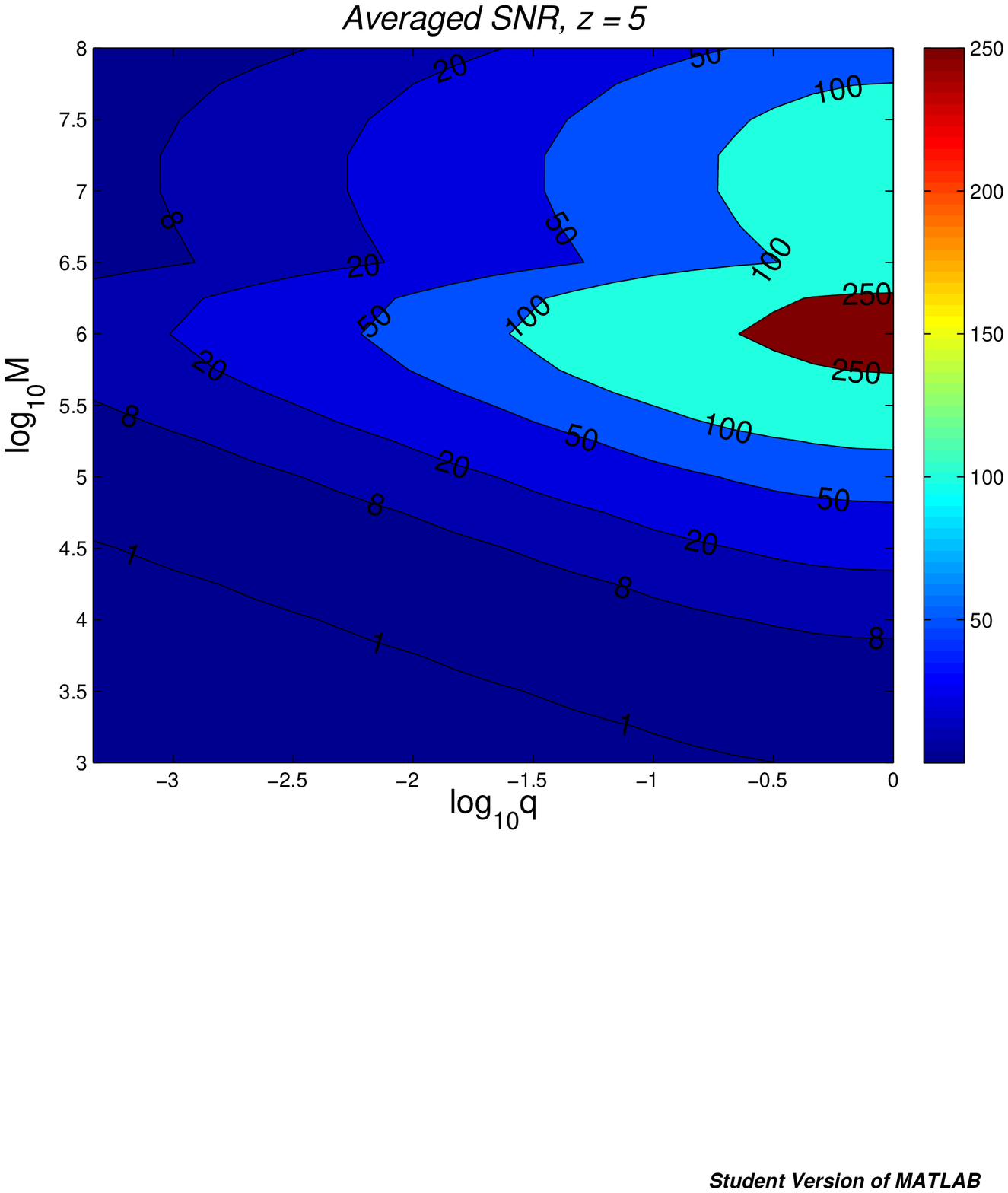}&
\includegraphics[scale=0.33,clip=true]{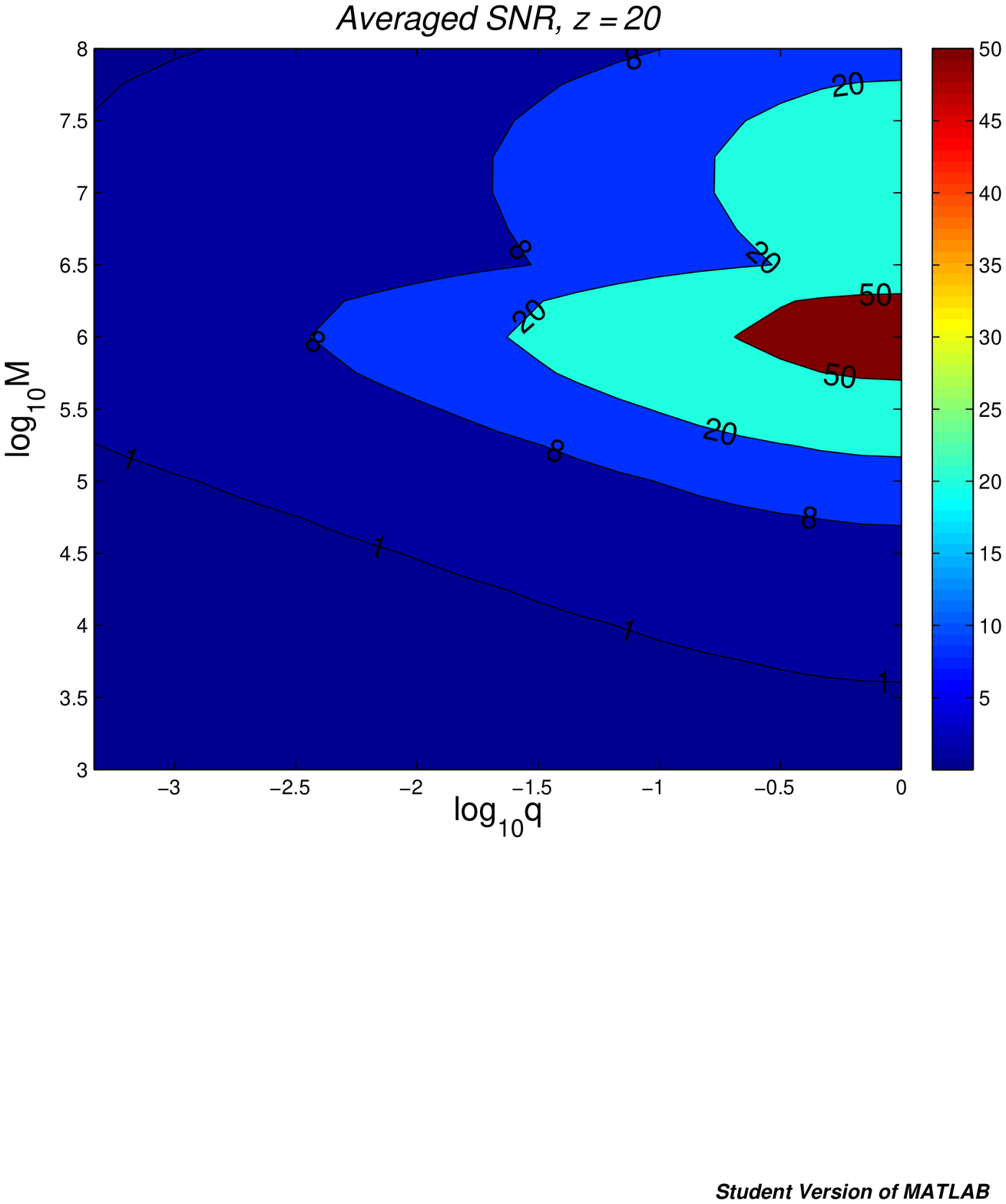}\\
\includegraphics[scale=0.33,clip=true]{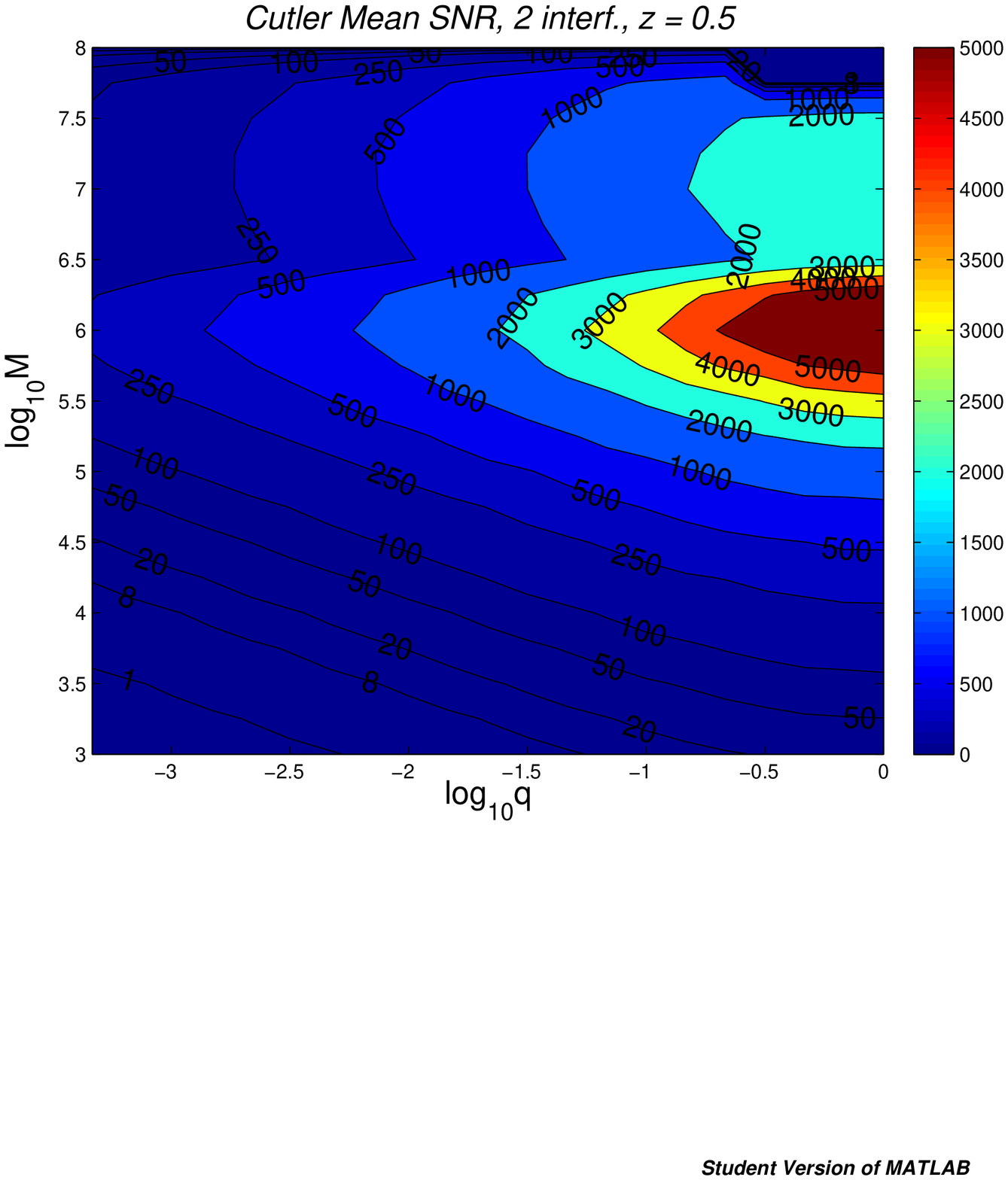}&
\includegraphics[scale=0.33,clip=true]{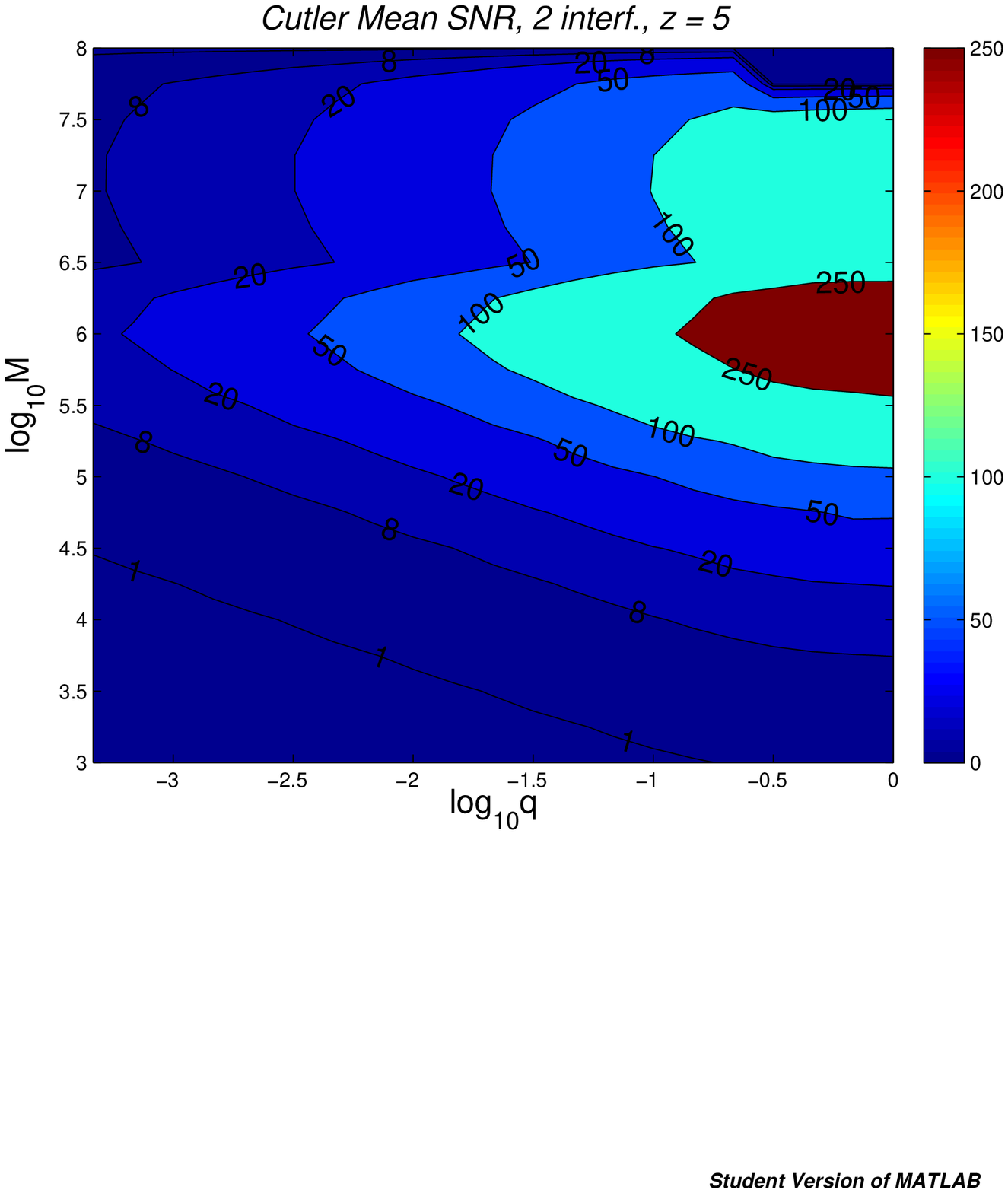}&
\includegraphics[scale=0.33,clip=true]{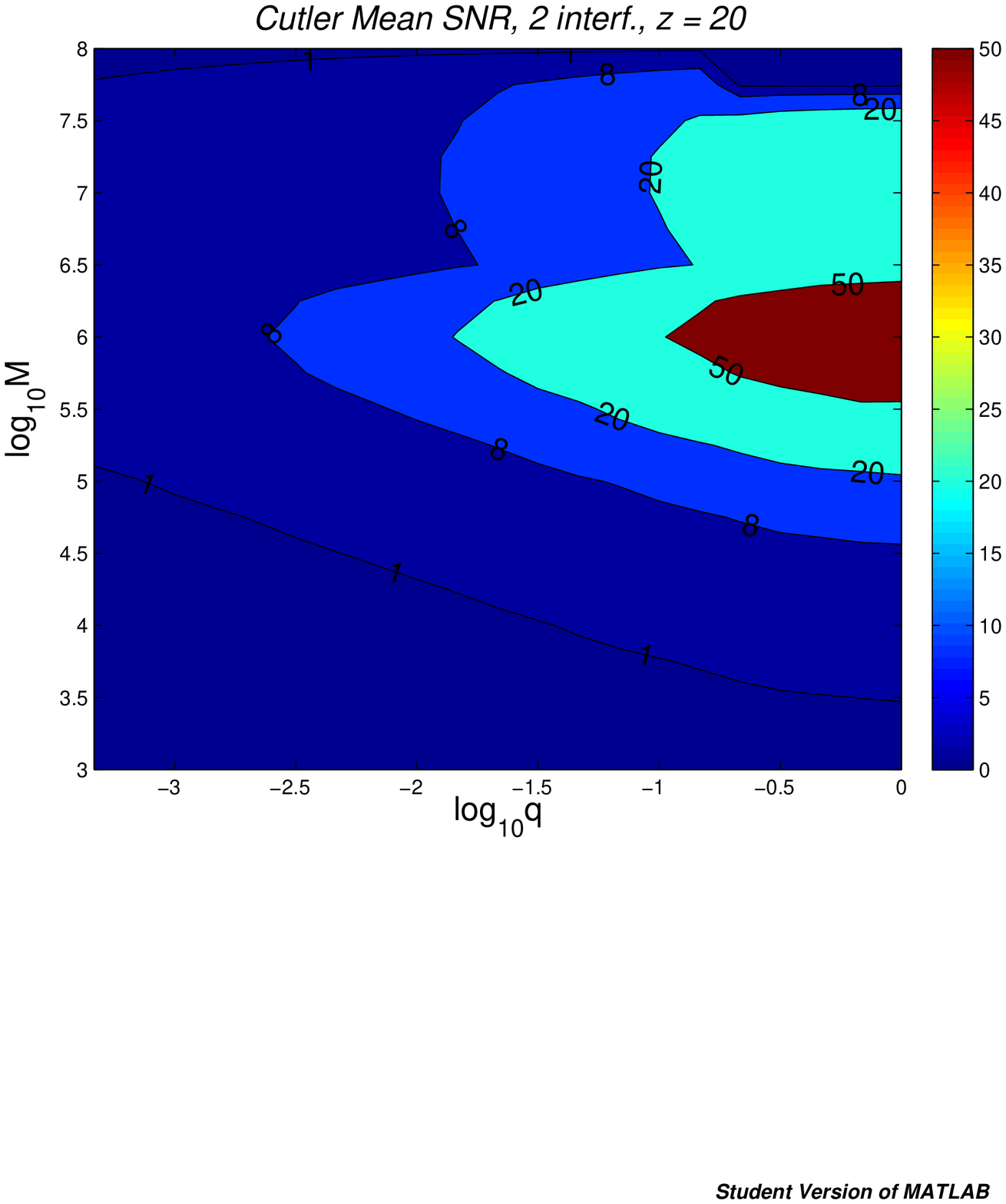}\\
\end{tabular}
\caption{SNR contours in the $(M\,,q)$ plane at different redshifts (from left
  to right: $z=0.5\,,5$ and $20$). The top row refers to the averaged code,
  the bottom row to the Cutler code with $T_{\rm obs}=3$~yrs and two
  interferometers.\label{fig:SNR} }
\end{figure*}

We note that there is a singularity in the transformation between $\eta$ and
$q$ when $q=1$ ($\eta=0.25$, which is indicated by the divergence of
$\left(\Gamma^x\right)^{-1}_{qq}$ there). The probability that two galaxies
with exactly the same black hole mass merge is zero, and so this was not a
problem in practice.

Additional errors arise from the effects of weak lensing. This means that the
apparent luminosity distance, $d_A$, of a source at the Earth is changed by a
factor $\mu$ from the true luminosity distance, $d_T$, so that $d_A = \mu
d_T$. If we assume that the weak-lensing (de-)magnification is drawn from a
Gaussian distribution (which is a reasonable approximation in the weak-lensing
limit, although not for stronger lensing), we can use the preceding arguments
in this case as well. The parameters $\vec{x}$ are the parameters we assign to
the source, which are the same as the measured parameters $\vec{y}$. However,
for a fixed value of $\mu$, the distribution of $\vec{y}$ is centred at a
luminosity distance $\mu d_T$. When we marginalize over possible values of
$\mu$, we end up with an integral of the form~(\ref{errtrans}), but the
dependence of the integrand on the unknown parameter is through the centre of
the distribution, $\vec{y}_0$, rather than through the mapping to
$\vec{x}$. We can follow the same arguments as before, but replace the
derivatives in equation~(\ref{errderiv}) by derivatives of
$\vec{y}-\vec{y}_0$. The end result then takes the same form. If the
distribution of $\mu$ is a Gaussian $\exp(-\Gamma_{\mu\mu}(\mu-1)^2/2)$, we
find the $(\Gamma^x)^{-1}_{dd}$ component must be corrected by addition of
$d_T^2/\Gamma_{\mu\mu}$. In practice, we take the weak-lensing error estimate
$\Delta z_{\rm wl}$ from Ref.~\cite{wangholz}, and directly modify the $zz$
component as $(\Gamma^{-1})_{zz} \rightarrow (\Gamma^{-1})_{zz}+(\Delta z_{\rm
  wl})^2$.

While we formulated the above in terms of computing the distribution of errors
in the parameters we measure from our observation, the same framework can be
applied to the analysis of the real LISA data set. Once we obtain a posterior
PDF for the measured waveform parameters, $Y(\vec{y})$, we can combine this
with a posterior on the cosmological parameters and on the lensing
magnification distribution through Eq.~(\ref{errtrans}) to derive the
posterior on the inferred source parameters $\vec{x}$. With the assumption
that these measured posteriors are Gaussians, the final result will take the
same form.

\section{\label{app:average}Effect of LISA motion on SNR and estimation of
  intrinsic parameters}

\begin{figure*}[htb]
\begin{tabular}{ccc}
\includegraphics[scale=0.33,clip=true]{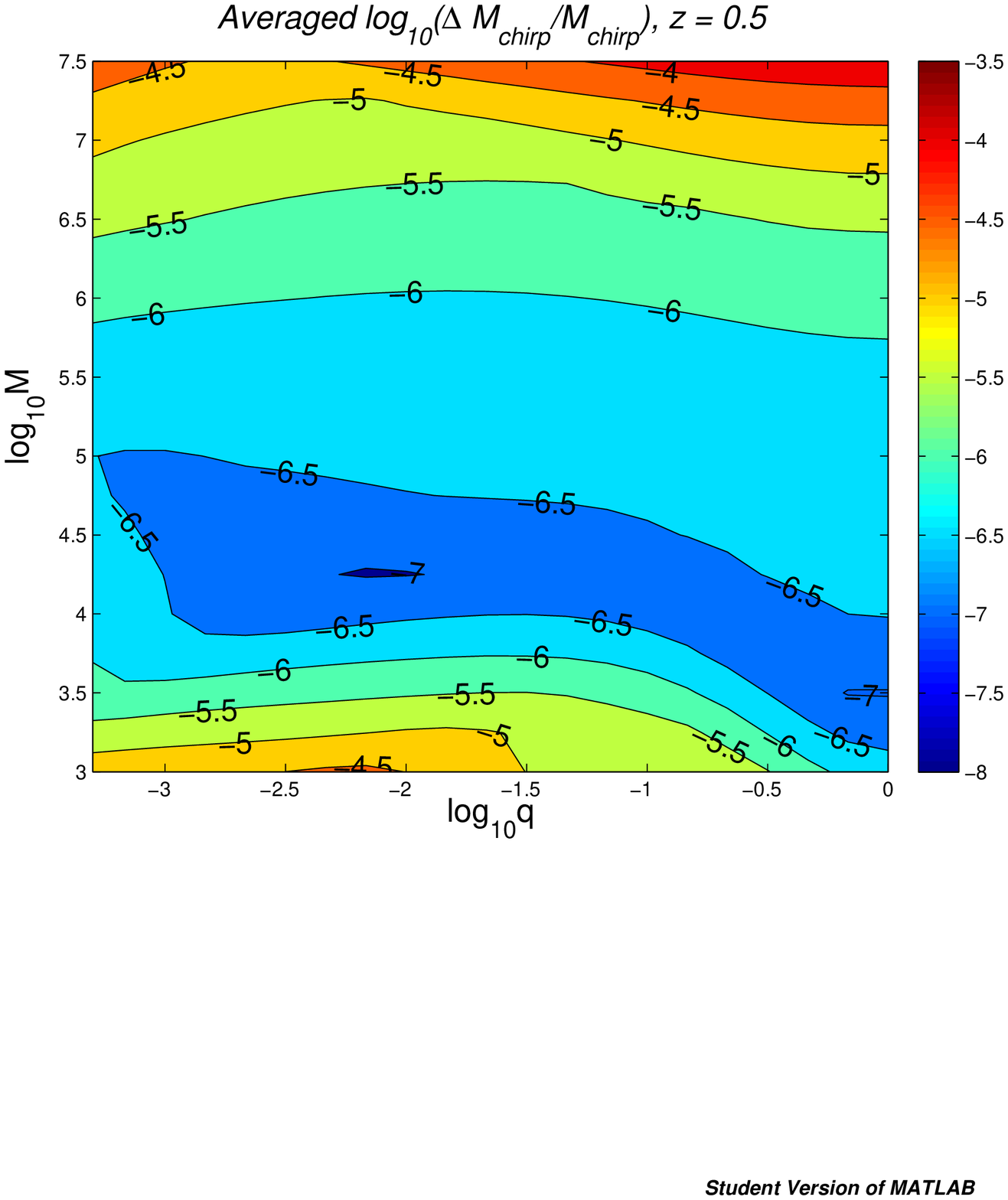}&
\includegraphics[scale=0.33,clip=true]{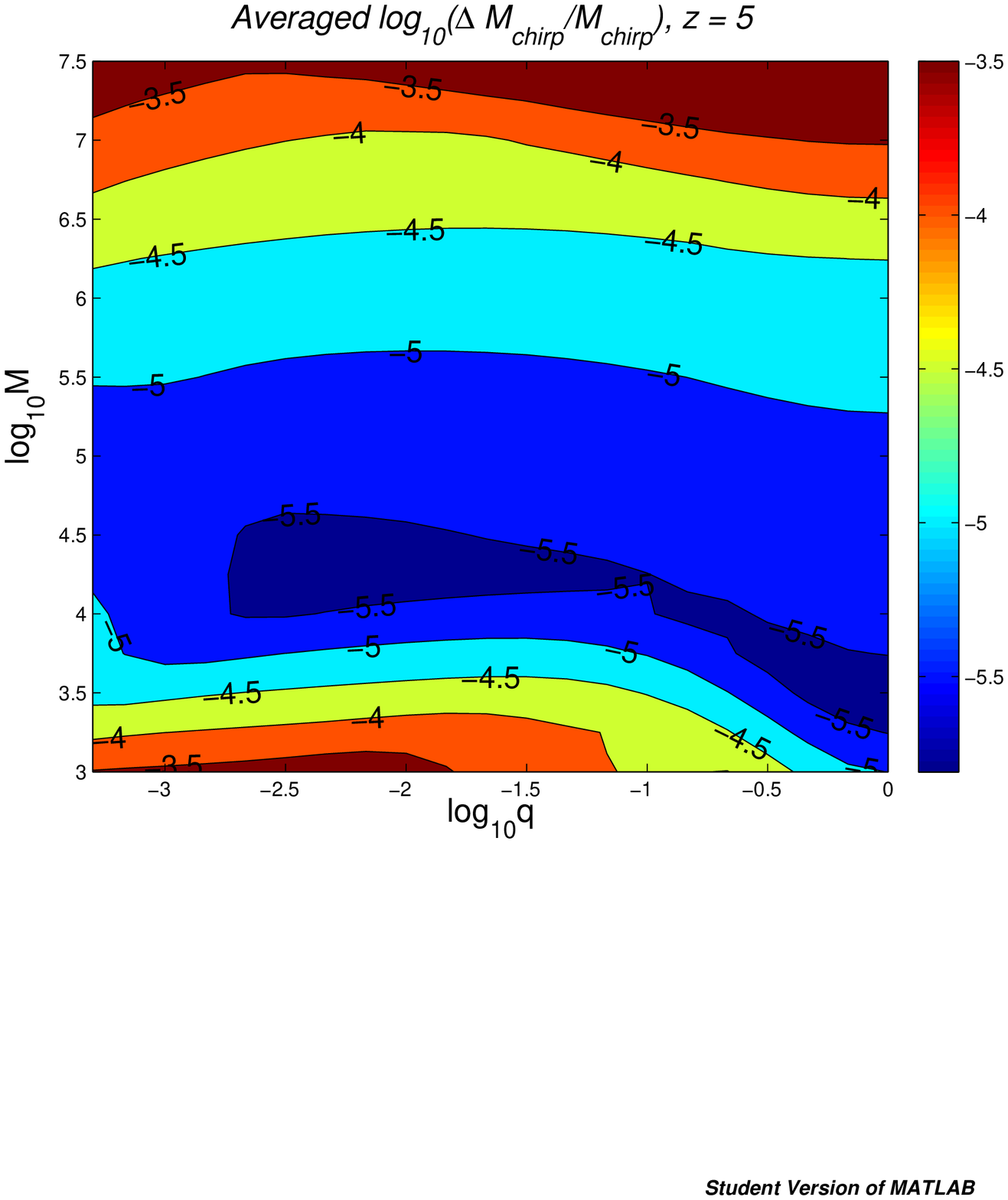}&
\includegraphics[scale=0.33,clip=true]{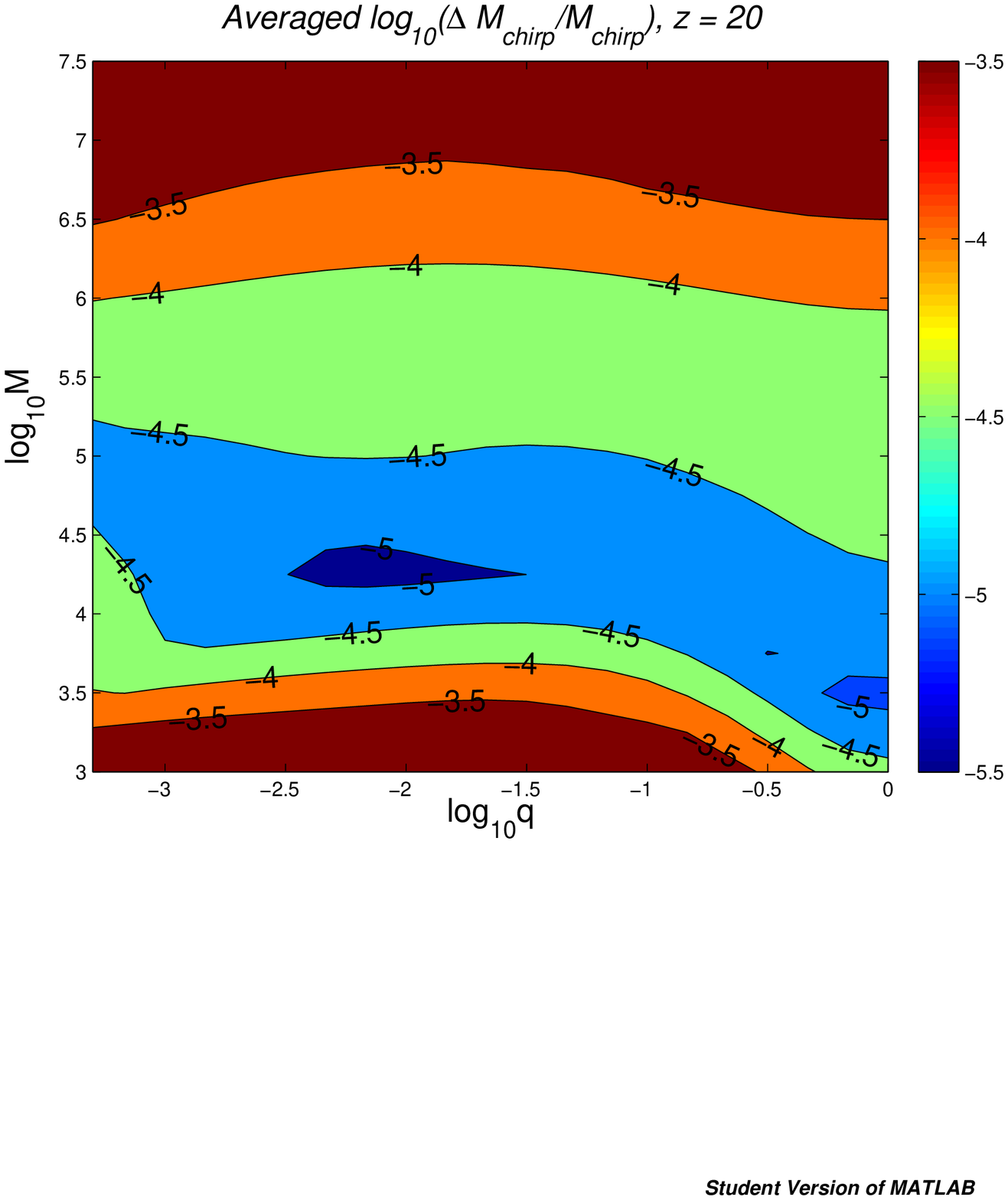}\\
\includegraphics[scale=0.33,clip=true]{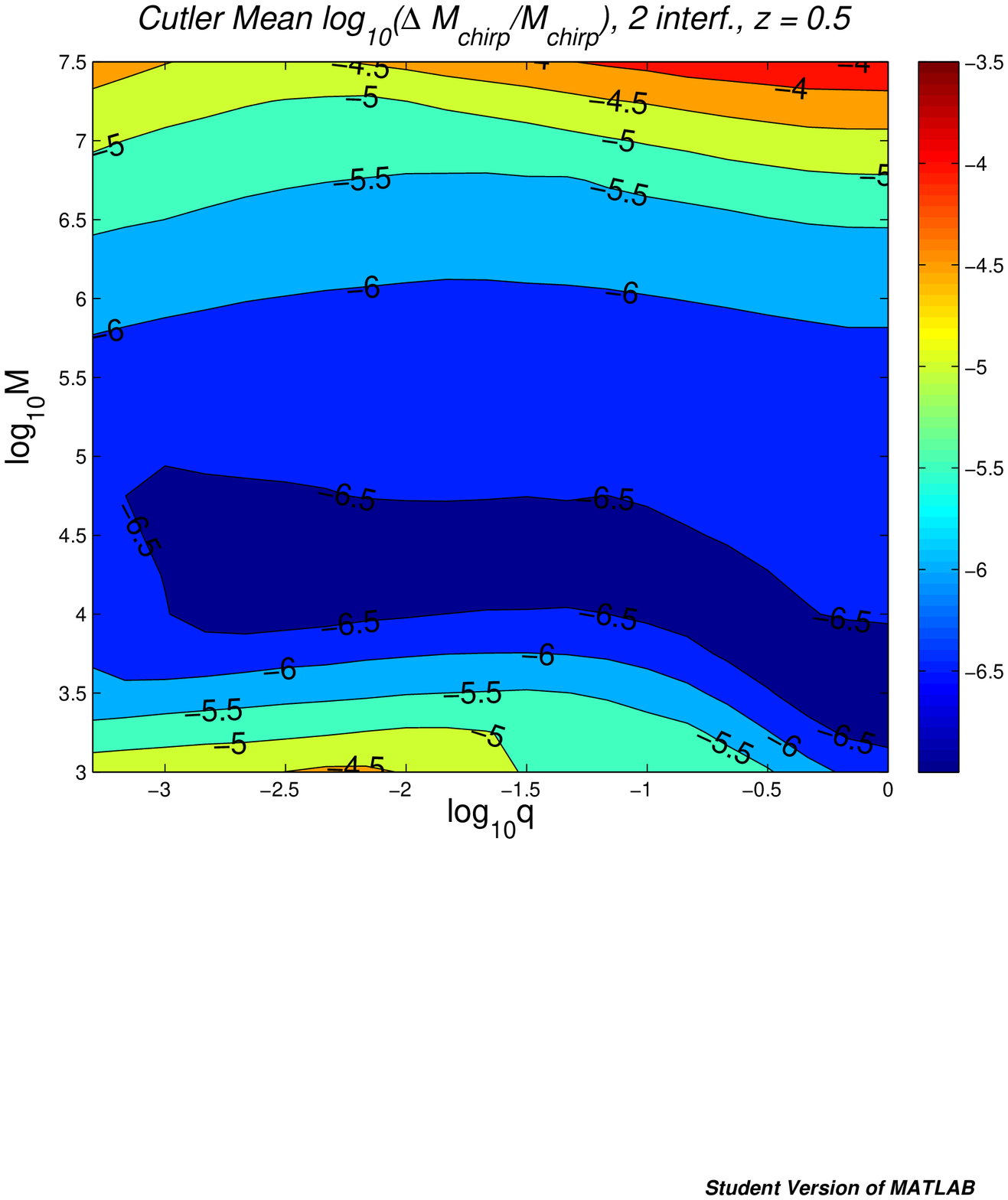}&
\includegraphics[scale=0.33,clip=true]{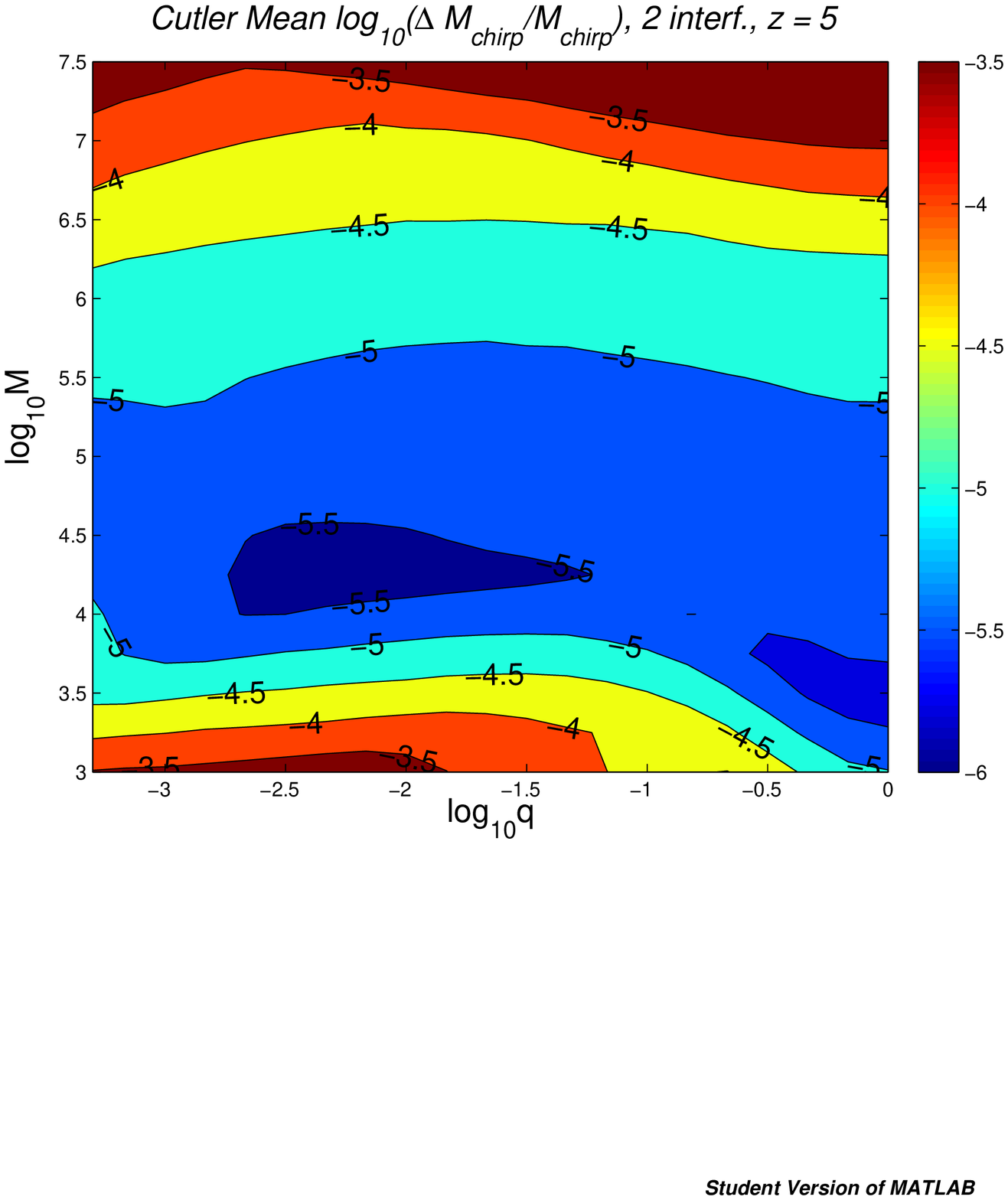}&
\includegraphics[scale=0.33,clip=true]{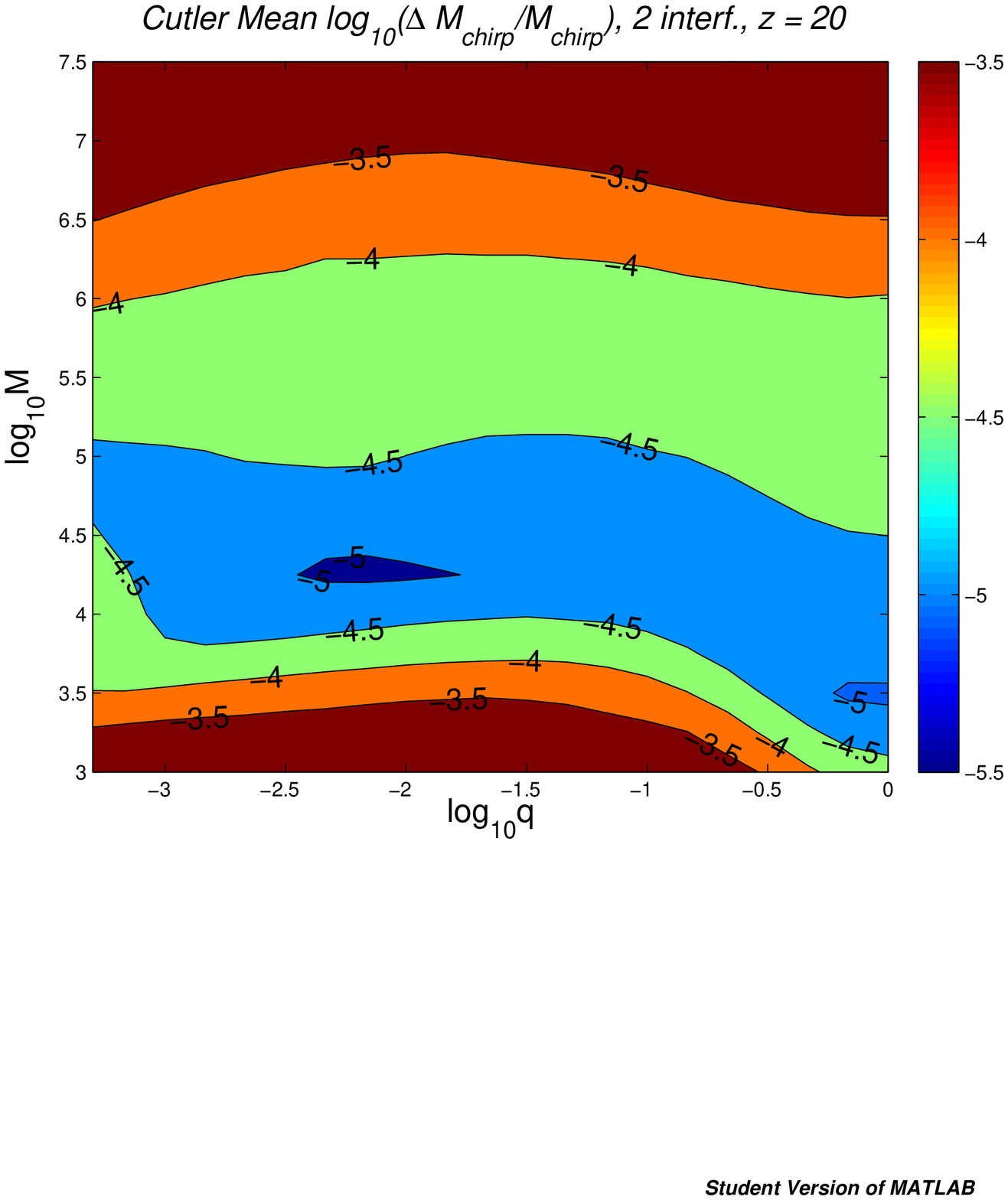}\\
\end{tabular}
\caption{Contours of the error on the chirp mass in the $(M\,,q)$ plane at
  different redshifts (from left to right: $z=0.5\,,5$ and $20$). The top row
  refers to the averaged code, the bottom row to the Cutler code with $T_{\rm
    obs}=3$~yrs and two interferometers.\label{fig:Mc} }
\end{figure*}

In this paper we tried to provide conservative estimates of the astrophysical
potential of LISA through observing MBH binaries. For this reason we only
considered MBHB inspiral waveforms in the restricted post-Newtonian
approximation. The choice of simple, frequency-domain waveforms has the
advantage that it significantly speeds up parameter estimation calculations:
we can easily compute the SNR and parameter estimation accuracy of $\sim 10^6$
binaries in one day on a single processor, something that would be impossible
if we used time-domain waveforms including spin precession. Computational
requirements were indeed a limiting factor in the analysis by Plowman et
al. \cite{plowman10}, who used an advanced parameter estimation code developed
by the Montana group \cite{Arun:2008zn}.

In Ref.~\cite{Berti:2004bd}, the potential of LISA to estimate binary
parameters was assessed using two independent formalisms. In one case (``non
angle-averaged'') the motion of LISA was taken into account using the
formalism developed by Cutler \cite{Cutler:1997ta}; in the other case
(``angle-averaged'') the authors averaged over the LISA beam-pattern
functions. The angle-averaging procedure removes all information related to
the Doppler shift due to the motion of LISA around the Sun, so in the
angle-averaged formalism it is impossible to estimate the distance and angular
location of the source in the sky. However, it is still possible to obtain an
``angle-averaged'' estimate of the SNR and of the intrinsic parameters of the
source (for our nonspinning binary model there are only two of them: $M$ and
$q$, or equivalently, ${\cal M}$ and $\eta$).

In summary, there are two ways of assessing the parameter estimation
capabilities of LISA. In the first method we angle-average over pattern
functions, {\it then} we estimate the SNR and the accuracy in determining
(say) ${\cal M}$ and $\eta$. In the second method we perform a Monte Carlo
sampling of the pattern function (by assuming that the source location and
angular orientation in the sky are isotropically distributed). In this way we
can estimate the SNR of each source, as well as the accuracy in determining
(say) ${\cal M}$, $\eta$, the luminosity distance $D_L$ and the angular
position of the source $\Delta \Omega$.

In this Appendix we address the question: when these two procedures can be
compared at all (i.e., in the case of ${\cal M}$, $\eta$ and the SNR), how are
they related? If the angle-average over pattern functions provides a sensible
estimate of SNRs and of the intrinsic binary parameters, it could provide a
significant saving in terms of computational time for future studies of MBH
populations.

In Figure~\ref{fig:SNR} we show contour plots of the SNR in the $(M\,,q)$
plane at selected values of the redshift, for both the averaged and the
non-averaged cases. This plot is encouraging: it shows that the shape of these
contour plots is essentially identical. Indeed, a more careful analysis
reveals that the ratio between the averaged and non-averaged SNRs is SNR$_{\rm
  A}$/SNR$_{\rm NA}\simeq 1.3$ and it is (to a very good approximation)
independent of $(M\,,q\,,z)$.

Fisher matrix calculations are inaccurate when the SNR is not much larger than
unity (see e.g.~\cite{Vallisneri:2007ev}).  Figure~\ref{fig:SNR} can be used
to identify regions where the SNR becomes so small that the Fisher matrix
approach will fail, and other parameter estimation techniques (such as
Markov Chain Monte Carlo) will become necessary. For example, by looking at
the contour line with SNR $\rho=8$ we see that Markov Chain Monte Carlo
calculations will be necessary to estimate the parameter estimation errors for
mergers of low-mass black holes at high redshift. In this context, recall that
$M=(1+z)M_r$, so (for large redshifts) the total mass in the source frame
$M_r$ is {\it smaller} than the mass $M$ appearing on the $y$ axis of the
contour plots.

Can we find more empirical relations between the pattern-averaged formalism
and the Cutler approach? Figure \ref{fig:Mc} shows contour plots of the chirp
mass determination accuracy in the two formalisms. Once again, we see that
there {\it is} indeed an approximate proportionality between mass estimation
accuracies in the two approaches. The ratios of the chirp mass errors show
small random fluctuations consistent with the Poisson noise in the $10^3$
Monte Carlo realizations at each point, but it is clear that the
angle-averaged approach does a good job at predicting the SNR and the
intrinsic parameter errors. This is true at any redshift.
Indeed, we find that ratios in the errors on ${\cal M}$ and $\eta$ are pretty
much redshift independent, and they show a very mild variation (in the range
$\sim 1-1.2$) in the $(M\,,q)$ plane.
If a similar proportionality applies also to estimates of the MBH spins, the
pattern-averaging may turn out to be a very useful simplication for future MBH
population studies.

\begin{figure*}[htb]
\begin{tabular}{ccc}
\includegraphics[scale=0.33,clip=true]{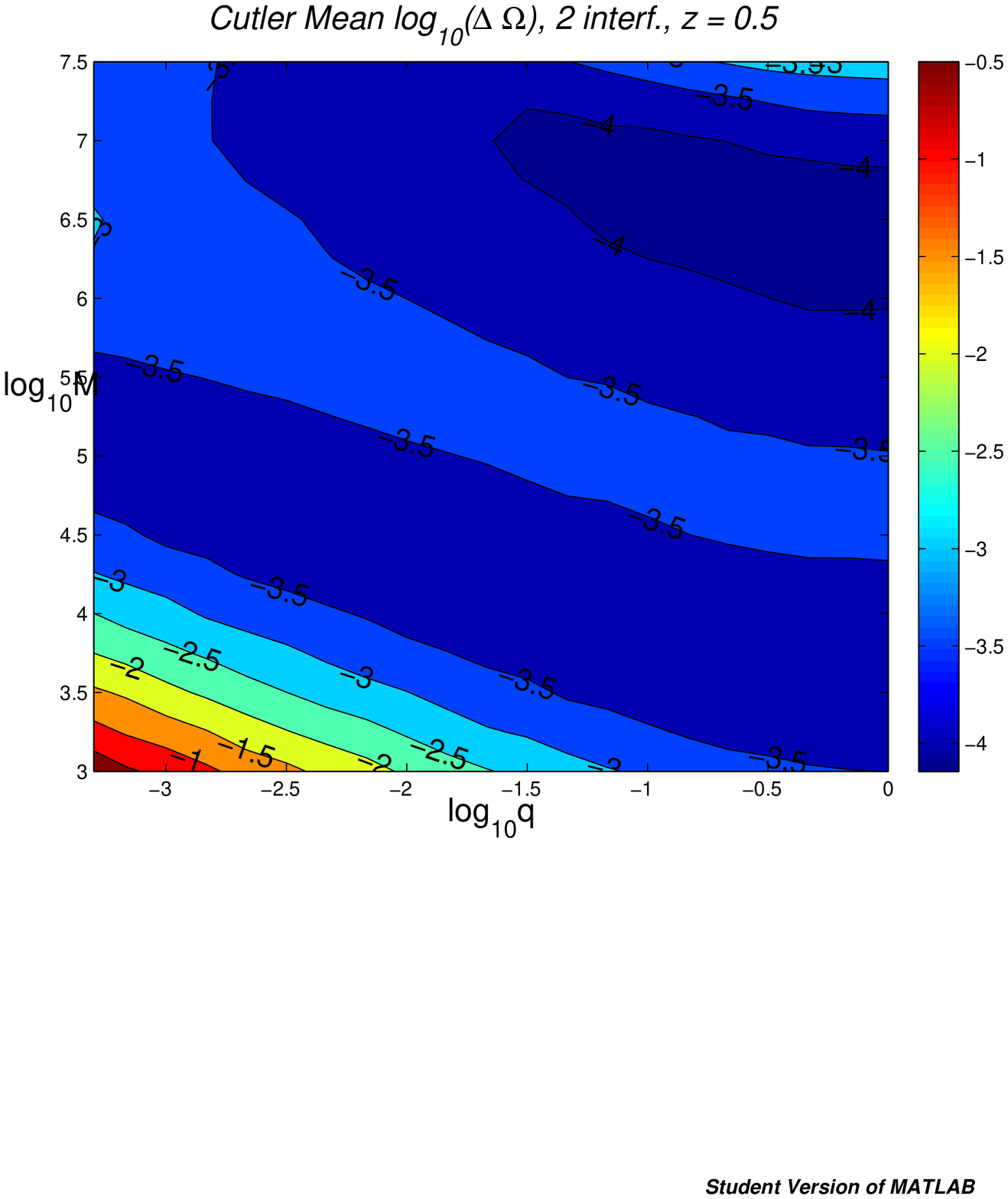}&
\includegraphics[scale=0.33,clip=true]{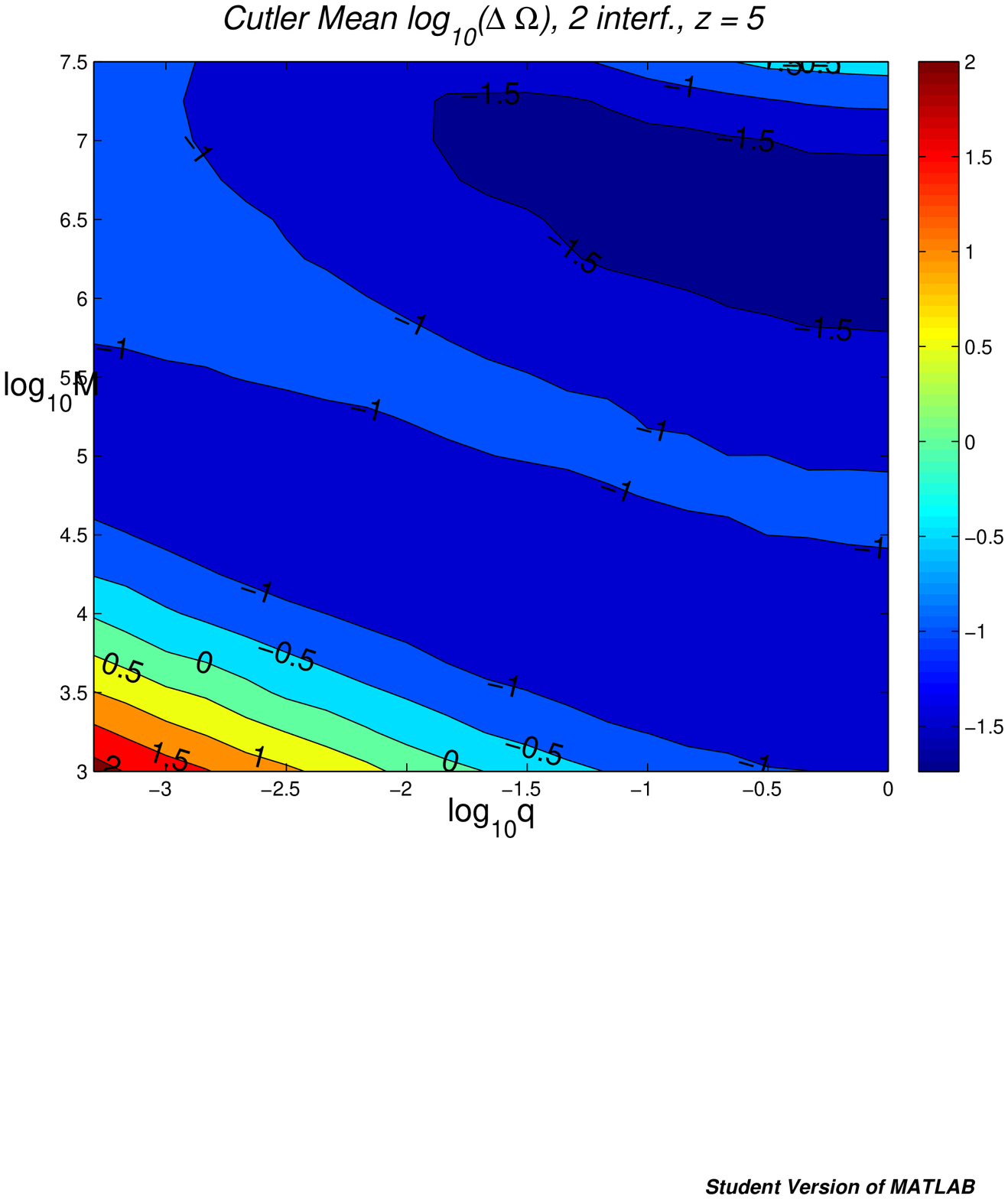}&
\includegraphics[scale=0.33,clip=true]{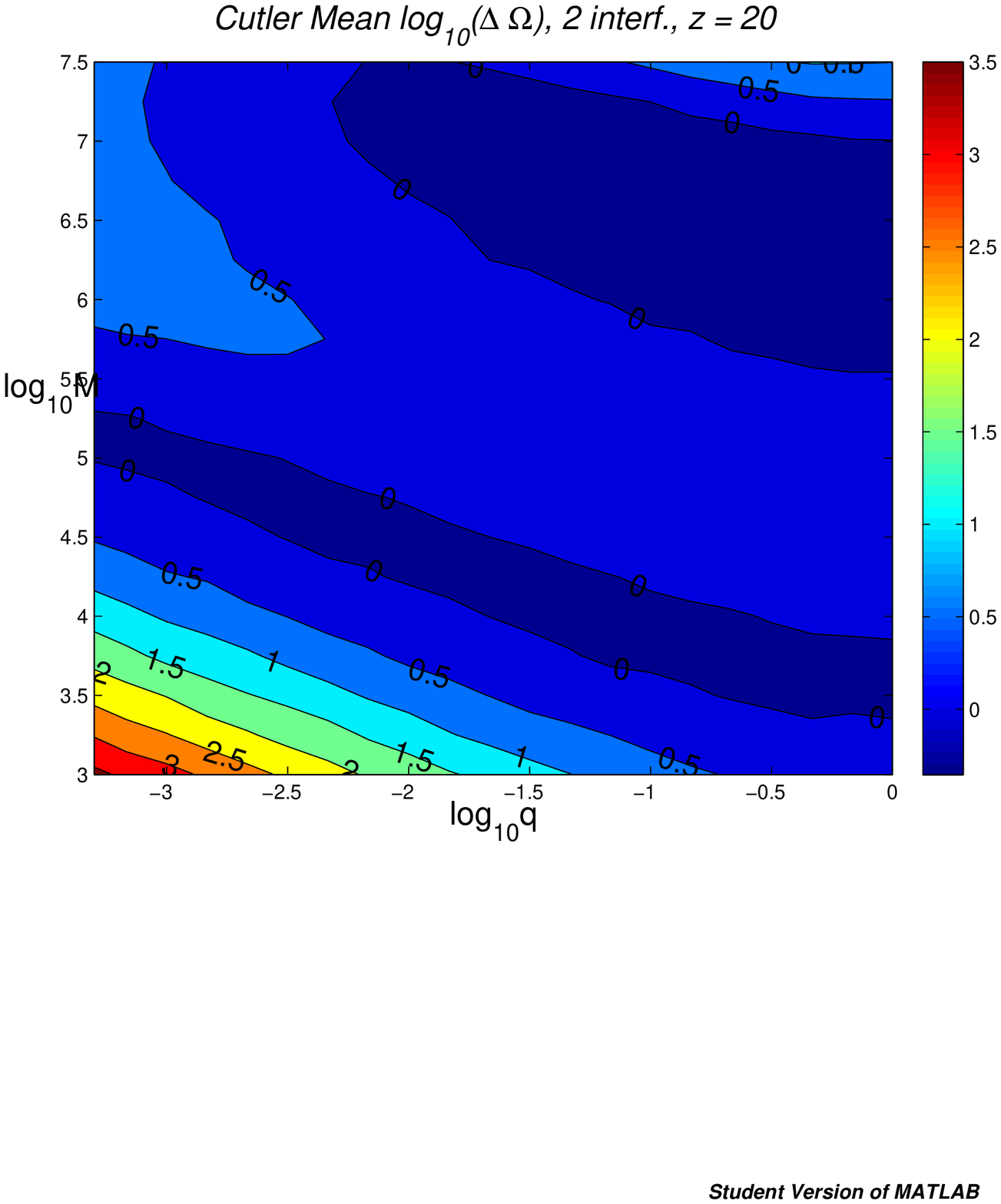}
\end{tabular}
\caption{Contours of the angular resolution error in the $(M\,,q)$ plane at
  different redshifts (from left to right and from top to bottom:
  $z=0.5\,,2\,,20$). The calculation uses Cutler's formalism with $T_{\rm
    obs}=3$~yrs and two interferometers.\label{fig:dOm} }
\end{figure*}

To finish, we will point out an interesting trend in the expected accuracy of
angular resolution.  We are not aware of systematic calculations of the
angular resolution in the three dimensional $(M\,,q\,,z)$ parameter space, so
we present such a study in Figure \ref{fig:dOm}. Angular resolution degrades
with redshift, as expected. The plot shows that, for any given redshift, the
angular resolution accuracy has a bimodal distribution -- i.e., there are {\it
  two} islands of good angular resolution accuracy in the $(M\,,q)$ plane. In
hindsight, this is not too surprising: the ``lower'' island corresponds to
low-mass binaries from which the GW emission is in the optimal sensitivity
bucket of LISA; the ``upper'' island correspond to higher-mass binaries that
merge at lower frequency, but have SNR large enough to compensate for the
relatively small number of cycles spent in band. It will be interesting to
verify if such a bimodal distribution persists when the merger/ringdown signal
is also included in the analysis.

\bibliographystyle{h-physrev4}

\bibliography{qnmcqg}

\end{document}